\documentclass[twocolumn, times, tighten,fleqn,usenatbib,useAMS]{aastex62}
\usepackage{graphicx}
\usepackage{xcolor}
\usepackage{mathtools}
\usepackage[version=4]{mhchem}
\usepackage{amsmath}

\def \degr         {\text{$^{\circ}$}}
\def \arcmin      {\text{$^\prime$}}
\def \arcsec      {\text{$^{\prime\prime}$}}

\DeclareOldFontCommand{\rm}{\normalfont\rmfamily}{\mathrm}
\submitjournal{ApJSupplSer}

\shorttitle{Giant radio sources} \shortauthors{Bhukta et al.}

\begin{document}

\title{Discovery of giant radio sources from TGSS ADR 1: radio, optical, and infrared properties}

\correspondingauthor{Sabyasachi Pal}

\author{Netai Bhukta} 
\affil{Department of Physics, Sidho Kanho Birsha University, Ranchi Road, Purulia, 723104, India}

\author{Souvik Manik}
\affil{Midnapore City College, Kuturiya, Bhadutala, Paschim Medinipur, West Bengal, 721129, India}

\author{Sabyasachi Pal}
\affil{Midnapore City College, Kuturiya, Bhadutala, Paschim Medinipur, West Bengal, 721129, India}
\email{sabya.pal@gmail.com}

\author{Sushanta K. Mondal} 
\affil{Department of Physics, Sidho Kanho Birsha University, Ranchi Road, Purulia, 723104, India}

\begin{abstract}
Giant radio sources (GRSs) are the single largest astrophysical objects known in the universe that have grown to megaparsec scales ($\ge$ 0.7 Mpc). GRSs are much rarer compared with normal-sized radio galaxies. Still, the reason for the formation of their gigantic sizes is under debate. We systematically search for GRSs from the TIFR GMRT Sky Survey Alternative Data Release 1 (TGSS ADR 1) at 150 MHz. We have newly identified 34 GRSs from this study. We have also studied the multi-wavelength properties (radio, optical, and infrared) of these GRSs. We have used the likelihood ratio method to identify highly reliable multi-wavelength counterparts of GRSs from Pan-STARRS (optical) and WISE (mid-IR) data. We have classified GRSs based on their accretion mode of the central black holes using optical and mid-IR data. For all sources, we also discuss the principal characteristic parameters (redshift distribution, angular and projected linear size, total integrated radio flux density, spectral index, and radio power). We show the radio evolution track and the location of the GRSs in the P-D diagram. Using a radio-optical luminosity diagram, we identify GRSs in the Fanaroff-Riley classification. Only two GRGs in our sample reside close to the centres of galaxy clusters.
\end{abstract}

\keywords{Active galactic nuclei; Catalogs; Jets; Quasars; Radio continuum emission; Surveys}

\section{Introduction} 
Radio-loud active galactic nuclei (AGN) with projected linear size greater than 0.7 Mpc are traditionally known as GRSs \citep{Wi74, Is99}. Typically, GRSs are hosted by galaxies and quasars known as giant radio galaxies (GRGs) and giant radio quasars (GRQs), respectively. In the universe, GRSs are among the longest-singular formations of radio sources (RSs, \citealt{Ro97, Ba06}). The radio emission from radio galaxies (RGs) is primarily caused by synchrotron radiation \citep{Sh55, Bu56}. Various studies claimed that the core engine of RGs is a mass-accreting supermassive black hole (SMBH) of mass $10^{8}-10^{10} M_{\odot}$ \citep{Ly69} which causes the collimated, bi-directional relativistic jets to be ejected perpendicular to an accretion disc \citep{Ly69, Be84}. In RGs, jets are narrow, well-collimated beams of relativistic particles that originate from the central region and are elongated towards the outer components. The beams transfer energy from the core to the outer components, known as lobes. Morphologically, RGs have been traditionally classified into two classes: the Fanaroff-Riley I (FR-I) class and the Fanaroff-Riley II (FR-II) class \citep{Fa74}.  RGs are also categorized based on their radio luminosity ($L_{178 ~\textrm{MHz}}$). Galaxies with $L_{178 ~\textrm{MHz}} \lesssim 2\times$10$^{25}$ W Hz$^{-1}$ sr$^{-1}$ are designated as FR I, characterized by lower radio power, a prominent core, and a diffused jet traversing the intergalactic medium. On the other hand, those with higher luminosities at 178 MHz fall into the FR-II  category, distinguished by their higher radio power and relativistic jets extending from the central AGN to the endpoints of the hotspots in the lobes \citep{Fa74}. Similar to regular RGs, GRSs are typically hosted by bright elliptical galaxies. However, massive rapidly rotating spiral galaxies sometimes host twin relativistic jets and lobes, which may be extended up to Mpc scales \citep{Ho11, Ba14}.

GRS is a relatively uncommon sub-group of RGs. 3C 236 was the first GRG, reported in 1974 by \citet{Wi74}. At that time the Hubble constant was assumed to be $H_{0}$ = 50 km s$^{-1}$Mpc$^{-1}$ which resulted in the physical linear size of GRG beyond 1 Mpc. The first two GRQs (1146--037, 1429+160) were discovered in 1983 \citep{Hi83}. The projected linear size of 1146--037 was 1.06 Mpc, at a redshift of $z=0.341$ and the projected linear size of 1429+160 was 1.37 Mpc at a redshift of $z=1.016$.  Many of the RSs were classified as GRSs based on their linear sizes which were calculated using the Hubble constant ($H_{0}$) known at that time (values between 50 to 100 km s$^{-1}$Mpc$^{-1}$). As a result, the linear sizes of these sources were over or under-estimated, and eventually, this led to improper statistical results of their population.

With the development of precision of cosmology evolved from the cosmic microwave background radiation seen with the Wilkinson Microwave Anisotropy Probe (WMAP; \citealt{Hin13}) and Planck mission \citep{Ad16}, the value of $H_{0}$ was set to $\sim$ 68 km s$^{-1}$Mpc$^{-1}$. Recent research \citep{Dab17, Da20b, Urs18, Bh22c} has accepted 0.7 Mpc as the lowest linear size limit of GRSs with the updated $H_{0}$ value obtained from observations. The total number of GRSs known to date is comparatively small with respect to radio sources (RSs), despite the fact that their number has significantly expanded during the past seven years \citep{Dab17, Ku18, Da20a, Da20b, Da20c, Ku21, An22, Ma22, Si22, Da23}. The longest known GRG to date is J081421.68+522410.0 in the LOw-Frequency ARray (LOFAR) Two-metre Sky Survey second data release (LoTSS DR2) field at 144 MHz, which extends 4.69 Mpc \citep{Oe22}.

Numerous studies were conducted to determine the cause of long radio jets of these sources.
The proposed hypotheses are:

(1) The dynamical ages of GRSs indicate that these sources have evolved over a long time \citep{Ka97}. GRGs are extremely old RGs that have had the time to spread out over vast distances. However, \citet{Ma98} determined the spectral ages of GRGs that are comparable with the RGs of normal sizes.

(2) Numerous authors have also looked into the nuclear host power contribution to the formation of these gigantic sources. Moreover, GRSs have significantly powerful radio jets when compared with normal radio sources that provide the necessary push to extend the projected linear size in Mpc scales \citep{Wi89}. But, \citet{Is99} showed that the GRGs have identical central engine strengths compared to the other normal-sized RGs, with similar radio luminosity.

(3) The radio lobes of GRS go far away from the host galaxy under the impact of the duty cycle of activity in radio wavelengths, radio galaxy evolution, and the influence of the surrounding inter-galactic medium (IGM). The GRGs are thought to represent the final stage of RG evolution because they are known to grow to gigantic sizes. Their study may enable us to offer significant restrictions on many evolutionary models of RGs.

(4) Additionally, it is notable that several GRS exhibit recurrent jet activity \citep{Kon13}. It is still not well understood the mechanism of the stopping and restarting of the jets of RGs.

(5) \citet{Ma08} assumed that a combined effect of jet speeds of about 0.1$c$ -- 0.2$c$ and low-density environment help RGs to grow to the large structure of GRG.

The availability of large-area and deep radio surveys, such as Sydney University Molonglo Sky Survey (SUMSS; \citealt{Bo99}), Westerbork Northern Sky Survey (WENSS; \citealt{Re97}), NRAO VLA Sky Survey (NVSS; \citealt{Co98}) and Faint Images of the Radio Sky at Twenty-cm (FIRST; \citealt{Be95}) helps to detect a significant number of GRSs.  However, the detection of GRS is not very common due to their low-brightness lobes. Using the above-mentioned surveys, many authors conducted scientific studies of GRSs. Different previous studies found that various radio properties of GRGs and RGs are identical in nature \citep{Su96, Sa96, Ma98, Da20a, Da20b, Da20c}. \citet{Ka97a} suggested the significance of self-similarity in the evolution of GRS. However, many researchers found that the radio spectra of GRGs are steep (0.75 $<\alpha<$ 1.2; $\alpha:  S_{\nu} \propto \nu^{-\alpha}$) which is due to diffuse lobe components.  Corresponding dynamical ages of lobes of GRGs are 10--100 Myr \citep{Sc00a, La00, Ma09}. \citet{Su08} and \citet{Sa09}  claimed that GRGs of such enormous size are most probably guided by the inhomogeneities of the underdense intergalactic medium. At high redshift ($z$), large extendable GRGs are not easily detected at high frequency. The lobes of GRGs usually have steep radio spectrum, making them easier to detect at low radio frequencies. Detection of GRGs has been reported using four large low-frequency surveys in recent years: (1) 119--158 MHz Multifrequency Snapshot Sky Survey (MSSS; \citealt{Hea15}), (2) 150 MHz TIFR GMRT Sky Survey- Alternative data release-1 (TGSS-ADR 1; \citealt{In17}), (3) 72--231 MHz Galactic and Extragalactic All-sky Murchison Widefield Array survey (GLEAM; \citealt{Hu17}), (4) 120--168 MHz LOFAR Two-metre Sky Survey (LoTSS; \citealt{Sc17, Sc17b}). 

\begin{figure*}
	\includegraphics[width=17cm,angle=0,origin=c]{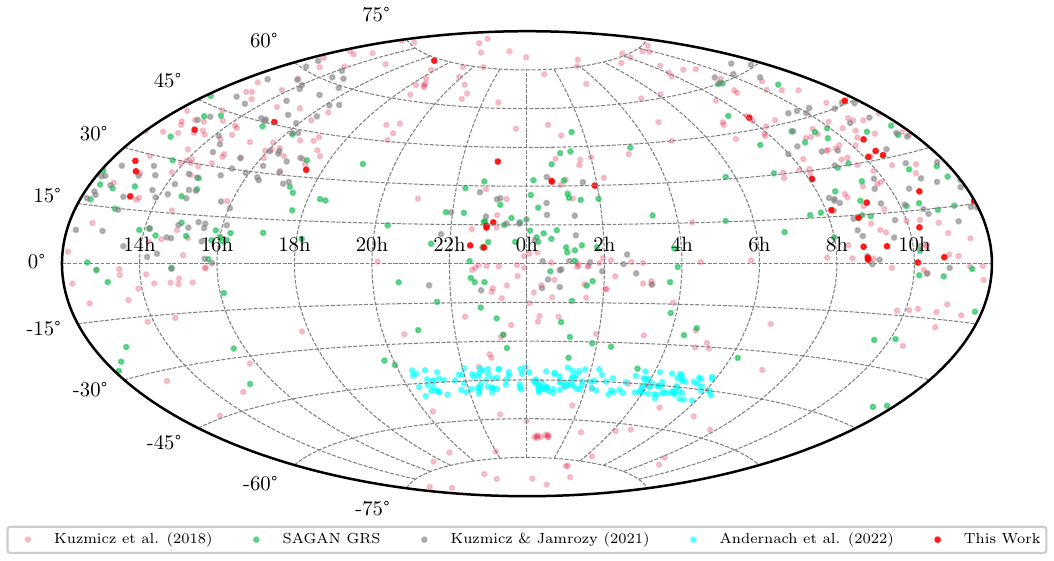}
	\caption{All-sky projection of the distribution of newly discovered GRS is shown with red circles. We also added GRS from four large samples, namely \citet{Da20c} (in green), \citet{An22} (in cyan), \citet{Ku21} (in grey) and \citet{Ku18} (in crimson).}
	\label{fig:sky-dist}
\end{figure*}

In this paper, we present 34 new GRSs and their principal characteristics from the TIFR GMRT Sky Survey (TGSS). We arrange the paper in the following ways: In section \ref{sec:data} and \ref{sub:counterpart}, we present the method for the identification of sources and their host galaxies. In the next section (section \ref{sec:result}), we describe different radio properties of the detected sources. In section \ref{sec:disc}, we discuss the general features and overall properties of sources. We summarize the study in the final section (section \ref{sec:conclusion}). We assumed the flat $\Lambda CDM$ cosmology with $H_0=67.8 \, \rm km \, s^{-1} \,Mpc^{-1}$, $\Omega_{\rm M}=0.308$, and $\Omega_\Lambda=0.692$ \citep{Ag18}.

\section{Identification of giant radio sources}
\label{sec:data}
\subsection{The TGSS alternative data release one}
For the search of GRGs, we used the TGSS first alternative data release (ADR 1; \citealt{In17}) at 150 MHz. The TGSS ADR 1 includes continuum stokes I images of 99.5 percent of the radio sky north of --53$\degr$ DEC which is 90 percent of the total sky. The resolution of the survey is $25\arcsec \times 25\arcsec$ north of 19$\degr$ declination and $25\arcsec \times 25\arcsec / \cos(\textrm{DEC}-19\degr)$ south of 19$\degr$ declination. To improve uv coverage, each pointing was spread among 3--5 scans with nearly 15 minutes of exposure time on each pointing. The survey has a median noise of 3.5 mJy beam$^{-1}$. Taking advantage of high sensitivity and large survey area in low radio frequencies, earlier TGSS ADR 1 was used to search for faint high-redshift RGs \citep{Sa18}, winged radio galaxies \citep{Bh22a} and tailed radio galaxies \citep{Bh22b}.

\subsection{Search from TGSS ADR 1}
\label{subsection:search}
We have searched for GRGs and GRQs from TGSS ADR 1 radio maps at 150 MHz. After searching systematically, we discovered a sample of 34 GRS including 24 GRGs and 10 GRQs. All the GRSs in our sample are detected above 0.4 Jy. We selected these sources using the following methods:
	
\begin{enumerate}
\item We closely examined each of the 5536 TGSS ADR-1 images to identify large radio structures. To achieve this, we divided a single image field into 32 subfields and conducted a manual visual search (MVS). After the MVS, we compiled a list of 3752 potential double-lobe radio sources.

\item In the following process, we superimposed TGSS maps over r-band optical images from Panoramic Survey Telescope and Rapid Response System (Pan-STARRS; \citealt{Fe20}) and SDSS (where Pan-STARRS are unavailable) and high-angular resolution maps from the LoTSS DR2, FIRST and the Very Large Array Sky Survey (VLASS) \citep{La20} surveys in order to determine the positions of the core and hotspots. We visually examined whether the positions of optical hosts matched the LoTSS DR2/FIRST/VLASS radio core emissions. During this step, we measured the flux density of low-frequency 150 MHz from TGSS radio maps, as well as the 1.4 GHz flux density from NVSS radio maps. We also measured the angular size from FIRST/VLASS high-resolution images using the hotspot-hotspot method as stated in \citet{Ku21}. If the host section of the radio source is not visible in the radio frequency domain, then the geometrical center of the radio galaxy is detected by AGN colour. In this step, we were able to locate the radio cores of 1861 host galaxies. Almost all the sources have radio cores in the FIRST/VLASS survey catalogue separated by less than $1''$ from the optical galaxies.

\item After determining the core position from the radio-optical overlay, we employed the likelihood ratio method to find highly reliable multi-wavelength counterparts from optical and IR catalogues. Using the reliable spectroscopic and photometric redshift information available for the host galaxies, we converted the angular size of these sources to a linear projected size. We only considered sources with a size greater than or equal to 0.7 Mpc as GRSs. After this screening process, we found a total of 1425 sources with redshift information available, out of which only 673 sources met the criteria to be classified as GRSs. 

\item In the next step, we removed all previously discovered sources from our sample and studied the multi-wavelength properties of our newly discovered sample. After applying all the filtering criteria mentioned above, we discovered 34 new GRSs.

\end{enumerate}

The catalogue reveals that 2 out of 34 radio cores are not detected in high-frequency radio maps (FIRST/VLASS), but their possible cores were visually identified through multi-wavelength (radio, optical) overlay images and selected as highly reliable core counterparts using the LR method as described in Section \ref{sub:counterpart}. Eleven of our sources appeared in the LoTSS DR2 survey. We measured the flux density from LoTSS DR2 and compared it with TGSS-ADR1 flux densities. The mean ratio of the LoTSS-DR2 integrated flux densities to the TGSS-ADR1 integrated flux densities is ~0.9, which is close to the derived ratio at \citet{Sc17b}.

\section{Host galaxy identification}
\label{sub:counterpart}

This section concentrates on the identification of the optical and infrared counterparts of GRSs. We use a likelihood ratio (LR; \citealt{Ri75, de77, Su92, Ci03}) technique that is specifically useful when working with deep optical images to minimize the number of spurious associations. For all GRSs mentioned in the current article, we found a reliable counterpart. The LR technique helps us to take into account the position of the counterpart, background source magnitude distribution, and the presence of multiple possible counterparts for the same radio source. It is given by the relationship \citep{Su92}

\begin{equation}
LR=\frac{q(m)f(r)}{n(m)}
\end{equation}
where $n(m)$ is the surface density of background sources as a function of band magnitude $m$. The parameter $q(m)$ is {\it a priori} probability that the radio source has a counterpart of magnitude $m$. The parameter $f(r)$ represents the probability distribution of the offset $r$ between the catalogued positions of the radio source and its potential counterpart. LR is a measure used to determine the probability that a radio core of GRSs and their optical/IR counterparts are related or unrelated. It is employed to determine a more reliable optical/IR counterpart of the GRG host galaxies. It is therefore helpful to define the reliability of each association as:

\begin{equation}
\rho_{j}=\frac{(LR)_{j}}{\sum_{i}(LR)_{i}+(1-Q)}
\end{equation}
where the sum is over all the candidate counterparts for the same radio source \citep{Su92}.

It is essential to choose multi-wavelength data to determine the accurate counterpart of a radio source. Deep and wide optical and IR data are available over the TGSS-ADR 1 covered sky. We utilized two deep surveys, one optical survey using the Pan-STARRS and one infrared survey using the Wide-field Infrared Survey Explorer (WISE;  \citealt{Wr10}). Pan-STARRS1 \citep{ch17} has performed a set of independent synoptic imaging sky surveys, including the $3\pi$ steradian survey. There are five bands present in Pan-STARRS1 ($grizy$; 23.3, 23.2, 23.1, 22.3, 21.4 mag). The WISE is a mid-infrared survey of the entire sky in four infrared bands W1, W2, W3, and W4 in the 3.4, 4.6, 12, and 22 $\mu$m respectively.  The W1 and W2 bands have significantly better sensitivity than the other two WISE bands \citep{cu12, Cu13}. The completeness of the WISE catalogue varies across the sky. With W1 $<$ 19.8, W2 $<$ 19.0, W3 $<$ 16.67, and W4 $<$ 14.32 mag, the completeness is 95 percent for our GRS sample. 

We use the magnitude and colour in the LR method to cross-match the TGSS ADR 1 GRSs with the subsidiary WISE-Pan-STARRS combined catalogue. Sometimes, the optical host galaxy is not visible in the radio core position of radio sources due to various limitations of the survey. The AGN colour of the galaxy close to the geometrical centre of the GRSs strongly indicates the presence of the host galaxy.  
We built a list of possible counterparts for each GRS around geometrical centers from all-WISE in the $W1$ band and Pan-STARRS in the $i$ band. The LR ratios are then derived for all Pan-STARRS sources within 5$''$ of all WISE positions. Finally, we make a WISE-Pan-STARRS combined catalogue that should be used for counterpart identification for GRSs. We sort only those sources with a reliability greater than 0.8 (the threshold limit to ensure the expected number of spurious associations is 5 percent of the subsidiary catalogue). As a result, the number of identified optical/infrared counterparts increases. The choice of the best threshold value $LR_{\text{thr}}$ is necessary to differentiate spurious and real identifications of counterparts. Here, $LR_{\text{thr}}$ should be large enough to reduce the number of spurious identifications while increasing reliability. If more than one spurious counterpart exceeds those thresholds, the counterpart with the highest LR in either band is accepted, and the other is rejected. We found counterparts for all of our newly discovered GRSs in both the $i$ and W1 bands. In our sample, all the sources except one have reliability $\geq$0.9; only one source (J2231+0702) has reliability $\sim$ 0.83. After identifying the host galaxy, we searched for available redshifts (either spectroscopic or photometric) in multiple catalogues \citep{He91, We12, Bi14, Be16, We16, Co16, Ve18, Ji21, Dun22}.

\section{Results}
\label{sec:result}
The discovery of 24 new GRGs and 10 new GRQs from TGSS ADR 1 is reported. All the GRSs reported in the present article are situated in the Northern Hemisphere because for the double radio sources discovered in the Southern Hemisphere, either the probable hosts of the sources have no redshift information available or are already discovered previously, or they do not meet the GRSs criteria; it is also worth noting that there are fewer spectroscopic or photometric catalogues available in the Southern Hemisphere than in the Northern Hemisphere. In our analysis, we compare the radio features of the GRSs with earlier catalogues of GRSs and smaller radio sources (SRSs). For comparison, we used SRSs from \citep{Ku21, Ma22}. The distribution of GRSs is shown in Figure \ref{fig:sky-dist}. Table \ref{tab:GRG-GRQ} lists all newly discovered GRGs and GRQs. Pan-STARRS r-band optical images are overlayed with TGSS and FIRST images. For all images, contours are overplotted with ten levels spaced in log scale with the lowest level of 3$\sigma$. Overlaid images of all GRGs and GRQs are presented in Figure \ref{Fig:GRS}.

\subsection{Optical properties}
\label{sec:optical properties}
The active phase of SMBH is known as AGN, whose activities can be seen at almost all frequencies (radio to gamma-rays). The understanding of the energy spectrum of AGN is unclear over a long period. In the generally adopted model, energy is produced by the accretion of cold matter onto the SMBH near the active center of the galaxy. AGNs that primarily radiate at radio frequencies are referred to as radio-loud AGNs (RLAGNs). A smaller fraction and more powerful class of AGNs are known as quasars, which are among the most energetic and luminous objects. Quasar mode AGNs, that exhibit radio emissions on either side of the central SMBH are known as radio-loud quasars (RQs; based on the radio-to-optical brightness ratio).
GRGs hosted by quasars are known as GRQs. Based on the different accretion modes of AGNs, RGs can be classified into two classes: (1) low-excitation radio galaxies (LERGs), (2) high-excitation radio galaxies (HERGs) \citep{La94}.

\citet{La94} suggested a clear difference between two classes (LERGs and HERGs) from the study of the optical properties of the radio-AGN 3CR sample. They studied the flux ratio ([OIII]5007/ H$\alpha>0.2$) and equivalent emission line width of [OIII] (EW) $>3\r{A}$ for HERGs. Similarly, \citet{Th98} found LERGs with EW$<10\r{A}$ in 2 Jy radio sources. Especially, \citet{Be12} reported EW of [OIII] $>5\r{A}$ and optical flux ratio ([OIII]5007/H$\alpha\ge 1$) for HERGs. When the information of host galaxies of GRGs is not available in SDSS, the excitation index (EI) can be used to classify them as either HERGs or LEGRs \citep{Bu10}.

\begin{scriptsize}
	\begin{equation}
	\mathrm{EI= \log\left(\frac{[OIII]}{H\beta}\right)-{\frac{1}{3}}\left[\log\left(\frac{[N II]}{H\alpha}\right)+\log\left(\frac{[S II]}{H\alpha}\right)+\log\left(\frac{[O I]}{H\alpha}\right)\right]}
	\label{equ:EI}
	\end{equation}
\end{scriptsize}

If EI $>$ 0.95, the host galaxies are classified as HERGs, otherwise as LERGs \citep{Bu10}. The SDSS DR16 spectroscopic observation data \citep{Ah20} are available for 17 GRGs in our sample. We use a Gaussian fit on the SDSS spectroscopic data for the measurement of emission-line flux and EW. For the cases of J0848+0115 and J1331+1353, the [NII] and [SII] lines are not available. We use the flux ratio method of classification from \citet{Be12}. The GRG J1315+3831 is classified as HERG by the emission width method \citep{Be12}. 

\begin{figure}
	\includegraphics[width=8cm,angle=0,origin=c]{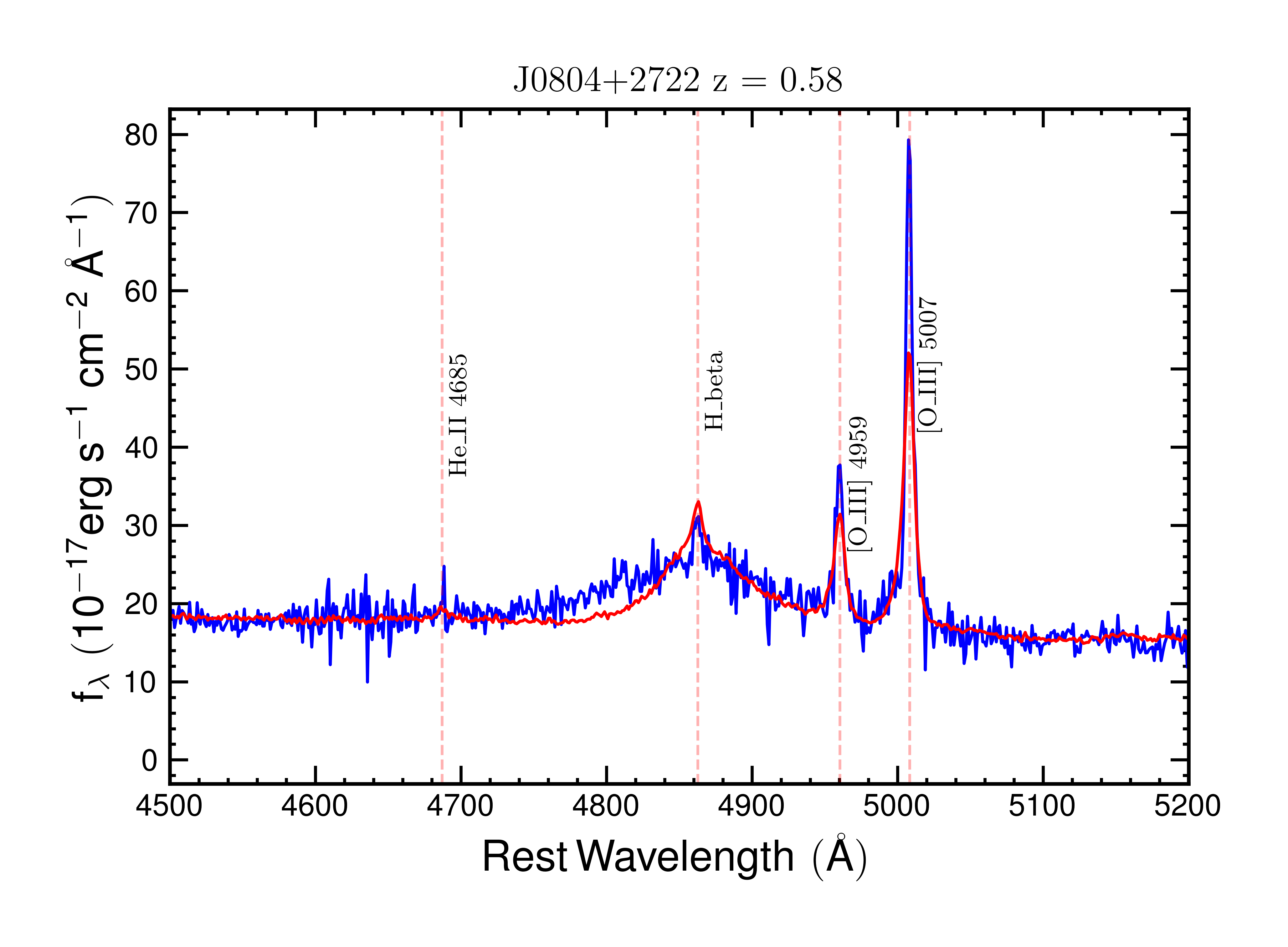}
	\caption{This figure shows a zoomed-in plot of the optical spectrum of the host of a GRQ (J0804+2722) obtained from SDSS. Gaussians are fitted to H$\beta$ and [OIII] doublet spectral lines. The broadness of H$\beta$ line is a clear indication of its quasar nature. The blue line represents the observed spectra, and the red line represents the best-fitted spectra.}
	\label{fig:GRQ-spec}
\end{figure}

\begin{figure}
	\includegraphics[width=8cm,angle=0,origin=c]{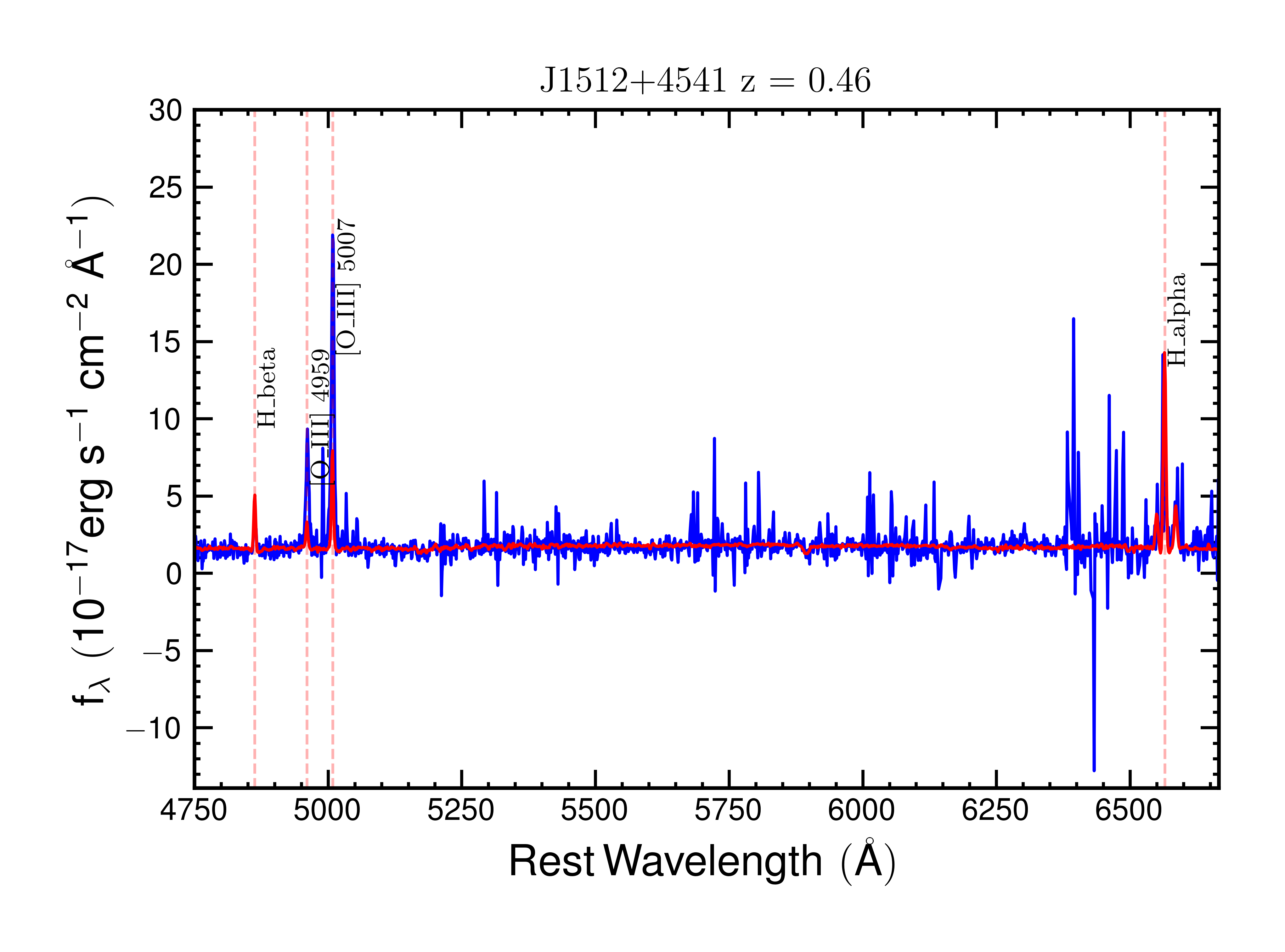}
	\caption{
 This plot depicts the optical spectrum of a HEGRG (J1512+4541) using SDSS. The observed spectra is represented by the blue line, whereas the best-fitted spectra is denoted by the red line. }
	\label{fig:GRG-spec}
\end{figure}

GRQs are known to be found at active/high excitation phases. We study the following optical properties of GRSs using SDSS DR16 \citep{Ah20}: (1) the different optical band magnitude, (2) clear doublet emission lines of oxygen at [OIII] and [OII], (3) broad H$\beta$ emission line. We have summarized the above-mentioned properties in Table \ref{tab:GRQ-optical}. We find that the optical u band magnitude is greater than other bands. Broad H$\beta$ emission line is a clear indication of GRQs as well as broad-line radio galaxies (BLRGs). Figure \ref{fig:GRQ-spec}  and Figure \ref{fig:GRG-spec} represent the SDSS spectrum of a GRQ and a GRG respectively. To obtain the emission widths of the H$\beta$ line and [OIII] doublet, Gaussians are fitted on the spectrum as can be seen in Figure \ref{fig:GRQ-spec} and \ref{fig:GRG-spec}.

\subsection{Mid-IR properties}
The mid-IR spectrum of host galaxies of GRSs is extremely useful for calculating radiative efficiency. The neighboring dusty torus absorbs the optical-ultraviolet radiation from the AGN accretion disc and it re-radiates at mid-IR bands. The Wide-field Infrared Survey Explorer (WISE) mid-IR magnitudes and colours \citep{Wr10} are used to analyze and identify different AGN types for GRGs and GRQs. WISE is an all-sky survey of four mid-IR bands [W1 (3.4 $\mu$m), W2 (4.6 $\mu$m), W3 (12 $\mu$m), W4 (22 $\mu$m)]. The angular resolutions of the corresponding bands are 6.1$''$, 6.4$''$, 6.5$''$, and 12$''$ respectively. RGs can be effectively classified into low-excitation radio galaxies (LERGs), high-excitation radio galaxies (HERGs), quasars, star-forming galaxies (SFGs), narrow line radio galaxies (NLRGs), and ultra-luminous infrared radio galaxies (ULIRGs) using WISE mid-IR colours \citep{St12, Gu14, Dab17, Da20b}. 

\begin{figure*}
	\includegraphics[scale=0.8]{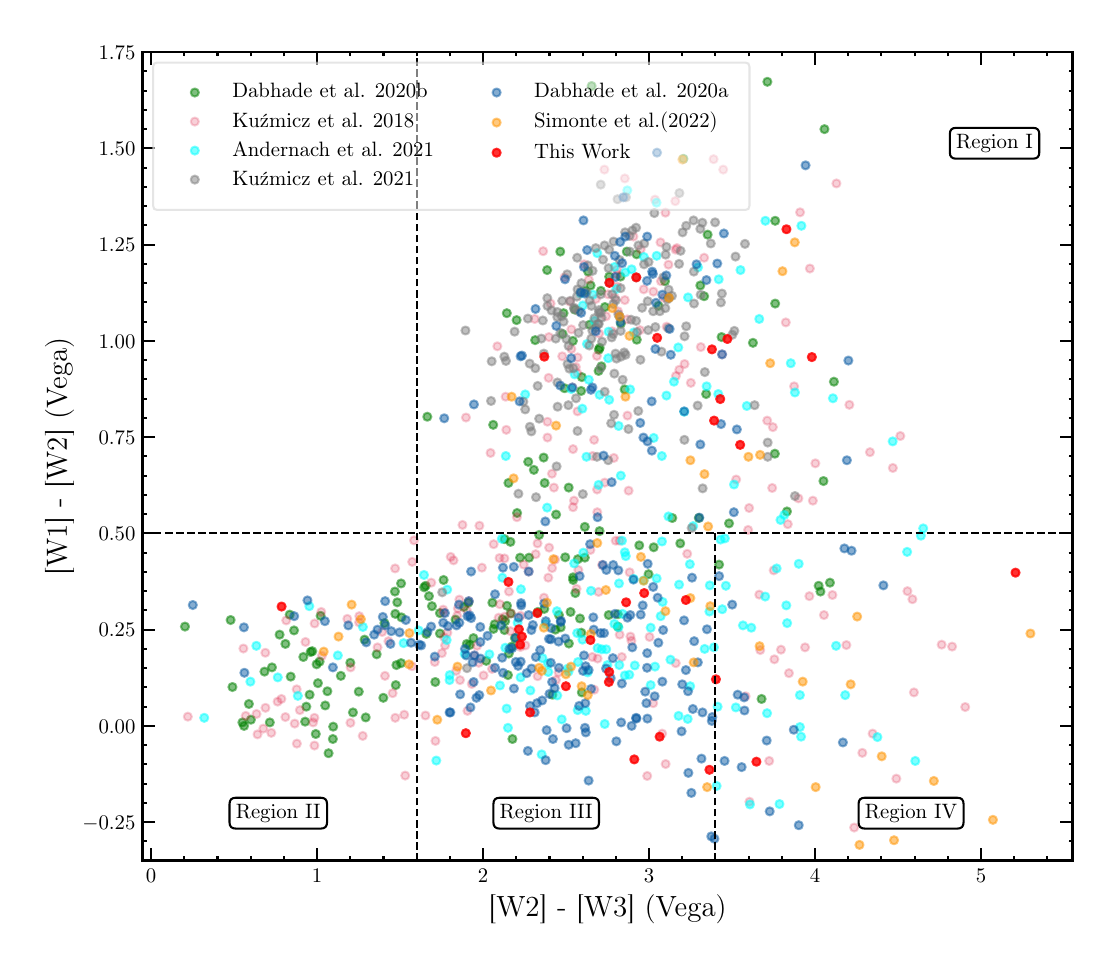}
	\caption{WISE colour-colour plot with the TGSS GRS sample is shown. This plot also includes 1007 sources from the previous GRG catalogues along with our newly discovered 34 GRSs. Here, 237 sources are from \citet{Ku18}, 137 sources are from \citet{Da20c}, 242 GRSs are from \citet{Da20b}, 165 are from \citet{An22}, 170 GRQs are from \citet{Ku21}, and 56 radio sources from \citet{Si22}.} 
	\label{fig:colour}
\end{figure*}

We classify the GRGs into HEGRGs, LEGRGs, and quasar categories using mid-IR colours. For HEGRGs and quasar, (W1--W2$>$0.5, W2--W3$<$ 5.1), and for LEGRGs (W1--W2$<$0.5, 0 $<$ W2--W3$<$1.6). We also compute the luminosities in the four respective bands and AGN class, which are shown in Table \ref{tab:IR-GRGs-GRQs}. Figure \ref{fig:colour} shows a colour-colour plot, where the y-axis is the difference in the magnitude of the W1 and W2 bands and the x-axis shows the difference in the magnitude of W2 and W3. It has been divided into four regions by vertical and horizontal lines. This plot also includes 1043 sources from the previous GRG catalogues along with our newly discovered 34 GRSs.  Region-I, quasars and HERGs ([W1]--[W2] $\ge$ 0.5 and [W2]--[W3] $<$ 5.1) is populated by 433 GRS. Region II, LERGs ([W1]--[W2] $<$ 0.5, 0 $<$ [W2]--[W3] $<$1.6), contains 121 sources. Region III, LERGs and star-forming galaxies ([W1]--[W2] $<$ 0.5, 1.6 $\le$[W2]--[W3] $<$ 3.4) are populated with 375 GRS. Region IV, ULIRGs ([W1]--[W2] $<$ 0.5, [W2]--[W3] $>$3.4) contains only 81 GRS. So, it can be concluded that more than 40 percent of GRS are in quasar mode and high-excitation radiative mode and 36 percent of sources are in either LERGs or star-forming galaxies.  

\subsection{Linear size}
All the GRSs in the current work show FR-II nature. We measured the largest angular size (LAS) from the high-resolution radio maps from VLASS \citep{La20} and FIRST \citep{Be95} (where VLASS data are not available) using the hotspot–hotspot method proposed by \citet{Ku21}. To measure the angular size, the position of the hotspot peak intensity is determined by fitting a 2D Gaussian function using CASA IMFIT.

The conventional physical projected linear size of GRG is greater than $0.7$ Mpc \citep{Dab17, Da20b, Urs18}. We compute the projected linear size of the selected GRSs using the following formula-

\begin{equation}
D({\textrm{Mpc}})=\frac{\Theta \times D_{\textrm{co}}}{(1+z)}\times\frac{\pi}{10800}
\end{equation}
where $D_{\textrm{co}}$ is the comoving distance of the galaxies in Mpc, $\Theta$ is the angular size of the radio galaxy in arcminutes, $z$ is the redshift of the host galaxy, and $D$ is the projected linear size \citep{Ke88}.

The uncertainty in linear size estimations due to redshift error and angular size measurement uncertainties is
	\begin{equation}
	\Delta D_{\textrm{Mpc}}=\sqrt{\left(\frac{D_{\textrm{co}}}{1+z}\Delta \Theta \right)^{2}+\left(-\frac{\Theta~ D_{\textrm{co}}}{(1+z)^2} \Delta z\right)^{2}}
	\end{equation}
	where $\Delta \Theta$ is the error in LAS measurement, $\Delta z$ is the error in redshift and $\Delta D_{\textrm{Mpc}}$ is the error in linear projected size.
 
Eight of the reported 24 GRGs have an angular size $\ge$ 3.0 arcmin. The corresponding linear sizes of the GRGs are in the range of 0.74 Mpc to 1.73 Mpc, with a median size of 1.04 Mpc and a mean size of 1.08 Mpc. Our sample consists of 10 GRQs, have a linear size ranging from 0.79 Mpc to  1.68 Mpc, with a mean size of 1.14 Mpc and a median size of 1.11 Mpc. GRGs and GRQs with sizes larger than 1.50 Mpc are rare.  We identify one GRG (J1046+0144) and one GRQ (J2255+1357) having sizes $\ge$ 1.5 Mpc. In the previous catalogues, the linear size of GRGs spanned 0.71 Mpc to 3.4 Mpc, and for GRQs, it was 0.70 Mpc to 2.87 Mpc. For example, \citet{Ku18} found GRSs with linear sizes in the range 0.71 Mpc--4.45 Mpc with median 1.14 Mpc, \citet{Da20a} detected GRSs at 150 MHz with linear sizes in the range 0.71 Mpc--3.44 Mpc with median 1.28 Mpc, \citet{Da20b} identified GRSs at 1400 MHz with linear sizes in the range 0.71 Mpc--2.20 Mpc with median 1.09 Mpc, \citet{An22} found GRSs at 888 MHz with linear sizes in the range 0.70 Mpc--1.87 Mpc with median 1.22 Mpc, \citet{Ku21} detected GRQs at 1400 MHz with linear sizes in the range 0.70 Mpc--1.87 Mpc with median 0.92 Mpc and \citet{Si22} found GRSs with linear sizes in the range 0.70 Mpc--3.40 Mpc with median 1.00 Mpc.

\begin{figure*}
	\includegraphics[scale=0.65]{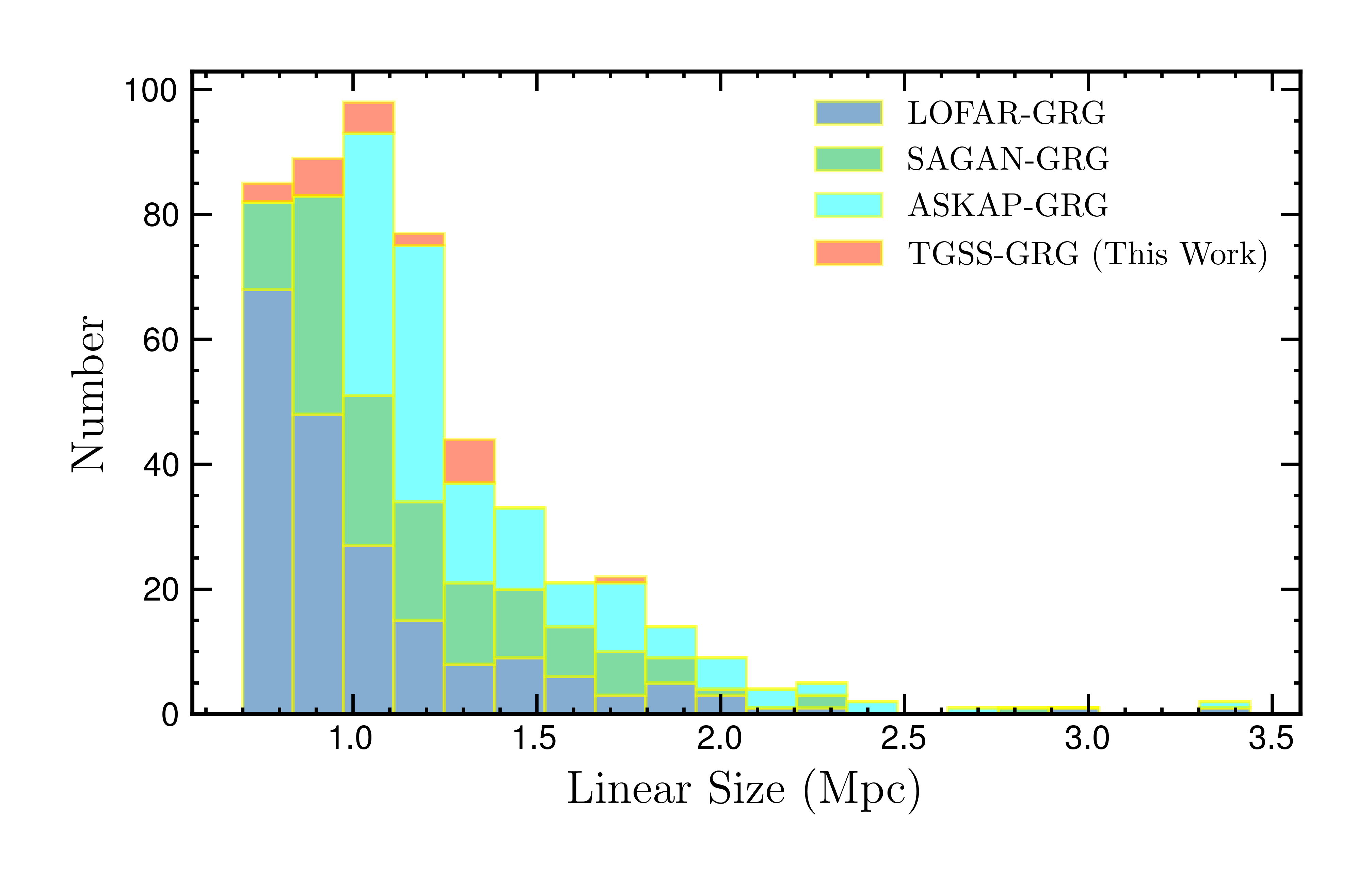}
	\includegraphics[scale=0.65]{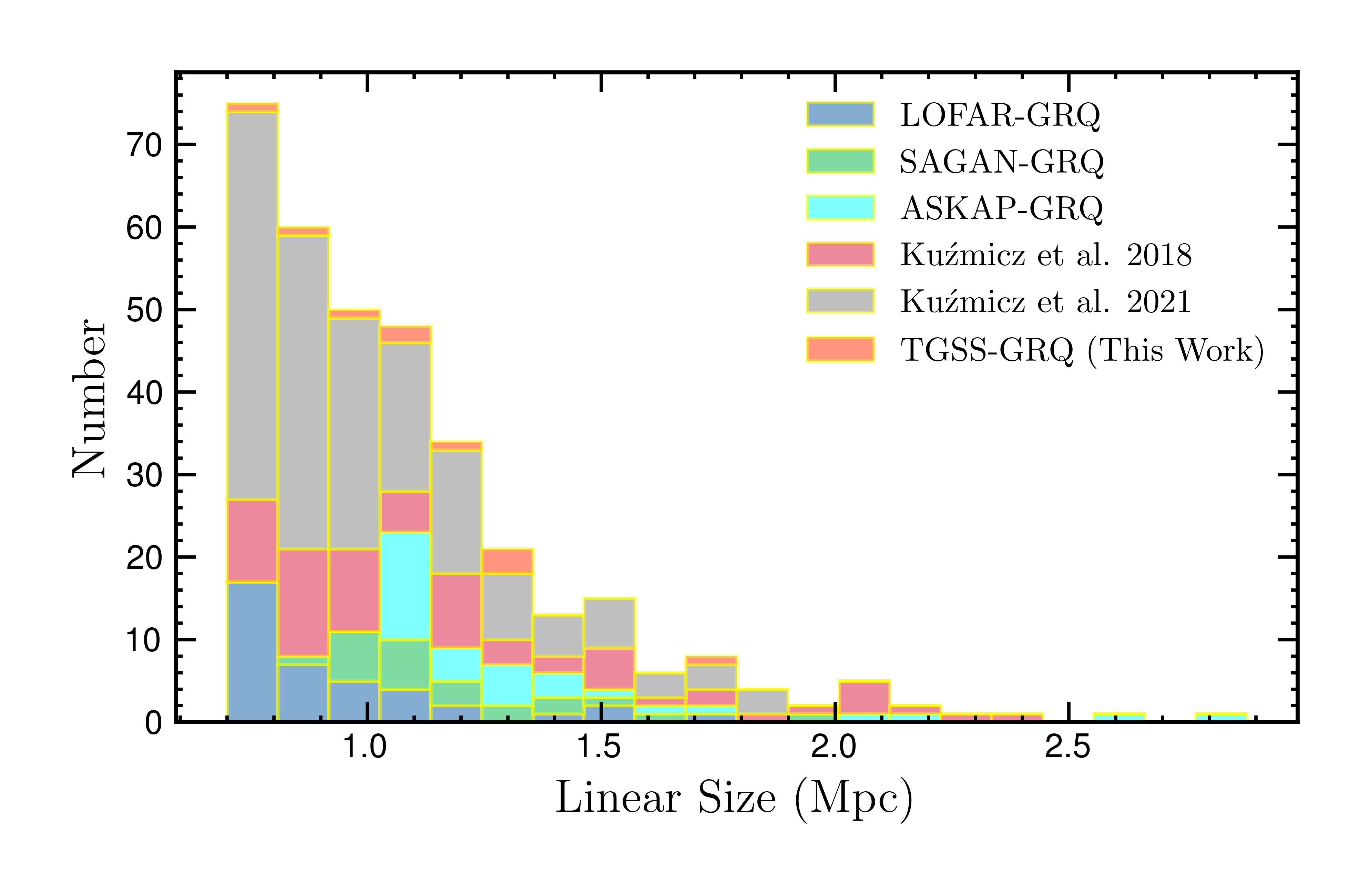}
	\caption{Histogram showing the projected linear size distribution of the reported GRGs (left) and GRQs (right). We also included GRGs presented in \citet{Da20b, Da20c, An22}, and GRQs presented in \citet{Ku18, Da20b, Da20c, Ku21, Si22}.}
	\label{fig:Linear size}
\end{figure*}

\subsection{Spectral index }
For a radio source, spectral index ($\alpha^{1400}_{150}$) represents the energy distribution of relativistic electrons, and its measurement should therefore ideally cover a wide frequency range. We made sigma-clipping at a level of 3$\sigma_{rms}$, where $\sigma_{rms}$ is the local RMS noise of images.

The two-point spectral index between 150 and 1400 MHz is calculated for the sources reported in the present paper and is mentioned in Table \ref{tab:GRG-GRQ}. These spectral indexes have been determined by integrating over the same aperture at both frequencies. Out of the 34 GRSs with spectral index information, no sources show flat-spectrum ($\alpha_{150}^{1400}<0.5$). All the GRSs show steep radio spectrum ($\alpha_{150}^{1400}>0.5$) which is the common property of lobe-dominated RGs.

The uncertainty in spectral index measurements due to flux density uncertainty \citep{Ma16} is
\begin{equation}
\Delta\alpha=\frac{1}{\ln\frac{\nu_1}{\nu_2}}\sqrt{\left(\frac{\Delta S_1}{S_1}\right)^{2}+\left(\frac{\Delta S_2}{S_2}\right)^{2}}
\end{equation}
where $\nu_{1, 2}$ and $S_{1, 2}$ refer to NVSS and TGSS frequencies and flux densities respectively. The flux density accuracies in TGSS ADR 1 and NVSS are $\sim10$ percent \citep{In17} and $\sim5$ percent \citep{Co98} respectively. Using equation 6, the spectral index uncertainty for the sources presented in the current paper is $\Delta\alpha=0.05$. For GRGs in our catalogue, the total span of $\alpha_{150}^{1400}$ ranges from $0.62$ to $1.19$ with a mean and median of 0.87 and 0.86 respectively. 10 GRQs from our sample have a spectral index in the range of 0.51 to 1.03 with a mean and median of 0.80 and 0.81, respectively.

The radio spectral index is commonly used to differentiate between different components of a radio galaxy like core and lobe. To ensure the location of the core, we compute the spectral index of the probable core region (found from optical/IR counterpart and radio morphology) using TGSS and NVSS; we found that all of them show flat spectral index ($\alpha_{150}^{1400}<0.5$) which is signature of the core of RGs. 

\begin{figure*}
	\includegraphics[scale=0.65]{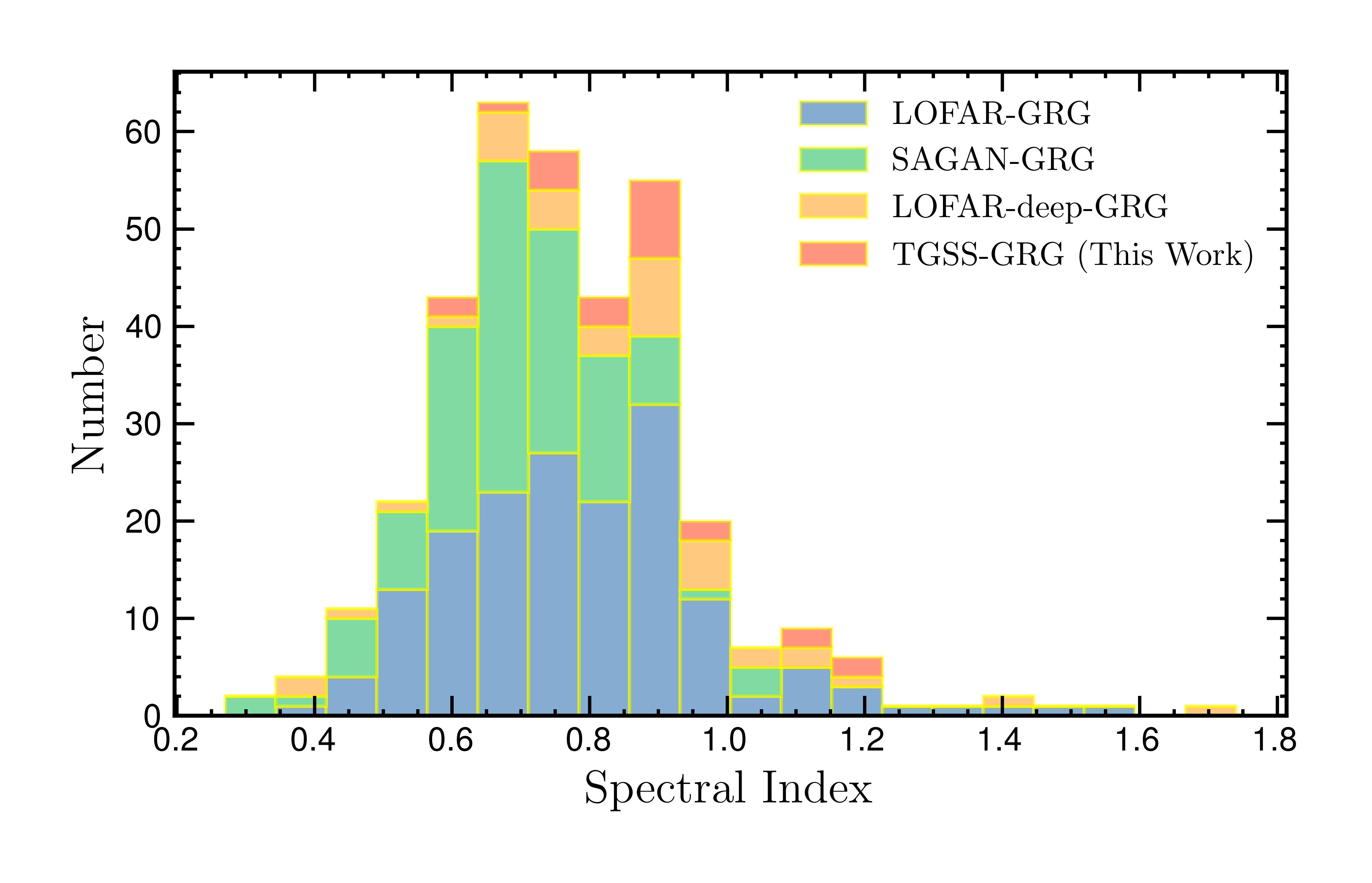}
	\includegraphics[scale=0.65]{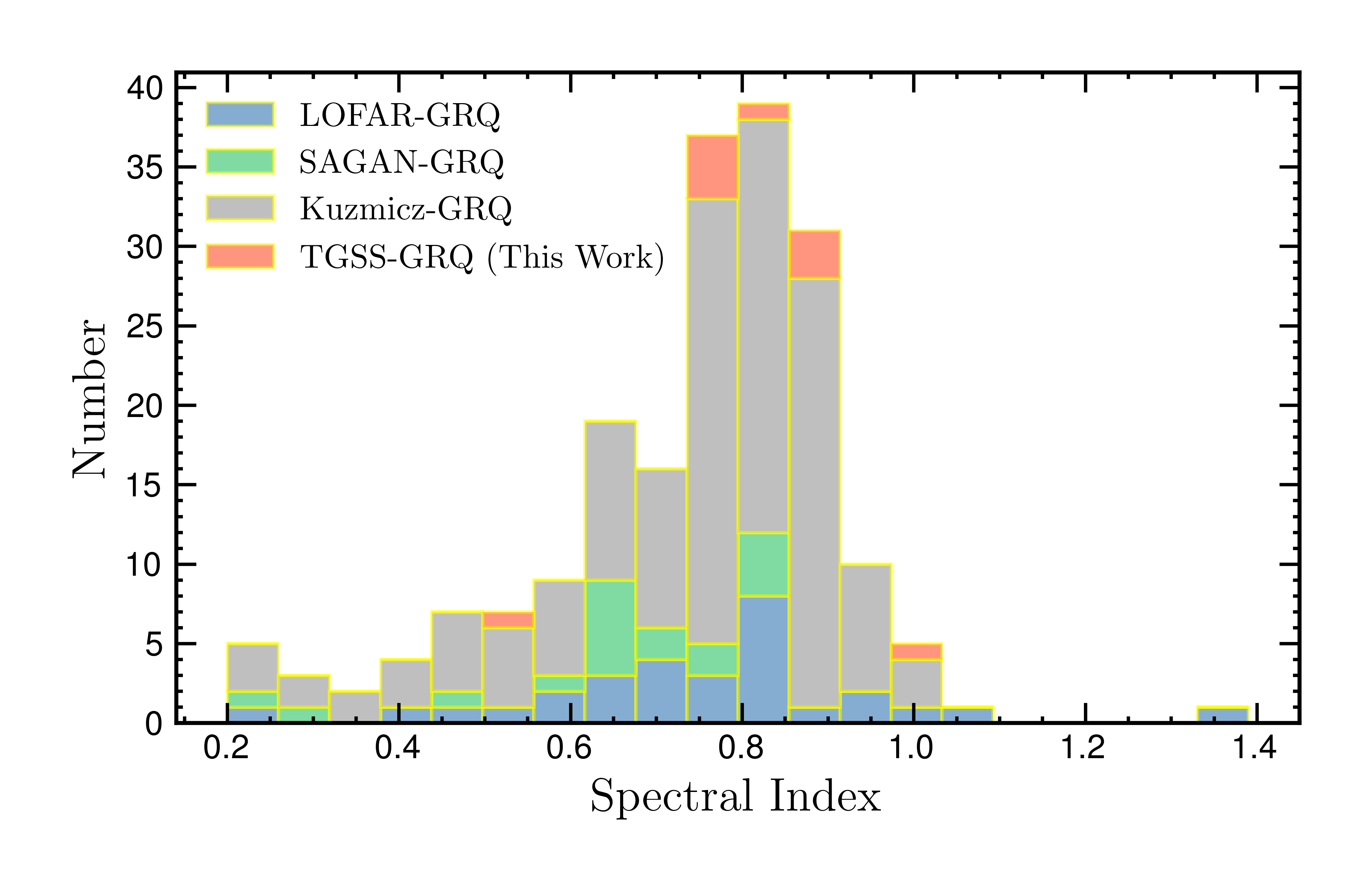}
	\caption{Histogram showing spectral index distribution of the reported GRGs (left) and GRQs (right). We also included GRGs presented in \citet{Da20b, Da20c, Si22}, and GRQs presented in \citet{Da20b, Da20c, Ku21}. For uniformity, we calculate the spectral index of GRS using TGSS ADR 1 flux and NVSS flux.}
	\label{fig:spec-hist}
\end{figure*}

\subsection{Radio Power}
We have calculated the radio power ($P_{150}$) of all the sources reported in the present paper (using spectroscopic and photometric $z$) with standard formula \citep{Do09}

\begin{equation}
P_{150}=4\pi{D_{L}}^{2}S_{150}(1+z)^{\alpha-1}
\end{equation}
where $z$ is the redshift of the radio galaxy, $\alpha$ is the spectral index, $D_{L}$ is the luminosity distance to the source (Mpc), and $S_{150}$ is the flux density (Jy) at a given frequency. 

For the sources presented in the current paper, the radio powers at 150 MHz are of the order of $10^{27}$ W Hz$^{-1}$, which is similar to a typical giant radio galaxy \citep{Da20b}. The mean value of $\log~P$ [W Hz$^{-1}$] for GRGs in the current paper is 27.04 (1$\sigma$ standard deviation $=0.72$, median $=26.98$), and for GRQs, the mean value is 27.17, with a median value of 27.09.

\begin{figure*}
	\includegraphics[scale=0.65]{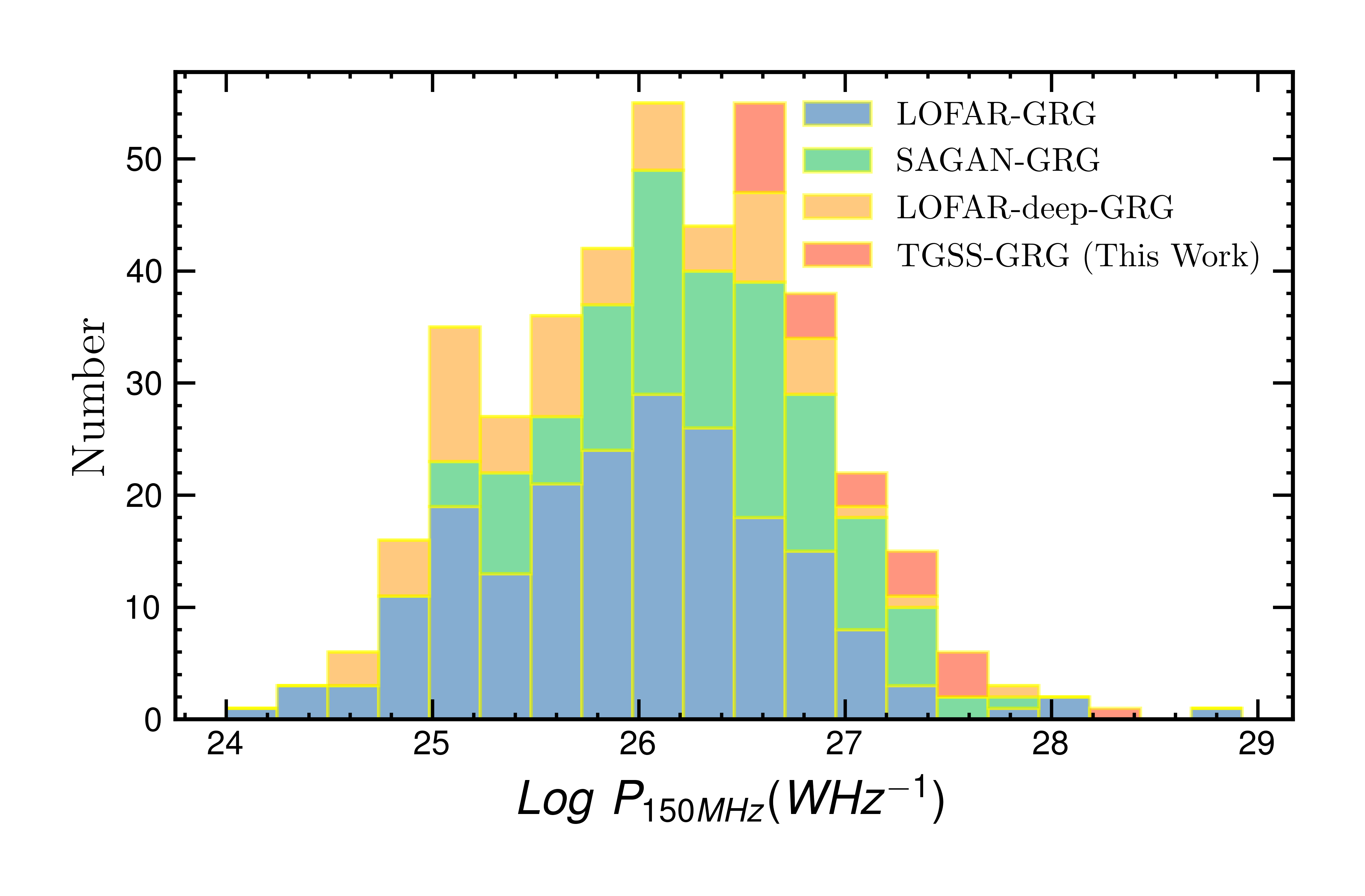}
	\includegraphics[scale=0.65]{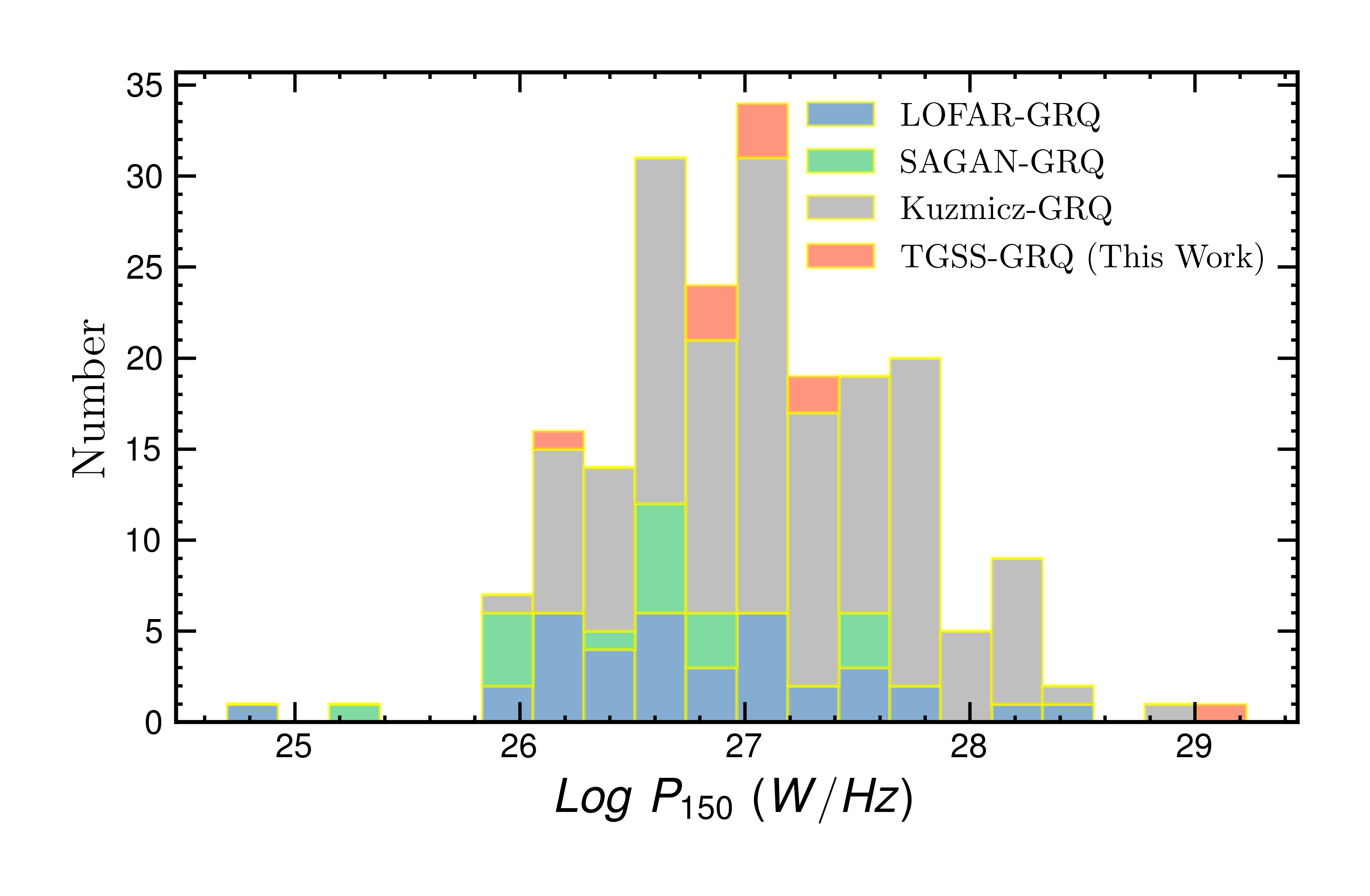}
	\caption{Histogram showing radio power ($P_{150}$) distribution of GRSs presented in the current paper for GRGs (left) and GRQs (right). For completeness of the study, the radio powers of GRGs and GRQs are calculated at 150 MHz in TGSS ADR 1. GRGs from \citet{Da20b, Da20c, Si22}, and GRQs from \citet{Da20b, Da20c, Ku21} are also included in the plot.}
	\label{fig:radio power}
\end{figure*}

\begin{figure*}
	\includegraphics[scale=0.68]{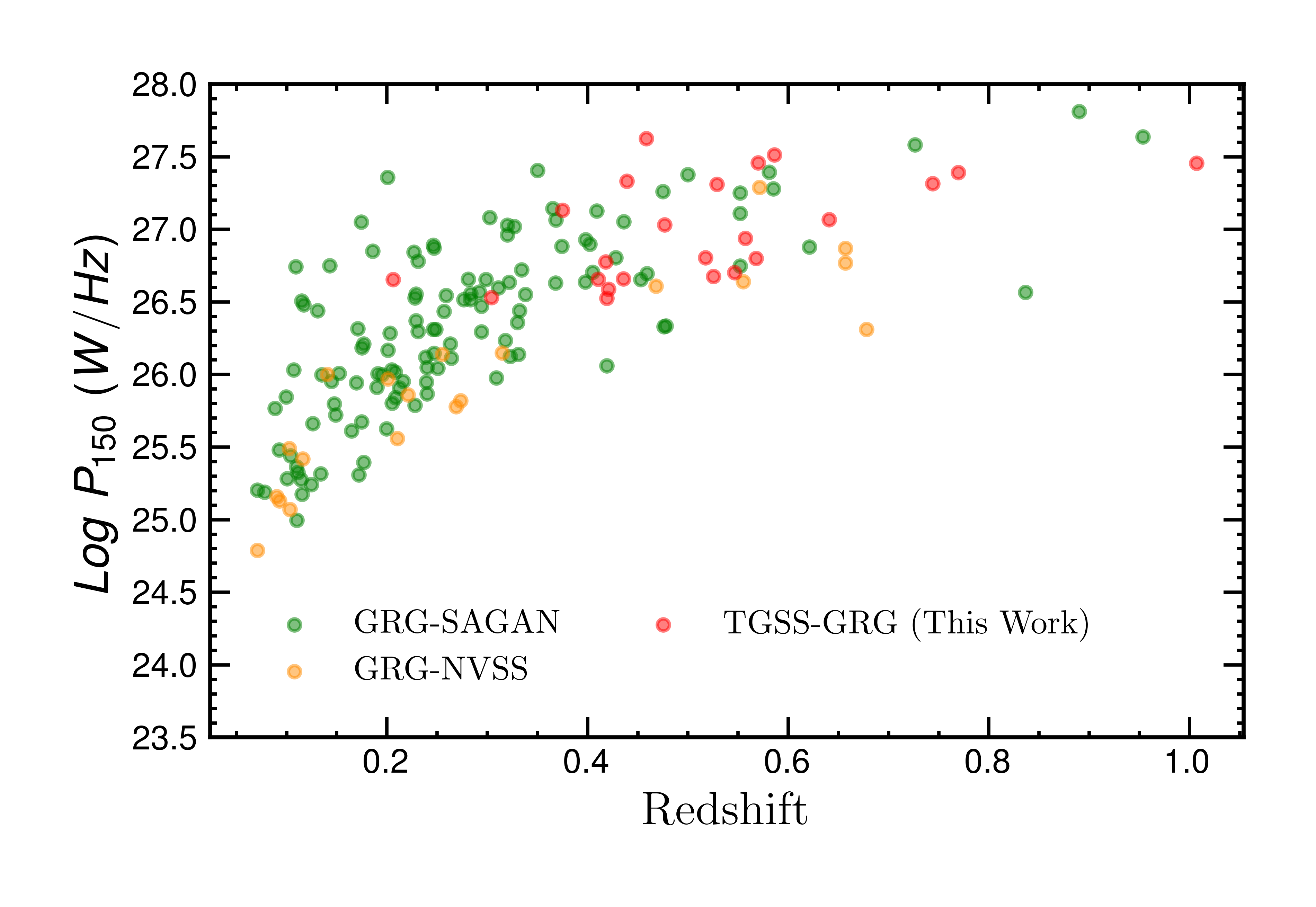}
	\includegraphics[scale=0.68]{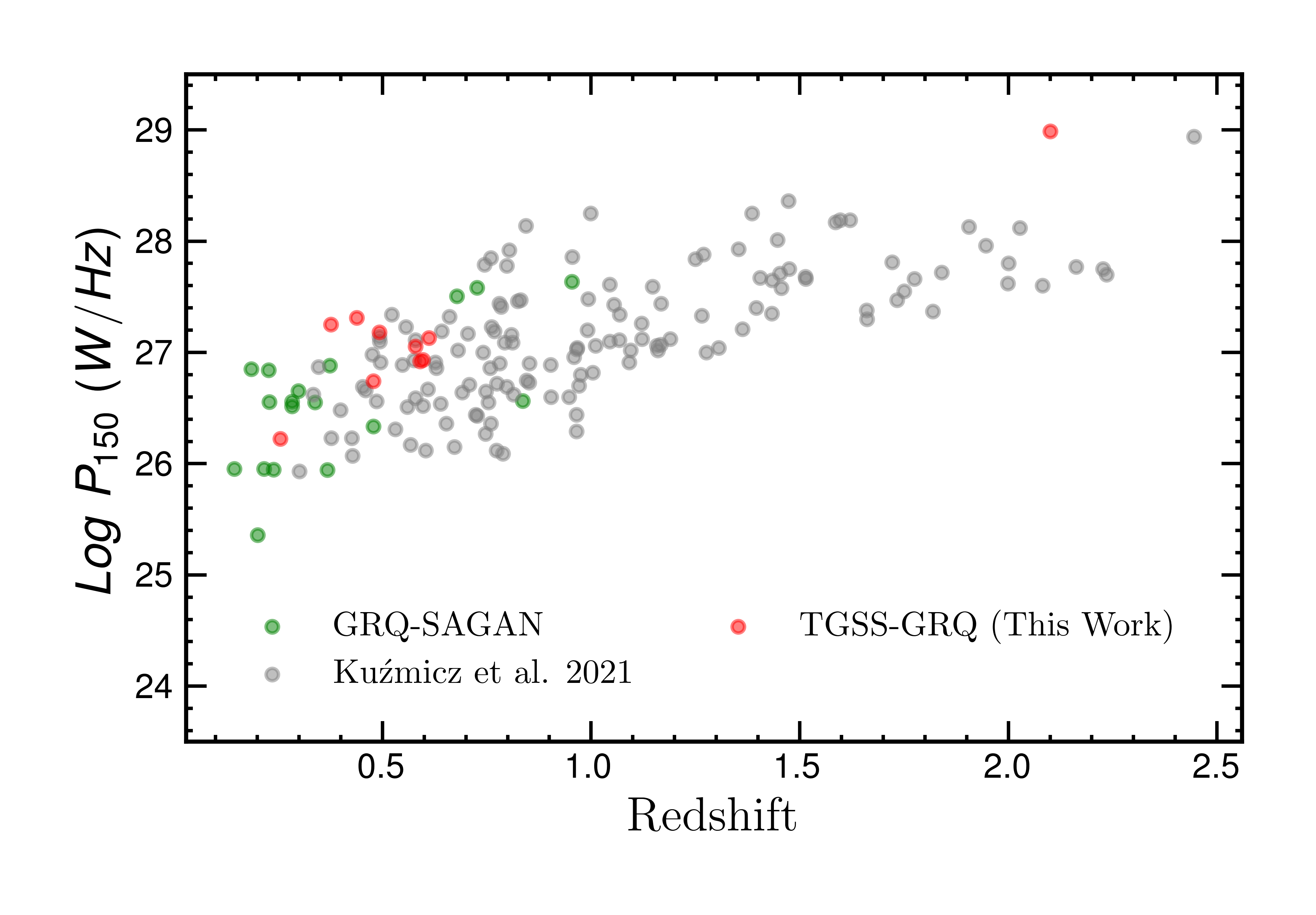}
	
	\caption{Figure shows  distribution of radio power (Log($P_{150}$ W/Hz)) with redshift ($z$). The 24 new GRGs (left) and 10 new GRQs (right) are shown with the filled red circles. The plot also includes GRGs from \citet{Dab17} (with dark orange points) and from \citet{Da20c} (with green points); GRQs from \citet{Ku21} (with grey points) and from \citet{Da20c} (with green points).}
	\label{fig:P-z}
\end{figure*}

\subsection{Black hole mass estimation of GRSs}
We compute the central black hole mass of the host of the GRSs using the well-known $M_{\textrm{BH}}-\sigma$ relation \citep{Tr02}. The central velocity dispersion of the optical hosts of the GRGs in the current paper (Table \ref{tab:GRG-optical}) is taken from SDSS DR16 which is made using fibre spectroscopy. 

\begin{equation}
M_{\textrm{BH}}=1.349\times 10^{8} M_{\odot} (\sigma/200 \textrm{km}~\textrm{s}^{-1})^{4.02}
\end{equation}

For GRQs, the $H_{\alpha}$ emission line width and luminosity can also be used to estimate BH masses. We estimate BH masses using the empirical relations from \citet{Gr05},

\begin{scriptsize}
	\begin{equation}
	\mathrm{M_{\textrm{BH}}=(2.0\pm0.4)\times10^{6}\left(\frac{L_{H_{\alpha}}}{10^{42~erg~s^{-1}}}\right)^{0.55 \pm 0.02}\times\left(\frac{\textrm{FWHM}_{H_{\alpha}}}{10^{3}~ km~s^{-1}}\right)\times M_{\odot}}
	\end{equation}
\end{scriptsize}
where $L_{H\alpha}$ is the luminosity of the H$\alpha$ and FWHM is the full width half maxima of $H_{\alpha}$ line.

Among the GRGs reported in the current paper, the dispersion measurement was available for 16 sources that are hosted by elliptical galaxies. We calculate the upper limit of mass of the host SMBHs of these galaxies, a summary of which is presented in Table \ref{tab:GRG-optical}. The range of masses of central black holes is $\sim 1.5\times10^{7}-2.1\times10^{10}M_{\odot}$. Two of our 18 GRGs (J1255+2507, J1315+1831) hosted by SMBH with mass more than $10^{10}M_{\odot}$. One GRG (J1512+4541) has a black hole mass below $10^{9}M_{\odot}$. It is seen that the megaparsec linear size scale of radio sources is mostly hosted by illuminated elliptical galaxies but for a small number of GRSs, the host galaxies are spiral \citep{Ho11, Ba14}. It is generally believed that GRG hosts have super-massive black holes of masses $\sim10^{8}-10^{9}M_{\odot}$ in their galactic cores which are accountable for strengthening their long-scale radio jets and lobes \citep{Be84}.

\section{Discussion}
\label{sec:disc}
\subsection{Physical features of GRGs}
Two GRGs (J0843+0513, J1138+4540) in our sample are present in the brightest galaxy cluster. In Figure \ref{fig:Linear size}, we present a histogram showing the distribution of the projected linear size for GRGs and GRQs using the sources presented in the current paper. GRSs have a linear size range between 0.74 Mpc and 1.68 Mpc, and from the histogram, it is evident that most GRGs in the present sample have a size of nearly 1 Mpc. The present sample also includes two high-redshift ($z>$1) GRSs. The radio power of GRSs ranges from $P_{150} \sim 10^{26}$ W~Hz$^{-1}$ to $10^{29}$ W~Hz$^{-1}$. 

\subsection{Comparision of results}
In this section, we compare the derived radio results between not only GRGs and GRQs, but also GRSs with SRSs at 150 MHz. We make a size comparison between GRGs and GRQs detected in the present work from the TGSS survey. In Figure \ref{fig:Linear size}, the distribution of linear size is shown for different surveys along with the current paper. The histogram shows the peaks for GRGs and GRQs between 0.8 Mpc and 1.0 Mpc. This histogram shows that the distributions of the projected linear size of GRGs are not significantly different from GRQs.

A histogram with spectral index distribution for GRGs and GRQs, presented in the current article, is shown in Figure \ref{fig:spec-hist}. GRGs from \citet{Da20a, Da20b, Si22} and GRQs from \citet{Da20a, Da20b, Ku21} are also included in this distribution. To maintain uniformity for the study of the spectral nature of GRGs and GRQs, we calculated the spectral index ($\alpha^{1400}_{150}$) of all GRGs and GRQs collected from the previous catalogue using corresponding fluxes from the TGSS ADR field, where available. 
The higher temperature and density of the circumgalactic medium may cause the material in the hotspot and the jet behind it to move at a slower pace. GRGs discovered at 1400 MHz have a median spectral index of approximately $\sim$ 0.7 \citep{Dab17, Da20b}. Since low-frequency surveys are more suitable for the detection of lobe-dominant diffuse GRSs, the low-frequency discoveries of GRGs have a steeper median value of the spectral index compared to the samples detected at higher frequencies. For example, in a low-frequency study, \citet {Da20a} found GRGs with a median spectral index of 0.78, and in the \citet{Si22} catalogue, GRGs have a median spectral index of 0.88. For our TGSS GRGs and GRQs, histogram peaks are observed near $0.85-0.90$ as shown in Figure \ref{fig:spec-hist}.

In Figure \ref{fig:radio power}, we show a histogram of radio power distribution of the GRGs and GRQs reported in the current paper as well as GRGs from \citet{Da20a, Da20b, Si22} and GRQs from \citet{Da20a, Da20b, Ku21} using corresponding measurements at 150 MHz, where sources are available. 
According to the figure and previous studies \citep{Da20b, Ma22}, it has been observed that GRQs have higher radio power at both 150 MHz and 1400 MHz as compared to GRGs. This is likely because the central engines of GRQs can produce stronger radio jets and luminosity than those of GRGs. It has also been found that the jet kinetic power ($\mathrm{Q_{jet}}$) of GRQs is greater than that of GRGs, indicating that they may host more massive black holes that accrete at a higher rate \citep{Ma22}. Furthermore, from the negative correlation between jet kinetic power and dynamical age ($\mathrm{t_{dyn}:~Q_{jet} \propto \frac{1}{t_{dyn}^{-2}}}$; \citealt{ito08}), it can be interpreted that, assuming identical sizes, the more powerful jets of GRQs could potentially traverse larger distances (Mpc) in a shorter timeframe compared to GRGs.
Figure \ref{fig:P-z} shows the radio power distribution of the sources presented in the current paper at 150 MHz with redshift. We also include GRGs and GRQs from \citet{Dab17, Da20b, Ku21} where we recalculate the radio power of all sources at 150 MHz. This plot indicates a positive correlation between radio power and redshift. The GRGs detected from TGSS ADR 1 at 150 MHz are more luminous compared to those detected from the NVSS survey at 1400 MHz. J1652+7408 is the least luminous GRG in our sample with $\mathrm{P_{150}=1.67\times10^{26}}$ W Hz$^{-1}$ ($z=0.25$). J2231+0702 is the most luminous GRS in our sample with $\mathrm{P_{150}=9.74 \times10^{28}}$ W Hz$^{-1}$ ($z\sim2.1$).

Based on the data availability from TGSS ADR 1, we can compare the radio features of 34 newly discovered GRSs with the sample of 488 SRSs, which are identified in TGSS radio maps. The mean value of the radio spectral index ($\alpha_{150}^{1400}$) of SRSs is close to 0.75. For GRGs in our catalogue, the total span of $\alpha_{150}^{1400}$ ranges from $0.61$ to $1.19$ with a mean and median of 0.87 and 0.86, respectively. In our sample, 10 GRQs have a spectral index in the range of 0.51 to 1.03 with a mean and median of 0.80 and 0.81, respectively. It implies that GRSs have a steeper spectral index compared with SRSs. For GRGs, the mean and median values of central black hole mass are $\mathrm{\sim 3.15\times10^{8}M_{\odot}}$ and $\mathrm{\sim 2.45\times10^{8}M_{\odot}}$. The mean and median value of $\mathrm{M_{BH}}$ for 122 SRSs are $\mathrm{2.23\times10^{8}M_{\odot}}$ and $\mathrm{2.51\times10^{8}M_{\odot}}$. The black hole mass distributions of RGs and GRGs are similar in nature. For the sources presented in the current paper, the radio powers at 150 MHz are of the order of $10^{27}$ W Hz$^{-1}$. The average value of $\mathrm{\log~P_{150}}$ [W Hz$^{-1}$] for GRSs in the current paper is 27.08 (1$\sigma$ standard deviation $=0.72$, median $=27.19$). For SRSs, the mean and median values of $\mathrm{\log~P_{150}}$ [W Hz$^{-1}$] are 26.70 and 26.69, respectively. The mean and median values confirm that GRSs and SRSs have a different distribution of radio power, which means that GRSs have a higher order of jet kinetic power.

\subsection{The Fanaroff-Riley classiﬁcation of giant radio sources}
\label{sub:FR-class}
\begin{figure}
	\includegraphics[width=9cm,angle=0,origin=c]{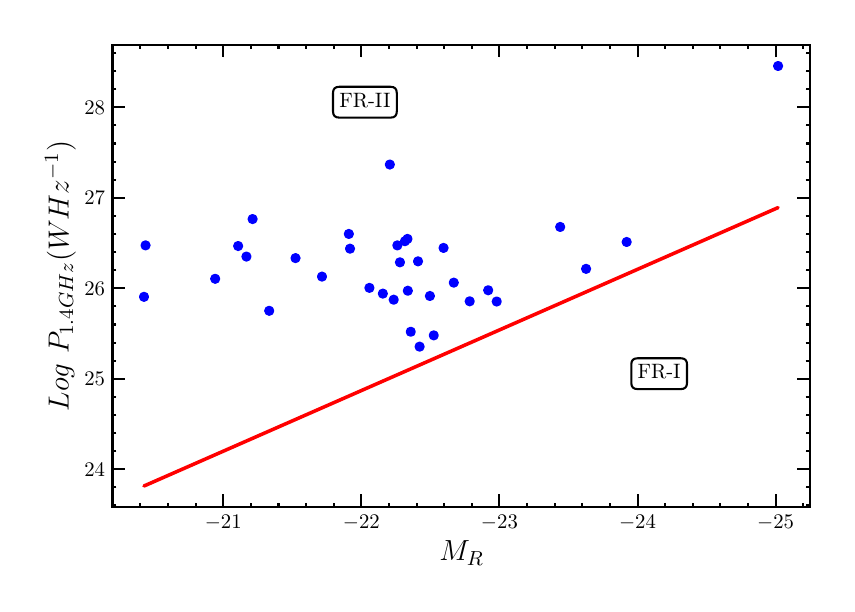}
	\caption{Plot shows the radio power at 1.4 GHz with absolute R-band magnitude for GRS. The Owen-Ledlow division line \citep{Wo07} is drawn by using $\mathrm{\log P_{1.4 ~ GHz}=-0.67M_{R}+10.13}$. }
	\label{fig:FR}
\end{figure}

Based on radio power, GRS has been found to fall into either the FR-I or FR-II classes. In the updated catalogue of GRSs \citep{Ku18} and GRGs with LoTSS’s \citep{Da20b}, the majority (90 percent) of sources showed the FR-II type of morphology. Out of 162 GRGs, only 8 exhibited an FR-I type radio morphology, indicating that the FR-II population is dominant in GRSs \citep{Da20c, Le96}. \citet{Le96} divided RGs into two classes (FR-I and FR-II) by radio--optical luminosity diagram which is known as the Owen-Ledlow plane. 
Figure \ref{fig:FR} shows the distribution of GRSs reported in the present paper on the Owen-Ledlow plane. The FR division line depicted in Figure \ref{fig:FR} can be parameterized as $\mathrm{\log P_{1.4 ~ GHz}=-0.67M_{R}+10.13}$, where $\mathrm{P_{1.4 ~ MHz}}$ is the radio luminosity at 1400 MHz and $M_{R}$ is the r-band absolute magnitude of the host galaxy. Among the sources presented in the current paper, all GRSs belong to the FR-II galaxies and fall above the division line, in the region known to be populated mostly by FR-II sources.

\subsection{P-D diagram of GRGs and GRQs} 
\begin{figure}
	\includegraphics[width=6cm,angle=270,origin=c]{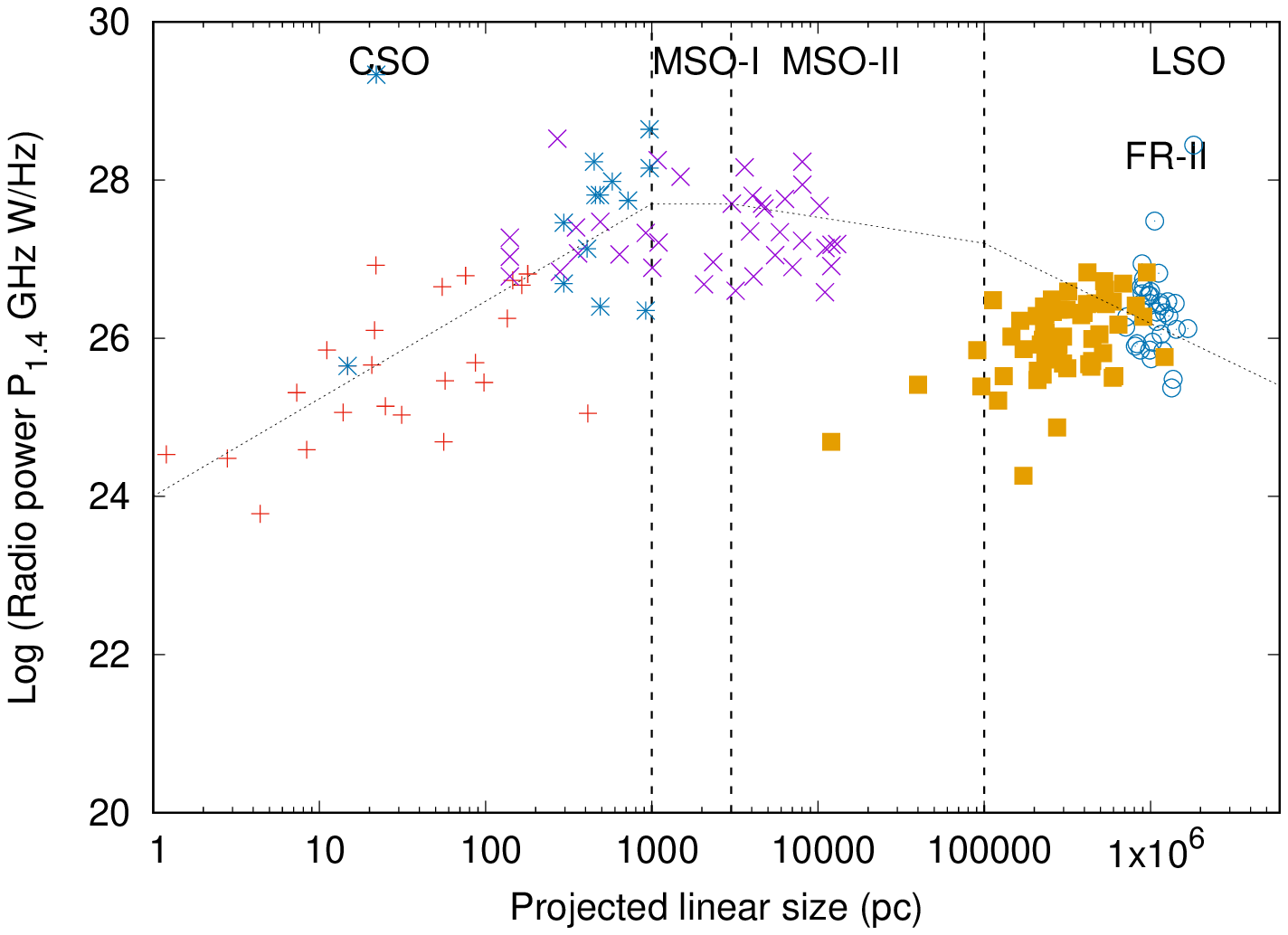}
	\caption{ Plot displays how radio power at 1400 MHz varies with the linear size of compact symmetric objects (CSOs), medium-sized symmetric objects (MSOs), and large symmetric objects (LSOs). Black dashed lines represent the evolutionary paths of high radio power sources. This plot includes TGSS-GRSs (with magenta open circles) (present work), large size FR-II sources from \citet{Bh22a} (with yellow squares), high-power CSSs from \citet{Fa01} (with the green plus), high-power GHz peaked-spectrum (GPS) sources from \citet{Xi06, Li07} (with the magenta asterisks), and high-power CSOs from \citet{Ku06} (with the red crosses).}
	\label{fig:P-D}
\end{figure}

A P-D diagram is a conventional representation \citep{Ba82} that shows the relationship between radio power ($P$) at a specific frequency and the linear size ($D$) of RGs. We may study the evolution of RGs using the P-D diagram \citep{Ka97, Is99, Ma02, Ma04}. In Figure \ref{fig:P-D}, we plot the P-D diagram using GRGs and GRQs presented in the current paper. For completeness, we also include compact symmetric objects (CSOs), medium-sized symmetric objects (MSOs), and large symmetric objects (LSOs).

GRGs and GRQs may grow larger than 1 Mpc, but after that, they become less radio-luminous. To study the dynamic evolution of radio double sources, a few researchers used parameters such as the lobe kinetic power, the total size of sources, the velocity with which the terminal hotspots are moving, and host galaxy ambient medium density gradients \citep{Bl99, An12}. The structural and spectral properties of the radio source are influenced by the power of the source, the local environment of the host galaxy, and the evolutionary age \citep{Ka07, An12}. According to \citet{An12}, RGs have four evolutionary stages: CSO, MSO-1, MSO-2, and LSO. GRSs belong to the last evolutionary stage. 

At the last stage of radio galaxy evolution, radio jets extend and begin to inflate. The radio source transforms into a normal FR-I or FR-II  radio galaxy based on the kinetic power of the jet. The radio jet eventually shuts down, and the radio jet kinetic power reduces as the electrons miss out on the energy, but the radio lobe continues to spread. After $\sim 10^{7}$ years, the radio galaxy may have enlarged enough to grow as a giant radio galaxy. The radio power decreases sharply as $P_{rad} \propto D^{-1.6}$, where inverse Compton loss originating from the cosmic microwave background (CMB) dominates over the synchrotron loss \citep{An12}.

We did not find any sources that have both large size and large radio power, i.e. at the top right corner of the P-D diagram. This supports the evolutionary model of the formation of GRSs. Moreover, no GRSs are detected in the bottom right corner of the diagram (extreme giant sizes and low radio powers). 

\subsection{$D-z$ relationship of GRGs and GRQs}
The projected linear size ($D$) of radio sources is commonly seen to vary with cosmological epoch, and redshift ($z$). Several research works have been carried out on this evolution scenario using different samples of radio sources \citep{Ku18}.  \citet{Ok82} suggested a linear size ($D$) evolution of the form $D\approx (1+z)^{k}$, depending on the value of the density parameter ($\Omega_{0}$). They reported a value of the linear size evolution parameter $k$ in the range of $1 \le k \le 2$, for both RGs and RQs and opined that extended steep spectrum (ESS) RGs and RQs undergo similar size evolution.  The variation of the linear size of extragalactic radio sources with redshift can be expressed as $D=D_{0}(1+z)^{k}$ \citep{Ka89}, where $D_{0}$ is the intrinsic linear size and depends on the assumed cosmology. In a recent paper, \citet{On18} presents $D$-$z$ plot of a sample consisting of both galaxies and quasars, where they report an increase in median size with redshift up to $z\approx1.0$ and for $z>1$ linear size of radio sources was decreasing. 

\begin{figure}
	\includegraphics[width=9cm,angle=0,origin=c]{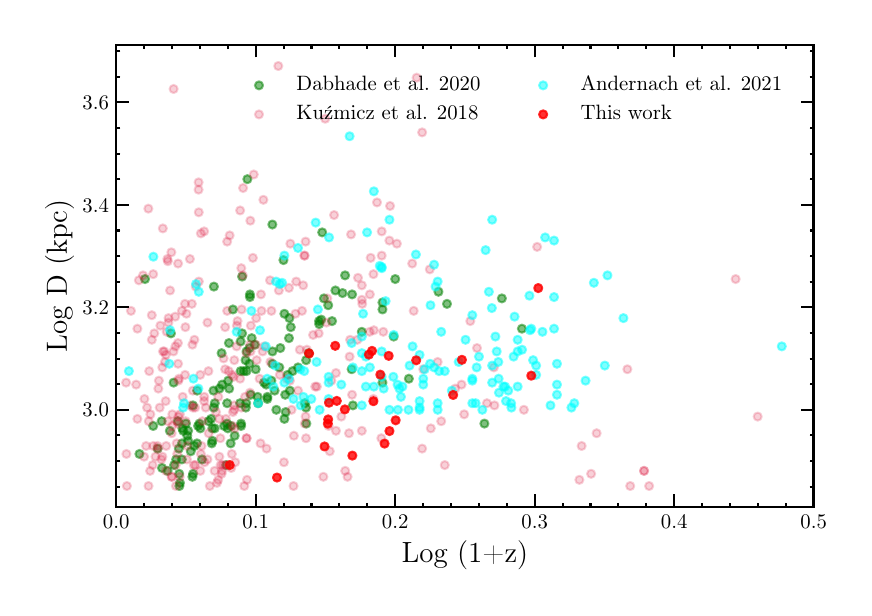}
	\caption{Plot showing linear size vs redshift of GRSs presented in the current paper. For completeness, we included GRS from \citet{Ku18, Da20c, An22}.}
	\label{fig:D-z}
\end{figure}

In our sample, for the redshift range of 0.21$<z<$0.50, $0.5<z<0.75$, and $z>0.75$, the median projected linear sizes are 1.0 Mpc, 1.21 Mpc, and 1.23 Mpc, respectively. For $z>0.75$, the linear size of GRGs and GRQs slightly decreases (Figure \ref{fig:D-z}). For a long time, it was believed that GRGs could not be found at high redshift. The density ($\rho_{\textrm{medium}}$) of the intergalactic medium (IGM) increases with redshift as $(1+z)^{3}$, hindering radio lobe development at high redshifts \citep{Go87, Ka89}. Moreover, the brightness of GRSs decreases as $(1+z)^{-4}$ due to the Tolman effect \citep{sandage61,sandlubin01}. This decrease in brightness is compensated for by the fact that the energy density of Cosmic Microwave Background radiation increases with redshift as $(1+z)^4$. This allows for the detection of inverse Compton/CMB (IC/CMB) X-rays even at high redshifts. The cosmological evolution of linear sizes of GRGs and GRQs are indistinguishable that implies variation of $D-z$ is similar in nature.

\subsection{GRGs in galaxy cluster}

\begin{figure}
	\includegraphics[width=8cm,angle=0,origin=c]{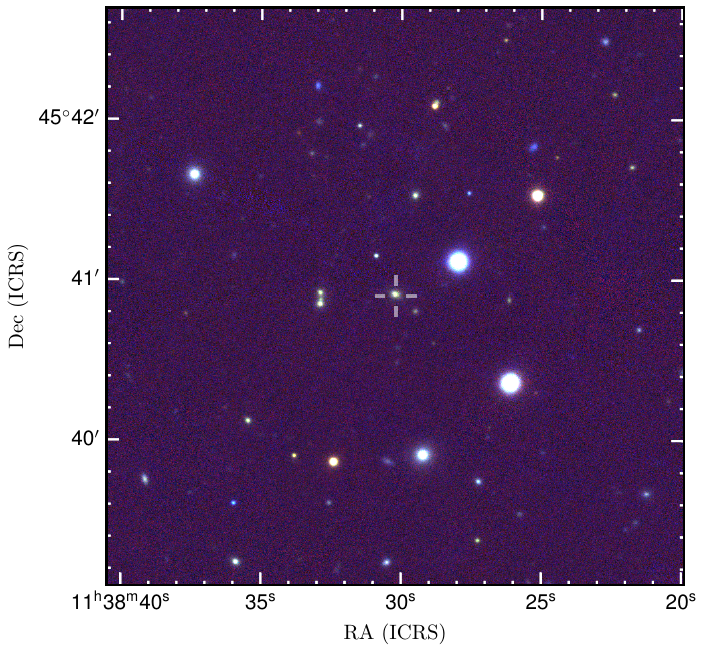}
	\caption{RGB image of the host galaxy of GRG J1138+4540 in BCG environment. This image uses Pan-STARRS with y-band (magenta), i-band (green), and g-band (red). The host galaxy is elliptical in shape which is marked in the figure.}
	\label{fig:BCG}
\end{figure}

The enormous size of GRGs is favoured in low-density environments. The presence of GRGs and GRQs in a dense cluster environment is unusual. For example, only 0.62 percent of GRGs are found in the brightest cluster galaxy (BCG; \citealt{Da20c}). We find from our sample that only two GRGs (J0843+0513, J1138+4540) are situated in a highly dense environment. The host galaxy of GRG J0843+0513 is present in the BCG in the cluster WHL \citep{We12}. The basic WHL cluster parameters are (1) $r_{200}$, the radius within which the mean density of a cluster is 200 times the critical density of the Universe, (2) $R_{L}$, cluster richness parameter, (3) $M_{200}$, the optical mass that has been computed from \citet{We12}, and (4) $N_{200}$, the number of galaxies within $r_{200}$. The host cluster for GRG J0042+3150 is WHL J004253.1+315028 for which these parameters are calculated to be $r_{200}$ = 0.87, $R_{L}$ = 13.91, $M_{200}$ = 0.61$\times 10^{14} M_{\odot}$ and $N_{200}$ = 10. The host galaxy of GRG J1138+4540 is the Gaussian Mixture Brightest Cluster Galaxy (GMBCG \citealt{Ha11}). The distances between the BCG centre and host galaxies for GRG J0843+0513 and GRG J1138+4540 are 3.98 pc and 4.84 pc respectively. Figure \ref{fig:BCG} shows the host galaxy of GRG J1138+4540.

The GRGs in the brightest galaxy cluster (BCG) possibly trace inhomogeneities in the intergalactic gas, which could be one of the determining factors for the growth and evolution of these giant sources. The forward propagation of jets, as well as backflow from hotspots in the lobes of GRGs, are influenced by the gas. These galaxies have expanded to a projected linear size of 0.71 to 0.88 Mpc despite being in a compact cluster \citep{Ku18}. A detailed study is required to fully comprehend their fundamental and spectroscopic evolution. For a powerful AGN, the environment has little impact on the giant growth of GRGs. Normally, GRGs in BCGs favor radio-loud AGN. But, only a small proportion of massive-GRGs are in BCGs \citep{Ba94, Be07, Ki08}. This study highlights the necessity of additional investigation into other parameters in GRGs' exterior surroundings, as well as different physical properties (mass, spin, and mass accretion rate) of the BHs that fuel these giant sources.

\section{CONCLUSIONS}
\label{sec:conclusion}
In the present work, a total of 24 GRGs and 10 GRQs (Table \ref{tab:GRG-GRQ}) are identified using TGSS at 150 MHz. This survey was performed by the Giant Metrewave Radio Telescope (GMRT), which covers 90 percent of the whole sky. The detection of a low number of GRGs from TGSS, despite its high sensitivity, is due to the lack of identification of the host galaxy with redshift for many RGs. The future optical study requires the identification of the redshifts of more prospective giant galaxies. In brief, our findings are as follows:\\
	
  $\bullet$ We detected 34 GRSs using TGSS at 150 MHz. So far, almost  3200 GRGs have been detected by the previous studies in the whole sky. Our GRGs and GRQs sample increase the number of known RGs by nearly one percent.\\ 
	
	$\bullet$ Our sample consists of 10 GRQs with a redshift range of 0.25 to 2.10 and only two GRGs (J0843+0513, J1138+4540) are found in  BCG. About Fifty-three percent of GRGs and GRQs of our sample lay at $z\le$ 0.5. 
	\\
	
	$\bullet$ A Study of the spectral index histogram shows that the spectral index of GRSs is the same as other radio sources.\\
	
	$\bullet$ From radio morphological studies, it is found that all GRGs and GRQs presented in the current paper are FR-II  RGs. \\
	
	$\bullet$ Among the sources presented in the current paper, optical spectra for 16 GRGs are available from the SDSS data. We classify GRGs into HEGRGs and LEGRGs using the SDSS spectral line flux ratios. We also calculate the emission linewidth of three GRGs (J1005+0020, J1315+3831, and J1512+4541). \\
	
	$\bullet$ We calculate mid-IR luminosity from WISE four-band data. GRSs are classified into LEGRGS, HEGRGs, and quasars using the WISE mid-IR colour-colour plots. \\
	
	$\bullet$ We study the optical spectral properties of hosts of the eight GRQs with H$_{\beta}$, doublet [OIII] emission line, and classify them as BLRGs.\\
	
	$\bullet$ 16 out of 24 GRGs in our sample have black hole mass above $6.0\times10^{7}M_{\odot}$.

\section*{Acknowledgements}
We thank the whole TGSS team and the staff of the GMRT behind this survey. GMRT is run by the National Centre for Radio Astrophysics of the Tata Institute of Fundamental Research.  This publication makes use of data products from the Two Micron All Sky Survey, which is a joint project of the University of Massachusetts and the Infrared Processing and Analysis Center/California Institute of Technology, funded by the NATional Aeronautics and Space Administration and the NATional Science Foundation.

\section*{Data Availability Statement}
The data that support the plots within this paper and other findings of this study are available from the corresponding author upon reasonable request. The TGSS ADR 1 images are available at \href{http://tgssadr.strw.leidenuniv.nl/doku.php}{http://tgssadr.strw.leidenuniv.nl/doku.php}.

\begin{table*}[b]
\begin{rotatetable}
\begin{scriptsize}
\begin{centering}
\caption{ Col. (1) is the name of GRSs. Col. (2) and Col. (3) present the right ascension (RA) and declination (DEC) of the host galaxy in h:m:s and d:m:s. Col. (4) is the host galaxy type; G presents the galaxy and Q presents the quasar. Col. (5) and Col. (6) are integrated radio flux densities ($F_{\nu}$) of GRS at 150 MHz and 1400 MHz. Col. (7) indicates the spectral index ($\alpha^{1400}_{150}$) of sources with the error value. Col. (8) is the redshift of host galaxies with error. Col. (9) is redshift type; S and P indicate spectroscopic and photometric redshift. Col. (10) is the linear angular size (LAS) with measurement uncertainties in arcminute. Col. (11) indicates the projected linear size (D) of GRSs in Mpc. Col. (12) is log10 of total radio power ($P_{\nu}$) of sources. Col. (13) is the other catalogue name. The last Col. is the redshift reference of sources.}
\label{tab:GRG-GRQ}
\begin{tabular}{cccccccccccccll}
\hline
Name &R.A.      & Decl.    &Typ.&$F_{150}$ &$F_{1400}$ &$\alpha^{1400}_{150}$ &$z$ &$z$-Typ  &LAS &$D$ &Log~(P)  &Other&$z_{ref}$\\
     &(hr min s)& ($\degr ~\arcmin~ \arcsec$) &   &(Jy)      &(mJy)       &($\pm$0.05)   &    &    &($'$)&(Mpc)& W Hz$^{-1}$ &Catalogue & \\   
(1)  & (2)      &(3)       &(4)  &(5)       &(6)        &(7)           &(8) &(9)&(10) &(11) &(12)   &(13)&(14) \\
					
\hline
J0042+3150 &00 42 53.1  &31 50 28  &G   &0.62$\pm$0.06      &51$\pm$2       &1.12    &0.5573$\pm$0.0705 & P &2.15$\pm$0.0076   &0.86$\pm$0.040 &27.36  &1, 2&19   \\
					
J0155+2956 &01 55 58.0  &29 56 23  &G   &3.31$\pm$0.33      &480$\pm$24     &0.86    &0.984$\pm$0.157&P  &2.36$\pm$0.0690   &1.16$\pm$0.090&28.20  &1 &11\\ 
					
J0803+4847 &08 03 32.9  &48 47 29  &G  &0.80$\pm$0.08      &73$\pm$4       &1.09    &0.769414$\pm$0.000341&S  &2.73$\pm$0.0082   &1.25$\pm$ 0.003 &27.39  &1 &11  \\ 
					
J0804+2722 &08 04 59.9  &27 22 49  &Q  &0.83$\pm$0.08      &120$\pm$6       &0.86    &0.578982$\pm$0.000049&S  &2.45$\pm$0.0079 &0.99$\pm$0.003 &27.05  &1 &11  \\ 
					
J0808+1708 &08 08 35.8  &17 08 32  &G  &0.83$\pm$0.08      &120$\pm$6       &0.86    &0.744$\pm$0.064&P  &2.36$\pm$0.0121 &1.06$\pm$0.039 &27.31  &1, 8 &20  \\ 
					
J0843+0513 &08 43 01.4  &05 13 26 &G  &0.57$\pm$0.05      &140$\pm$7       &0.62    &0.418920$\pm$0.000020&S   &2.80$\pm$0.0415   &0.96$\pm$0.014 &26.51  &1 &11\\
					
J0846+1413 &08 46 36.1  &14 13 08  &Q   &1.62$\pm$0.16      &230$\pm$11       &0.87    &0.492703$\pm$0.000034&S &2.83$\pm$0.0034   &1.06$\pm$0.001 &27.18  &8  &11  \\
					
J0847+0148 &08 47 59.7  &01 48 56  &G   &0.47$\pm$0.04      &60$\pm$3       &0.92      &0.568$\pm$0.037&P  &3.16$\pm$0.0071   &1.27$\pm$0.007 &26.80  &1  &20  \\
					
J0848+0115 &08 48 13.7  &01 15 02  &G   &2.16$\pm$0.21      &300$\pm$15       &0.88    &0.570023$\pm$0.000183&S &2.25$\pm$0.0035   &0.91$\pm$0.003 &27.45  &5  &11 \\ 
					
J0909+1832 &09 09 57.0  &18 32 03  &G   &2.73$\pm$0.27      &391$\pm$19       &0.87    &0.374955$\pm$0.000070&S  &4.03$\pm$0.0213   &1.28$\pm$0.006 &27.13  &--&11  \\

J0919+0510 &09 19 29.1  &05 10 46  &Q   &0.69$\pm$0.06      &220$\pm$11       &0.51    &0.596504$\pm$0.000059&S &3.13$\pm$0.0059   &1.29$\pm$0.002 &26.94  &6 &11  \\
					
J1005+0020 &10 05 20.3  &00 20 19  &G   &0.61$\pm$0.06      &90$\pm$4       &0.85    &0.517550$\pm$0.000077&S &3.33$\pm$0.02431   &1.28$\pm$0.009 &26.80  &1&11 \\
					
J1006+3114 &10 06 56.0  &31 14 42  &G   &0.42$\pm$0.04      &50$\pm$3       &0.95    &0.525389$\pm$0.000208&S &3.36$\pm$0.0053   &1.30$\pm$0.002 &26.67  &7  &11  \\
					
J1016+1012 &10 16 20.2  &10 12 58  &G   &0.98$\pm$0.09      &190$\pm$9       &0.73    &0.418064$\pm$ 0.000050&S &2.75$\pm$0.0316   &0.94$\pm$0.011 &26.77  &1, 6&11 \\
					
J1028+3221 &10 28 32.3  &32 21 53  &G   &0.42$\pm$ 0.04     &60$\pm$3       &0.87    &0.546441$\pm$0.000121&S &2.96$\pm$0.0536   &1.17$\pm$0.021 &26.70  &1 &20 \\ 

J1030+3554 &10 30 44.0  &35 54 51 &G   &0.75$\pm$0.07      &180$\pm$9       &0.63    &0.640830$\pm$0.000083&S &2.93$\pm$0.0053   &1.25$\pm$0.002 &27.06  &7 &11 \\
					
J1033+3054 &10 33 12.2  &30 54 21  &G   &1.30$\pm$0.13      &240$\pm$12       &0.75    &0.640830$\pm$0.000083&S &2.93$\pm$0.0053   &1.24$\pm$0.002 &27.06  &7, 9&11  \\
					
J1042+1957 &10 42 51.5  &19 57 25  &G   &0.58$\pm$0.05      &42$\pm$2       &1.19    &0.435545$\pm$0.000044&S &3.81$\pm$0.0062   &1.33$\pm$0.002 &26.65  &1 &11 \\ 
					
J1046+0144 &10 46 29.2  &01 44 57  &G   &0.52$\pm$0.05      &59$\pm$3       &0.96    &0.478614$\pm$0.000109&S &3.48$\pm$0.0039   &1.28$\pm$0.001 &26.67  &1 &11    \\ 
					
J1138+4540 &11 38 30.2  &45 40 54  &G   &1.02$\pm$0.10      &68$\pm$3       &1.19    &0.303983$\pm$0.000050&S  &2.65$\pm$0.0362   &0.74$\pm$0.010 &27.50  &4,7&11\\

J1158+1535 &11 58 40.3  &15 35 28  &Q   &0.59$\pm$0.05      &90$\pm$4       &0.84    &0.5899$\pm$0.0415&P &3.11$\pm$0.0012   &1.27$\pm$0.033 &26.92  &1&20 \\
					
J1239+2737 &12 39 46.0  &27 37 00  &Q   &0.66$\pm$0.06      &121$\pm$6       &0.76    &0.477824$\pm$0.000044&S &2.80$\pm$0.0743   &1.03$\pm$0.027 &27.34  &1, 7&11   \\ 
					
J1255+2507 &12 55 41.5  &25 07 45  &G   &0.62$\pm$0.06      &120$\pm$6       &0.76    &0.420848$\pm$0.000108&S &3.01$\pm$0.0913   &1.03$\pm$0.031 &26.57  &7 &11 \\ 
					
J1315+1831 &13 15 11.3  &18 31 16  &G   &1.85$\pm$0.18      &270$\pm$13       &0.86   &0.529$\pm$0.035&P &2.67$\pm$0.0049   &1.04$\pm$0.002 &27.31  &10  &20    \\ 

J1315+3830 &13 15 24.3  &38 30 44  &G   &2.32$\pm$0.23      &340$\pm$17       &0.85   &0.586206$\pm$0.000022&S &2.33$\pm$0.0731   &0.95$\pm$0.030 &26.67  &4, 7&11  \\ 
					
J1512+4541 &15 12 19.5  &45 41 10  &G   &5.32$\pm$0.53      &733$\pm$36       &0.88   &0.458497$\pm$0.000030&S  &2.78$\pm$0.0042   &1.00$\pm$0.002 & 27.62  &1, 4&11  \\ 
					
J1652+7408 &16 52 30.1  &74 08 28  &Q   &0.83$\pm$0.08      &150$\pm$7       &0.76    &0.2554$\pm$0.0127&P  &3.25$\pm$0.0023   &0.80$\pm$0.008 &26.01   &6&19 \\
					
J1731+3232&17 31 14.5  &32 32 49  &Q    &3.70$\pm$0.37      &690$\pm$34       &0.75    &0.376000$\pm$0.003000&S &2.73$\pm$0.0213   &0.87$\pm$0.007&27.24  &1&11      \\ 
					
J2231+0702&22 31 41.7  &07 02 21 &Q    &3.56$\pm$0.35      &601$\pm$30       &0.79    &2.10$\pm$0.02&P&2.36$\pm$0.0091   &1.21$\pm$0.005  &28.98   &1, 3&20    \\ 
					
J2253+0615&22 53 03.2  &06 15 13  &G    &0.78$\pm$0.07      &150$\pm$7       &0.73    &0.410389$\pm$0.000046&S&2.51$\pm$0.0073   &0.85$\pm$0.002  &26.65    &1 &11     \\ 
					
J2255+1357&22 55 59.6  &13 57 35  &Q    &0.80$\pm$0.08      &80$\pm$4       &1.03    &0.610961$\pm$0.000040&S &4.03$\pm$0.0132   &1.68$\pm$0.006  &27.12    &8 &11     \\
					
J2256+1433&22 56 21.9  &14 33 51  &G    &3.06$\pm$0.30      &502$\pm$25       &0.81    &0.439163$\pm$0.000044&S&2.96$\pm$0.0031   &1.04$\pm$0.001  &27.33    &1, 8 &11 \\ 
					
J2306+3927&23 06 04.6  &39 27 20  &G    &3.65$\pm$0.36      &770$\pm$38       &0.69    &0.2061$\pm$0.0013&P     &3.73$\pm$0.0098   & 0.78$\pm$0.002 &26.65    &1, 2  &18   \\
					
J2306+1556&23 06 30.3  &15 56 20  &Q    &2.88$\pm$0.28      &411$\pm$20       &0.87    &0.438621$\pm$0.000030&S &3.60$\pm$0.0072   &1.27$\pm$0.003&27.31   &1 &11  \\
					\hline
				\end{tabular}
			\end{centering}
			References--  
			1: NVSS \citep{Co98}; 2: VLSS \citep{Co07}; 3: 4C \citep{Pi65, Go67, Ca69};  4: 6C \citep{Ba85}, \citep{Ha88, Ha90, Ha91, Ha93a};  5: PMN \citep{Gr94};  6: 87GB \citep{Gr91}; 7: B2 \citep{Co70, Co72, Fa74};  8:  VFK \citep{Va15}; 9: FIRST \citep{Be95}; 10:  MG2 \citep{Be86}; 11: SDSS \citep{Be16}; 12: WHL \citep{We12}; 13: \citep{Ji21}; 14: \citep{We16}; 15: \citep{Ve18}; 16: \citep{Co16}; 17: \citep{Bi14}; 18: \citep{He91}; 19: \citep{Ah20}; 20: \citep{Dun22}.
			
		\end{scriptsize}
	\end{rotatetable}
\end{table*}

\begin{table*}
	\caption{This table shows the optical spectral properties of GRGs. Here, r and $M_{R}$ are apparent and absolute r-band magnitude. H$\alpha~\lambda6563$ and [OIII] $\lambda5007$ denote the emission line flux of H$\alpha$ and [OIII]. EW is the emission width calculated by using Gaussian fit on the SDSS optical data. EI is the excitation index determined by using equation \ref{equ:EI}. VD presents the velocity dispersion of the optical host galaxies. $M_{BH}$ is the central black hole mass derived by using $M_{\textrm{BH}}-\sigma$ relation. In the last column, the class represents the classification of GRGs using the optical scheme.}
	
	\label{tab:GRG-optical}	 
	\begin{tabular}{cccccccccccll}
		\hline
		Name &r &$M_{R}$  &H$\alpha~\lambda6563$ &[OIII] $\lambda5007$ &Flux ratio               &EW         &EI  &VD         & $M_{\textrm{BH}}$   &Class \\
		&   &         &Flux ($\times10^{-17}$)   &Flux ($\times10^{-17}$)&[OIII]/H$\alpha$ &[OIII]&   &($\sigma$) &            &        \\    
		&   &         &(erg /s cm$^{2}$\AA)      &(erg /s cm$^{2}$\AA)   &                         &(\AA)      &   &(km s$^{-1}$)& ($10^{9}M_{\odot}$)& \\
		(1)   &(2)&(3)      & (4)                      &(5)                    &(6)                      &(7)        &(8)&(9)        &(10)        &(11) \\
		\hline
		
		J0155+2956&20.94 &--22.03   &--&2.06&-- &--&--  &211.6      & 1.28           &--                 \\
		J0808+1708&20.52 &--21.98   &1.24&1.79&0.69 &--&0.02 &193.9      & 1.01      &LERG                  \\
		J0848+0115&20.59 &--22.25   &3.03&1.62&0.53&--&--  &237.3      & 1.61          &LERG                \\
		J1005+0020&20.08 &--22.48   &4.37&2.83&0.64&3.76&0.23  &198.5      & 0.61          &LERG                \\ 
		J1006+3114&21.12 &--21.48   &2.20 &1.80&0.81 &--&0.18  &190.3      & 0.50          &LERG                 \\
		J1028+3221&19.77 &--22.93   &3.13&1.61&0.51 &--&0.05  &235.3      & 1.47          &LERG               \\
		J1030+3554&20.38 &--22.74   &-- &2.61&3.55 &--&--  &206.4      & 0.72          &LERG                \\
		J1033+3054&19.79 &--22.55   &2.29&1.44 &0.62&--&0.02  &219.5      & 1.60          &LERG                \\
		J1042+1957&19.61 &--22.50   &4.60&3.55&0.77 &--&0.15  &157.2      & 0.25          &LERG                \\
		J1046+0144&20.26 &--22.45   &2.50&1.92&0.76 &--&0.03   &165.0     & 1.31          & LERG               \\
		J1255+2507&19.65 &--22.40   &3.87&2.05&0.52 &--&0.01  &346.8      & 10.2          &LERG                \\
		J1315+1831&19.90 &--22.71   &1.76&1.55&0.88  &--&0.33   &339.7      & 21.1          &LERG                \\
		J1315+3830&19.30 &--23.58   &-- &31.6&-- &436.0&--  &320.5      & 7.16          &HERG                \\
		J1512+4541&20.89 &--21.40   &7.70&13.9 &1.80&21.5 &1.01 &127.6      & 0.06          &HERG               \\
		J2253+0615&19.64 &--22.32   &3.75 &3.03&0.80 &--&0.06  &245.6      & 1.94          &LERG              \\
		J2256+1433&19.66 &--22.45   &7.37 &3.37&0.45&--&--0.09   &235.2       & 1.94          &LERG              \\
		
		\hline
	\end{tabular}
\end{table*}

\begin{table*}
	\caption{This table shows the optical properties of GRQs using doublet emission [OIII] and broad H $\beta$ lines. Gaussians are fitted to spectral lines H$\beta$ and [OIII] doublet to obtain the emission peak ($\lambda$), line width ($\Delta\lambda$), and flux. $M_{BH}$ is black-hole mass, which is calculated from optical continuum luminosity.}
	\label{tab:GRQ-optical}	 
	\begin{tabular}{cccccccccccll}
		\hline
		Name&   &H$\beta$   & & & Strong [OIII]  & & & Weak [OIII] &&\\ 
		&   &Line   & & &Line  & & &Line  & & \\
		\hline
		
		GRQs        &$\lambda$ &Flux &$\Delta \lambda$  &$M_{BH}$&$\lambda$ &Flux &$\Delta\lambda$ &$\lambda$ &Flux  &$\Delta \lambda$  \\
		& &$10^{-17}$& &$10^{8}$ & &$10^{-17}$& & &$10^{-17}$& \\
		&(\AA) & ergs s$^{-1}$ cm $^{2}$\AA&(\AA) &$M_{\odot}$ &(\AA) &erg s$^{-1}$ cm$^{2}$\AA & (\AA)&(\AA) &erg s$^{-1}$ cm$^{2}$\AA&(\AA) \\
		\hline
		
		J0804+2722  &7678   &~5.5    &134.6&18.6    &7906   &~38.4    &9.2   &7830   &13.5    &~9.3  \\
		J0846+1413  &7259   &~4.6    &10.0&0.15     &7477   &~49.4    &9.6   &7405   &16.4    &~9.8  \\
		J0919+0510  &7760   &~1.2    &23.8&0.51     &7994   &~15.7    &19.4  &7917   &~6.2     &20.0   \\
		J1158+1535  &7659   &~2.9    &-- &--        &7890   &~~3.5     &--    &7800   &~1.8    &--     \\
		J1239+2737  &7184   &~1.4    &13.6&0.30     &7401   &~15.8    &13.2  &7328   &~4.7     &13.2   \\
		J2255+1357  &7830   &~2.9    &13.6&0.28     &8064   &~35.2    &10.4  &7889   &10.8    &11.4   \\
		
		\hline
	\end{tabular}
\end{table*}

\begin{table*}
\caption{This table presents mid-IR band magnitudes of GRGs and GRQs reported in the present manuscript. Err is the error values in W1, W2, W3, and W4. L$_{W1}$, L$_{W2}$, L$_{W3}$, and L$_{W4}$ are WISE mid-IR luminosities. The last column presents the classification of giant radio sources using the mid-IR WISE scheme. }
\label{tab:IR-GRGs-GRQs}
\begin{tabular}{cccccccccccl}
\hline
GRS &W$_{1}+W1_{Err}$ &W$_{2}+W2_{Err}$ &W$_{3}+W3_{Err}$ &W$_{4}+W4_{Err}$ &L$_{W1}$         &L$_{W2}$         &L$_{W3}$        &L$_{W4}$        &Class  \\
    &                 &                 &                &                  &$\times10^{44}$  &$\times10^{44}$  &$\times10^{44}$ &$\times10^{44}$  &      \\
    &                 &                 &                &                  & (erg s$^{-1}$)  &(erg s$^{-1}$)   &(erg s$^{-1}$)  &(erg s$^{-1}$)   &    \\
		
\hline
J0042+3150 &14.66$\pm$0.039 &13.66$\pm$0.037 &10.18$\pm$0.077 &7.18$\pm$0.150  &2.98 &3.02  &5.71    &12.00    &HERG\\
J0155+2956 &16.42$\pm$0.066 &16.02$\pm$0.173 &10.82$\pm$0.105 &8.16$\pm$0.211  &2.86 &1.32  &1.31    &~3.58    &LERG\\
J0803+4847 &15.61$\pm$0.045 &15.64$\pm$0.116 &12.57$\pm$--    &9.12$\pm$--     &4.54 &1.78  &2.31    &~7.36    &LERG\\
J0804+2722 &14.02$\pm$0.027 &12.87$\pm$0.027 &10.11$\pm$0.075 &7.31$\pm$0.123  &9.85 &11.49 &11.17   &19.56    &Quasar\\
J0808+1708 &15.69$\pm$0.048 &15.78$\pm$0.145 &12.24$\pm$--    &8.74$\pm$--     &1.43 &0.62  &1.14    &~3.83    &LERG\\
J0843+0513 &15.10$\pm$0.037 &14.88$\pm$0.072 &12.23$\pm$0.459 &8.78$\pm$0.468  &1.68 &0.83  &0.73    &~2.33    &LERG\\
J0846+1413 &14.64$\pm$0.032 &13.48$\pm$0.034 &10.56$\pm$--    &7.31$\pm$--     &3.73 &4.39  &4.94    &13.11    &Quasar\\
J0848+0115 &15.25$\pm$0.041 &15.34$\pm$0.102 &12.43$\pm$--    &8.61$\pm$--     &3.02 &1.12  &1.25    &~5.61    &LERG\\
J0909+1832 &14.79$\pm$0.037 &14.58$\pm$0.065 &12.3$\pm$--     &8.86$\pm$--     &1.70 &0.83  &0.49    &~1.64    &LERG\\
J0919+0510 &15.27$\pm$0.040 &14.54$\pm$0.056 &11.0$\pm$0.114  &7.81$\pm$0.199  &3.30 &1.04  &5.22    &13.10    &Quasar\\
J1005+0020 &15.38$\pm$0.042 &15.50$\pm$0.124 &12.13$\pm$0.420 &8.28$\pm$--     &2.12 &0.76  &1.31    &~6.03    &LERG\\
J1006+3114 &15.53$\pm$0.044 &15.42$\pm$0.098 &12.66$\pm$--    &9.01$\pm$--     &1.91 &0.85  &0.83    &~3.19    &LERG\\
J1016+1012 &14.79$\pm$0.034 &14.65$\pm$0.061 &11.89$\pm$--    &8.95$\pm$--     &2.19 &1.01  &0.98    &~1.95    &LERG\\
J1028+3221 &14.96$\pm$0.034 &14.67$\pm$0.068 &12.34$\pm$--    &8.44$\pm$--     &3.56 &1.88  &1.23    &~5.93    &LERG\\
J1030+3554 &14.70$\pm$0.031 &14.45$\pm$0.049 &12.24$\pm$--    &8.63$\pm$--     &6.65 &3.38  &1.98    &~7.32    &LERG\\
J1033+3054 &15.11$\pm$0.036 &15.01$\pm$0.068 &12.51$\pm$--    &9.12$\pm$--     &2.33 &0.99  &0.75    &~2.28    &LERG\\
J1042+1957 &15.21$\pm$0.038 &14.19$\pm$0.081 &11.67$\pm$--    &8.56$\pm$--     &1.64 &0.89  &1.32    &~3.09    &LERG\\
J1046+0144 &14.23$\pm$0.029 &14.25$\pm$0.048 &12.36$\pm$--    &8.33$\pm$--     &1.24 &0.50  &1.00    &~2.85    &LERG\\
J1138+4540 &14.92$\pm$0.030 &14.54$\pm$0.049 &12.39$\pm$0.357 &8.73$\pm$--     &0.89 &0.51  &0.28    &~1.09    &LERG\\
J1158+1535 &15.05$\pm$0.035 &14.97$\pm$0.076 &12.68$\pm$--    &8.92$\pm$--     &3.76 &2.57  &2.71    &10.51    &Quasar\\
J1239+2737 &15.78$\pm$0.047 &14.77$\pm$0.074 &11.72$\pm$0.227 &8.91$\pm$0.426  &0.35 &0.44  &0.62    &~3.55    &Quasar\\
J1255+2507 &15.46$\pm$0.042 &14.97$\pm$0.099 &12.00$\pm$--    &8.70$\pm$--     &1.37 &0.76  &0.90    &~2.50    &LERG\\
J1315+1831 &14.30$\pm$0.028 &13.52$\pm$0.030 &10.10$\pm$0.054 &9.02$\pm$0.116  &2.14 &0.90  &1.08    &~3.11    &LERG\\
J1315+3830 &14.37$\pm$0.028 &13.52$\pm$0.030 &10.10$\pm$0.054 &9.02$\pm$0.116  &7.26 &6.42  &11.47   &20.65    &HERG\\
J1512+4541 &15.92$\pm$0.041 &14.94$\pm$0.050 &11.56$\pm$0.125 &9.05$\pm$0.358  &0.97 &0.96  &1.66    &~2.22    &HERG\\
J1652+7408 &13.84$\pm$0.024 &13.53$\pm$0.027 &12.74$\pm$0.362 &8.93$\pm$--     &1.91 &1.02  &0.16    &~0.72    &LERG\\
J1731+3232 &12.14$\pm$0.023 &11.18$\pm$0.021 &~8.81$\pm$0.028 &6.28$\pm$0.048  &19.57&19.15 &13.00   &17.75    &Quasar\\
J2253+0615 &15.10$\pm$0.036 &14.78$\pm$0.066 &11.91$\pm$--    &7.79$\pm$--     &0.47 &0.33  &0.50    &~1.80    &LERG\\
J2256+1433 &15.02$\pm$0.036 &14.79$\pm$0.068 &12.56$\pm$--    &8.81$\pm$--     &3.77 &3.69  &11.04   &26.94    &LERG\\
J2306+1556 &15.74$\pm$0.036 &15.50$\pm$0.033 &12.38$\pm$0.047 &8.44$\pm$0.089  &1.02 &0.51  &0.70    &~3.51    &Quasar\\

\hline
\end{tabular}
\end{table*}

\setcounter{figure}{12}
\begin{figure*}
\includegraphics[height=3.5cm,width=3.5cm,angle=0]{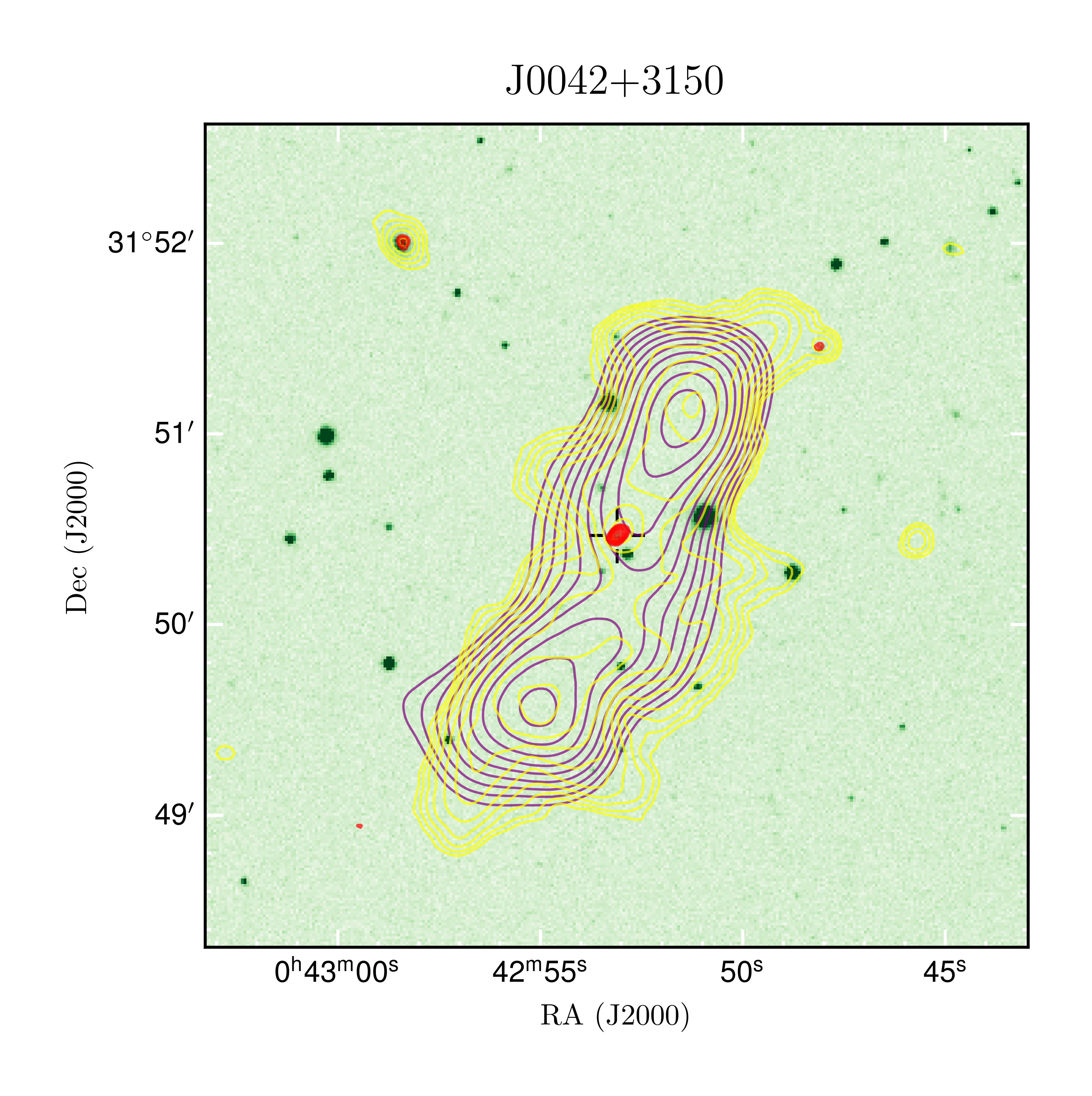}
\includegraphics[height=3.5cm,width=3.5cm,angle=0]{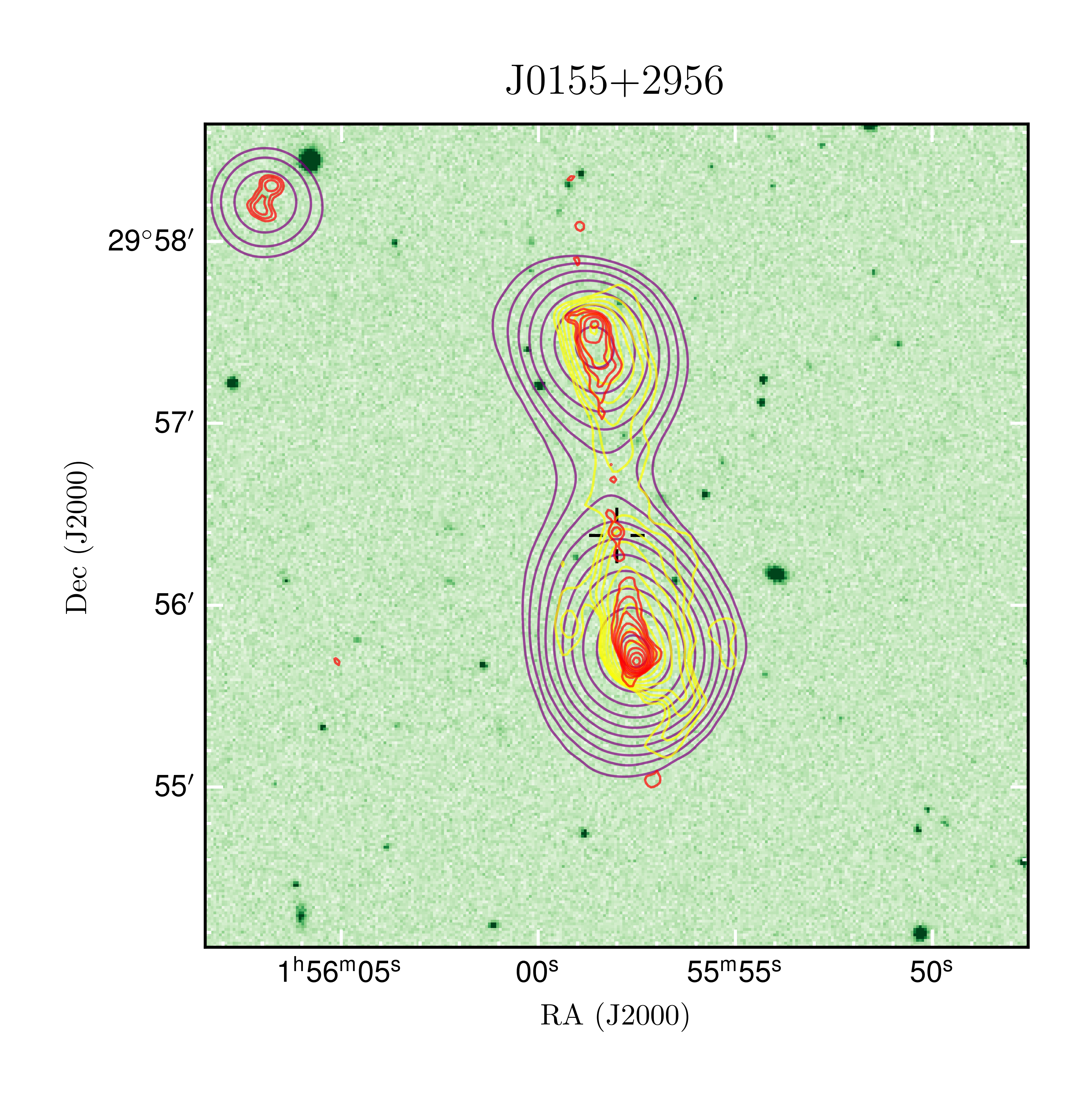}
\includegraphics[height=3.5cm,width=3.5cm,angle=0]{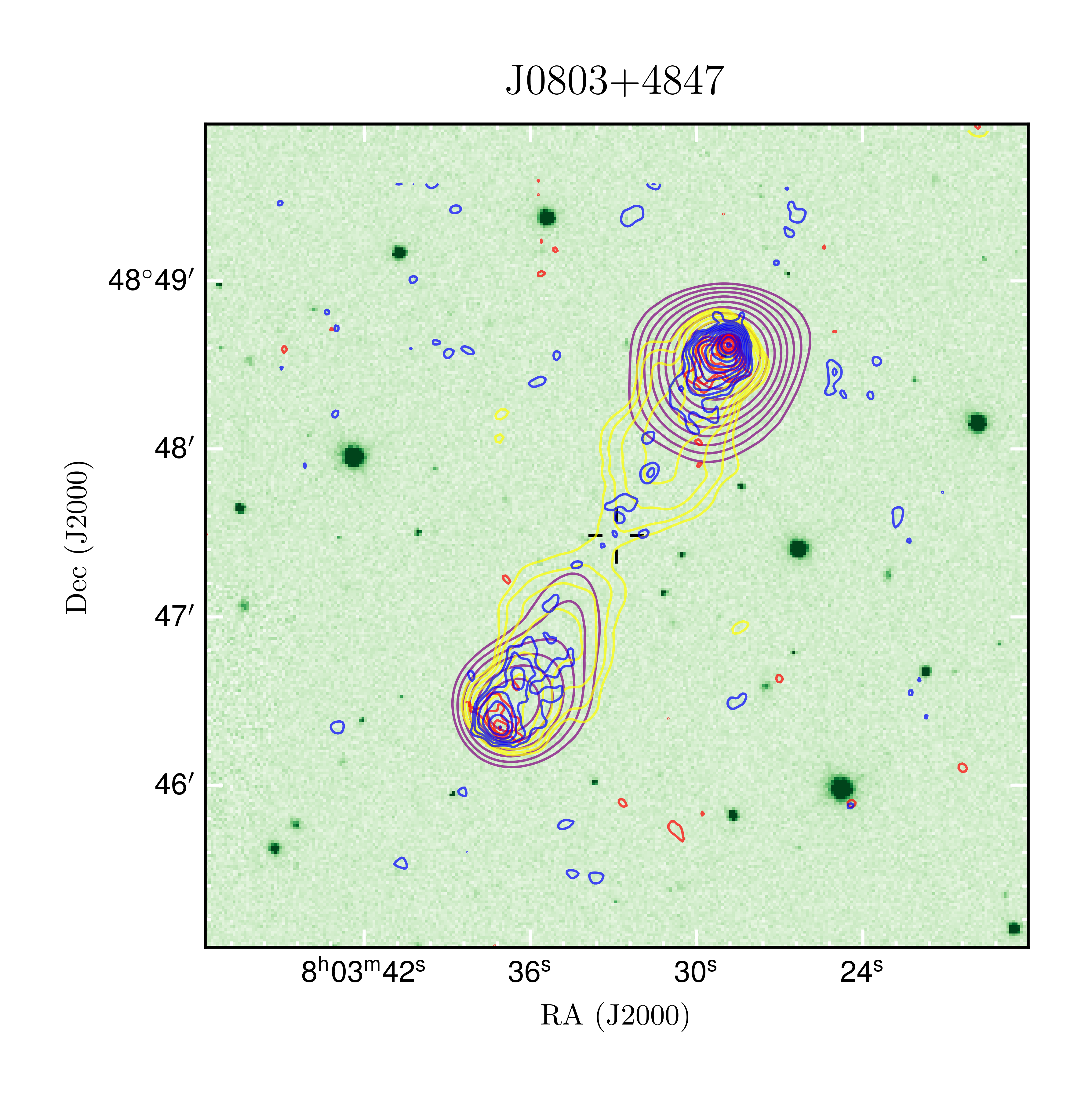}
\includegraphics[height=3.5cm,width=3.5cm,angle=0]{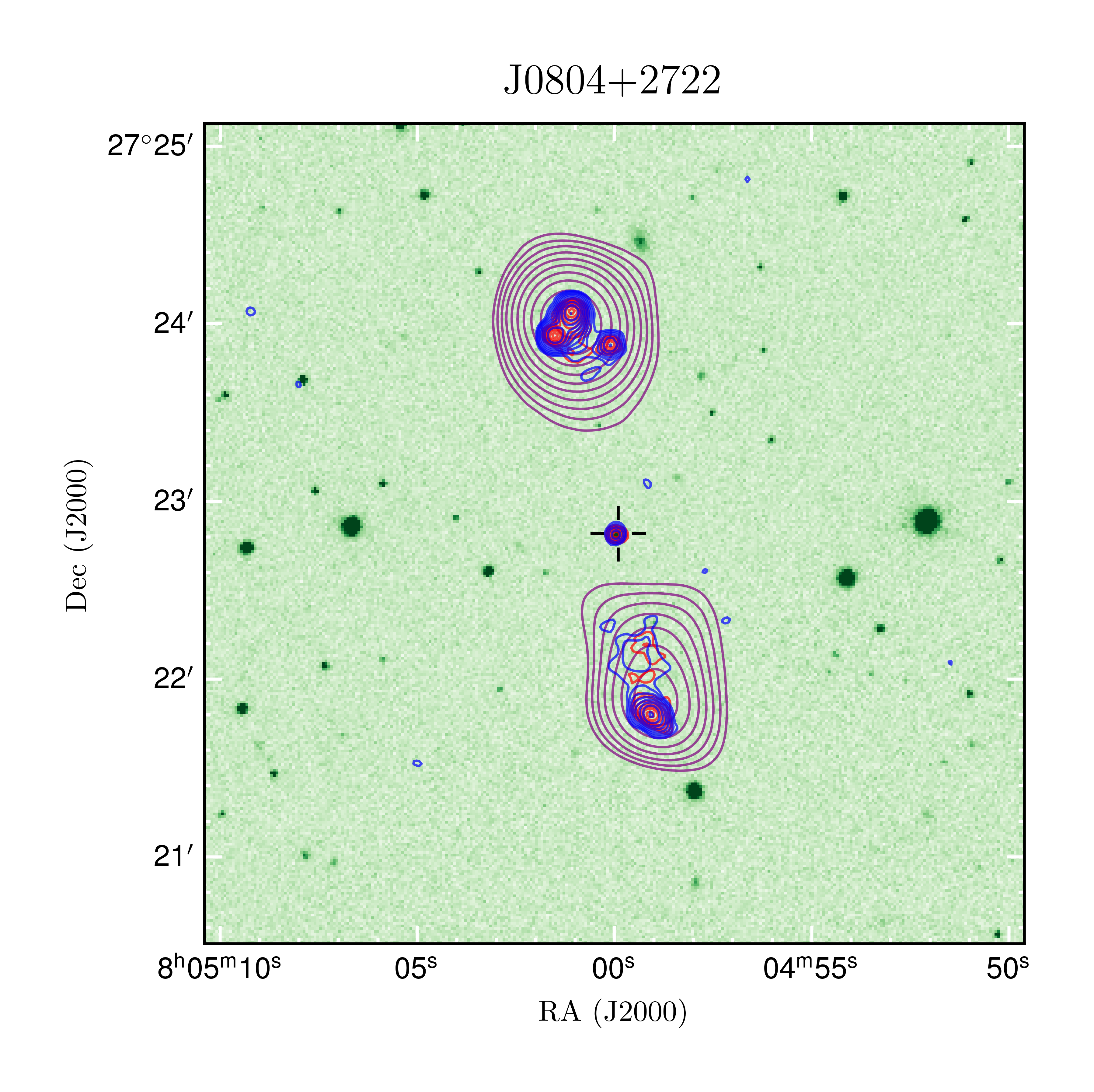}
\includegraphics[height=3.5cm,width=3.5cm,angle=0]{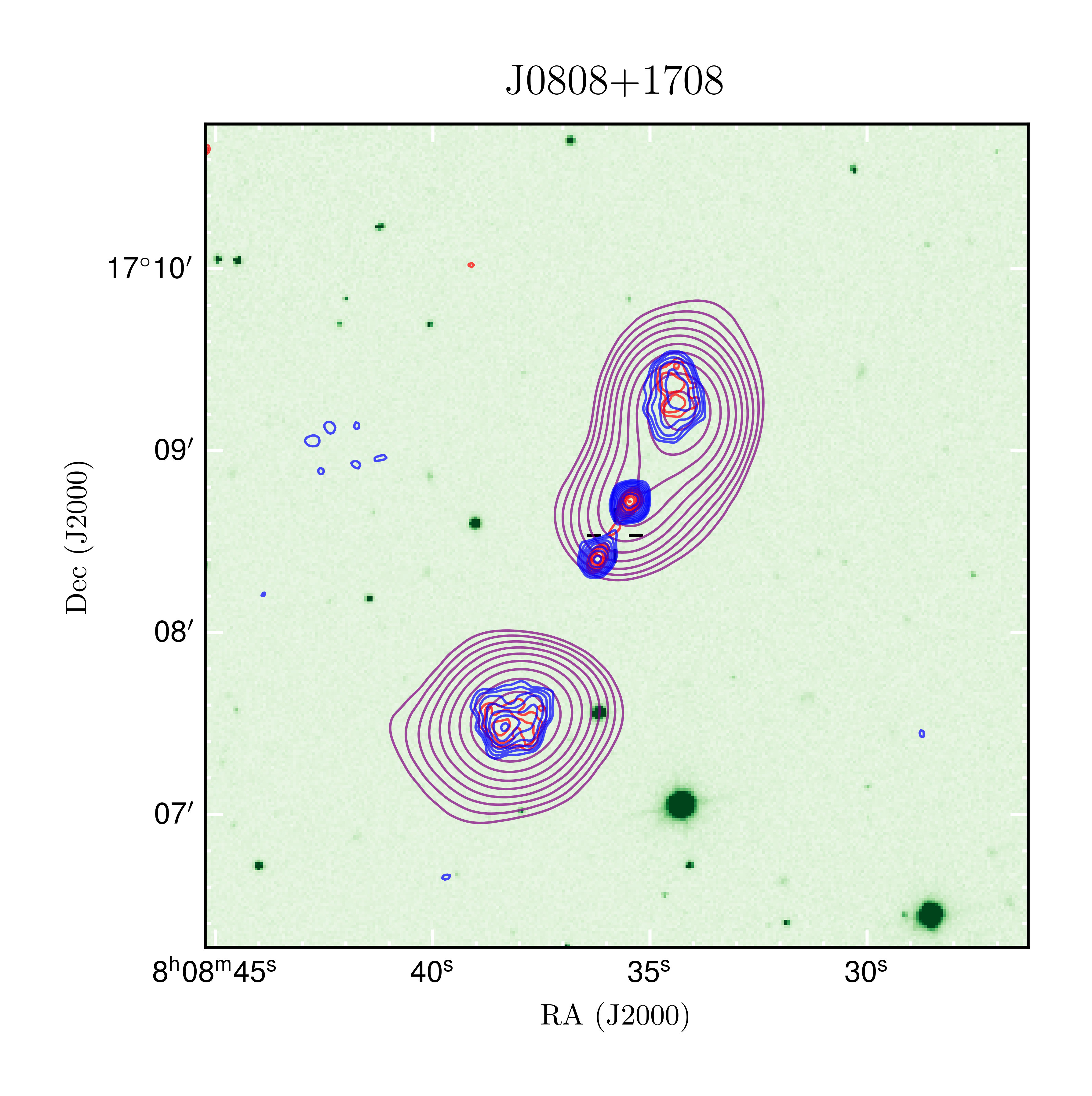}
\vskip 0.2cm 
\includegraphics[height=3.5cm,width=3.5cm,angle=0]{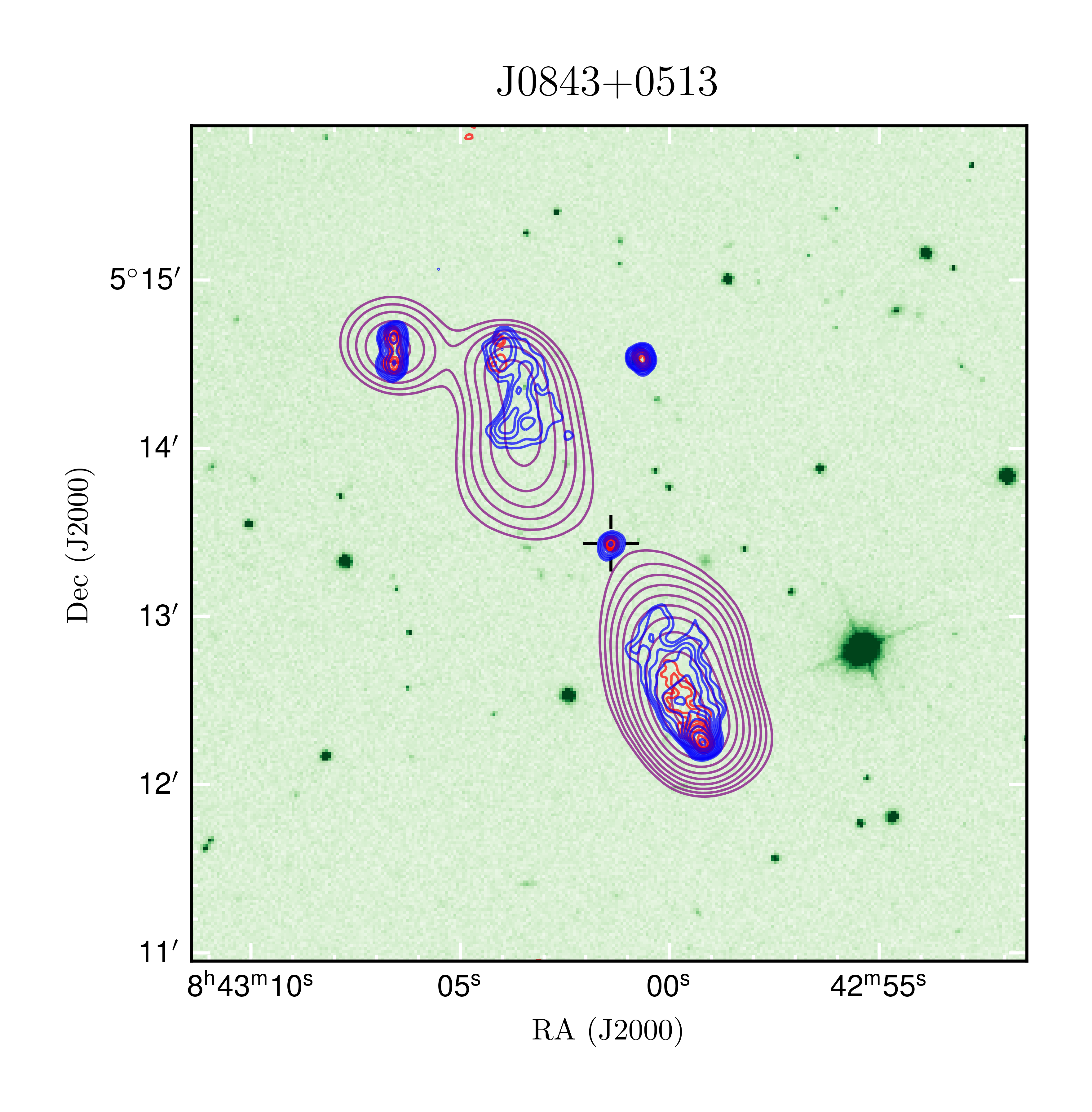}
\includegraphics[height=3.5cm,width=3.5cm,angle=0]{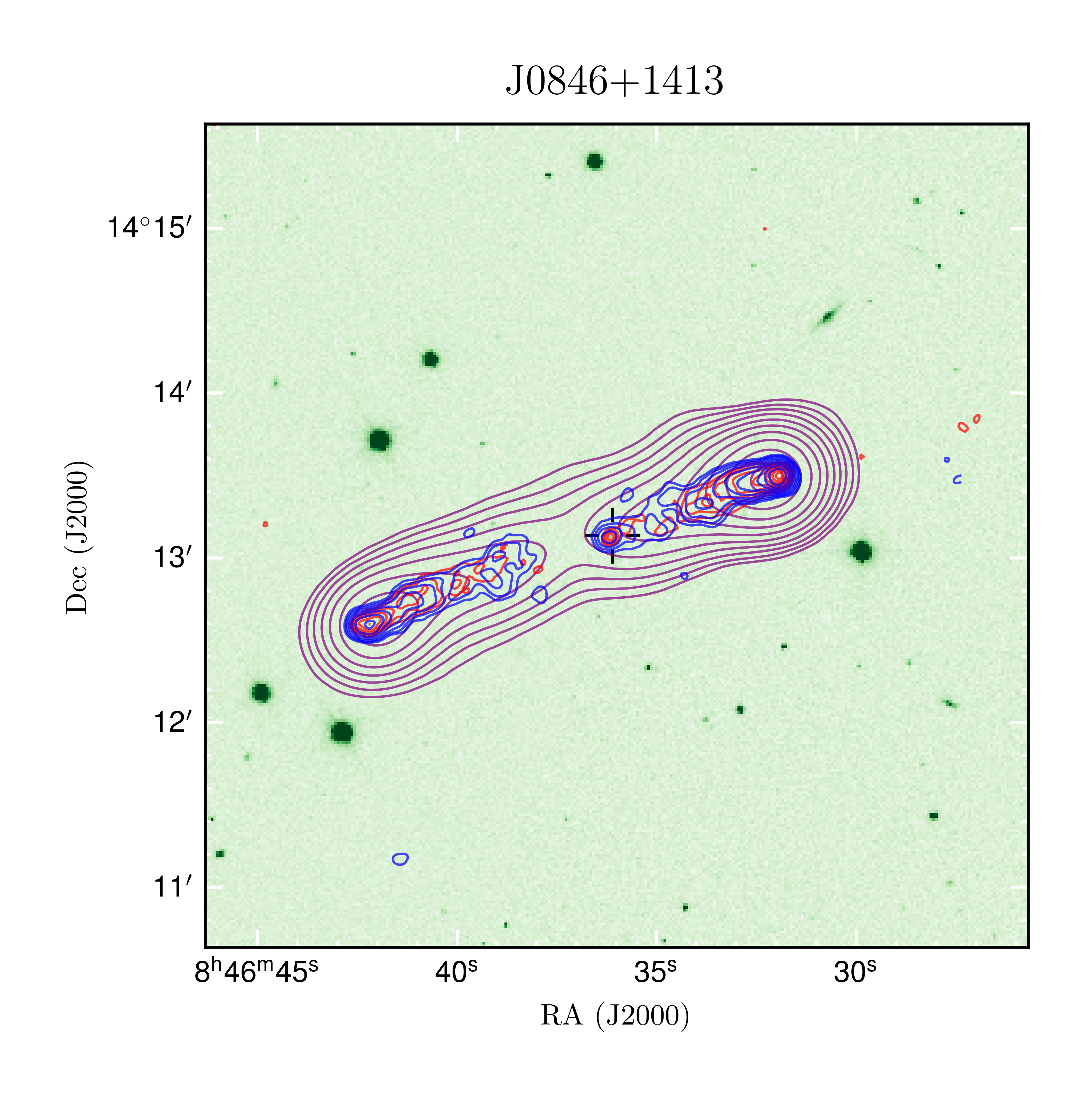}
\includegraphics[height=3.5cm,width=3.5cm,angle=0]{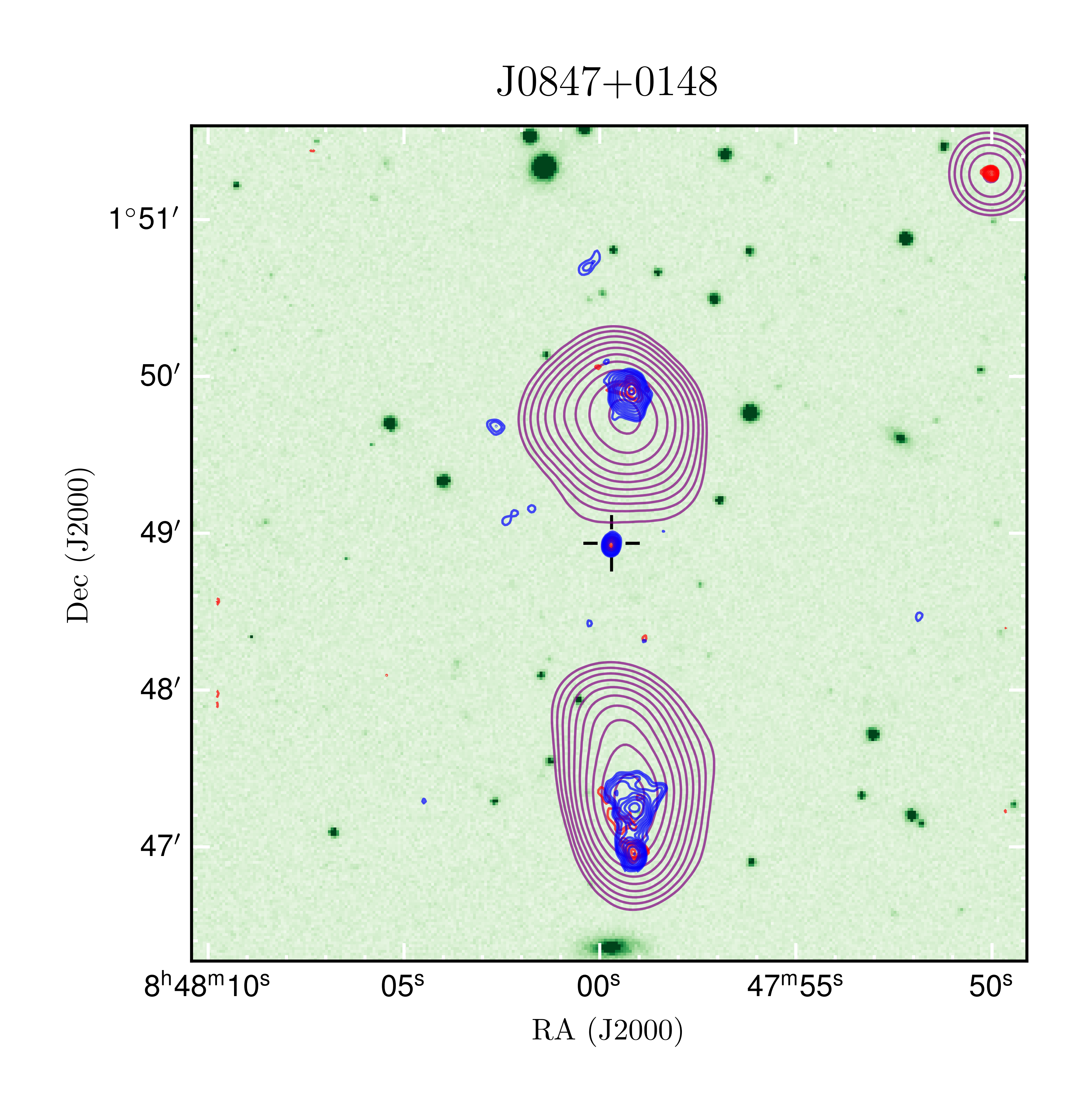}
\includegraphics[height=3.5cm,width=3.5cm,angle=0]{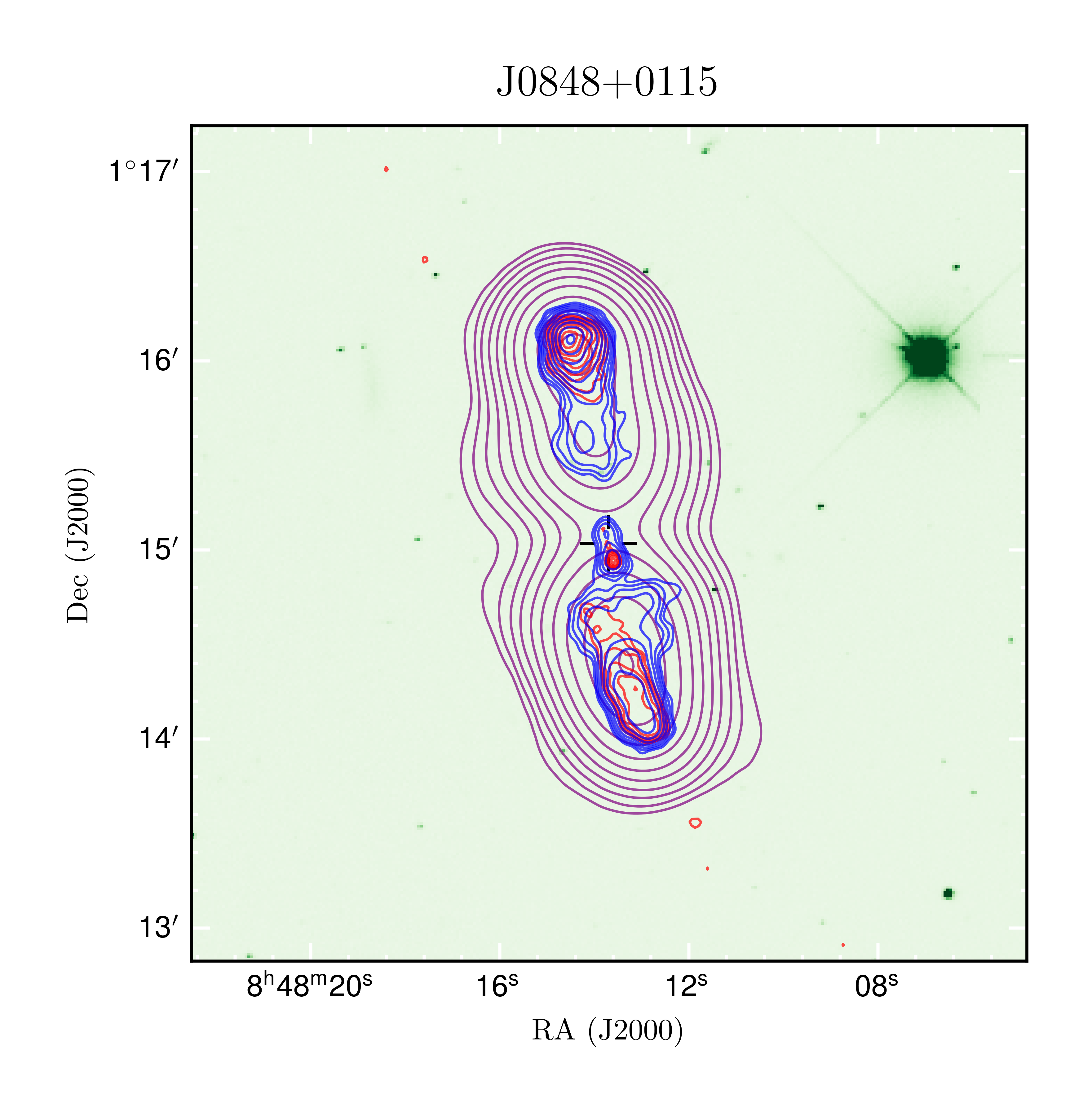}
\includegraphics[height=3.5cm,width=3.5cm,angle=0]{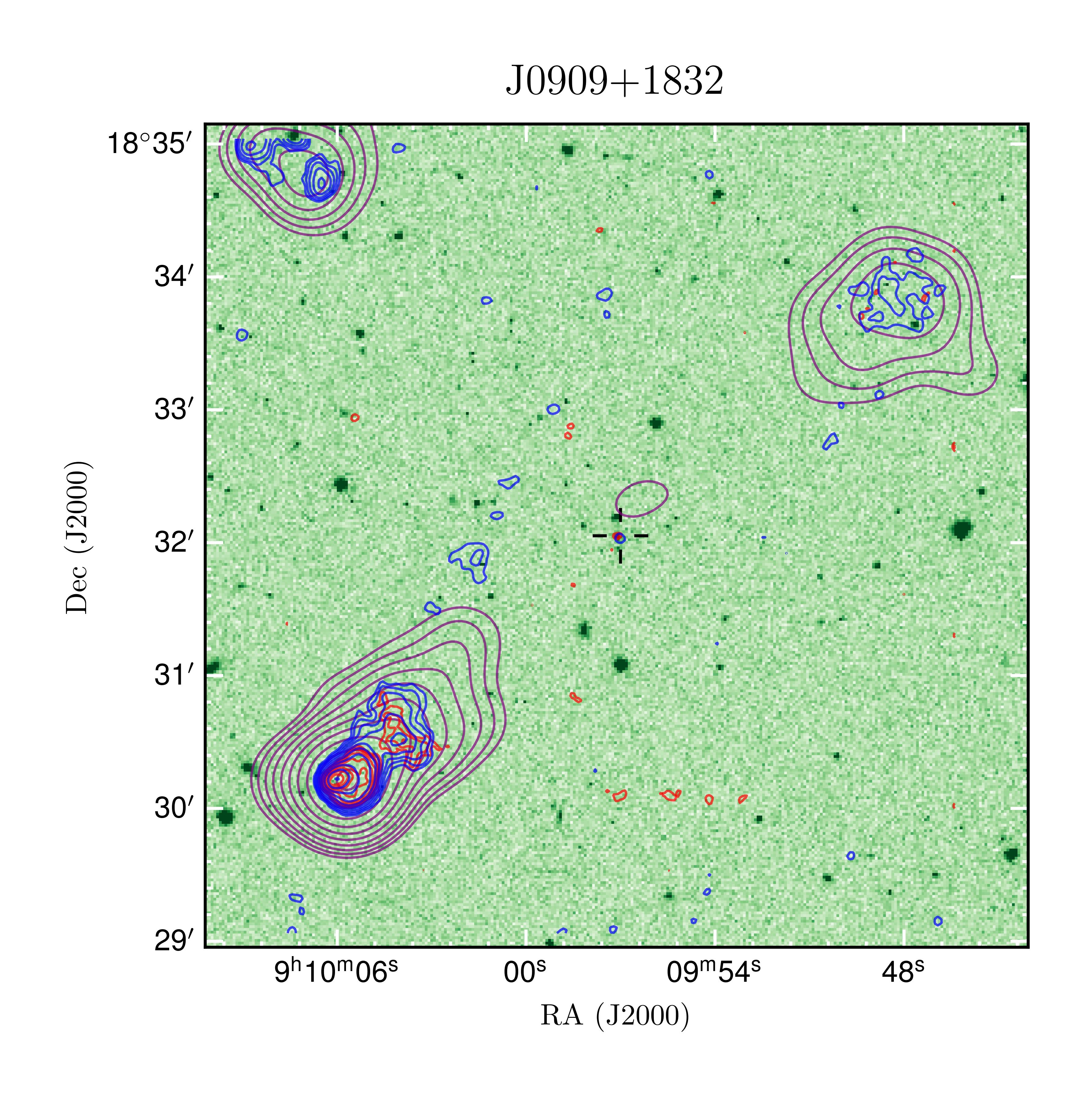}
\vskip 0.2cm
\includegraphics[height=3.5cm,width=3.5cm,angle=0]{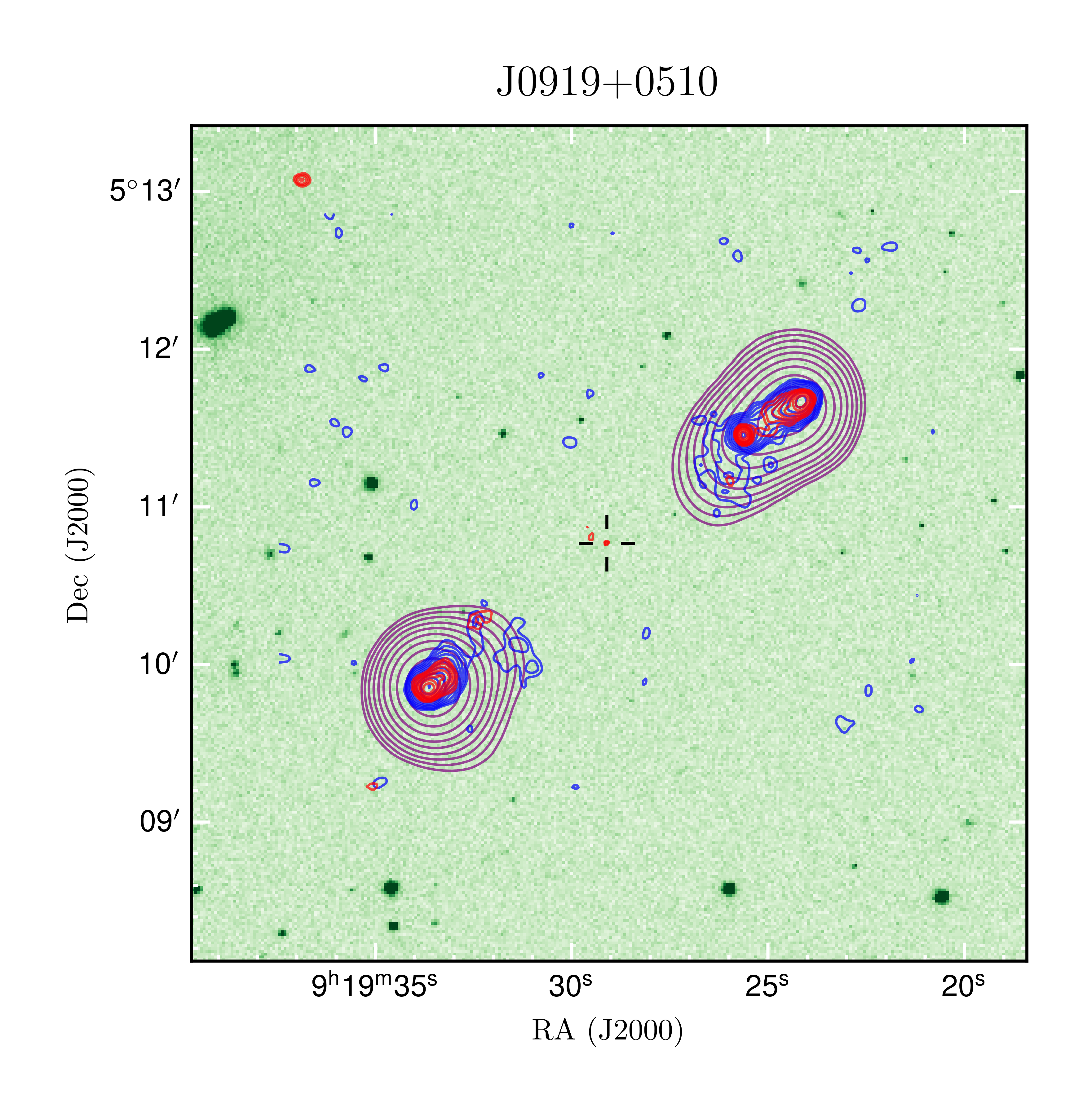}
\includegraphics[height=3.5cm,width=3.5cm,angle=0]{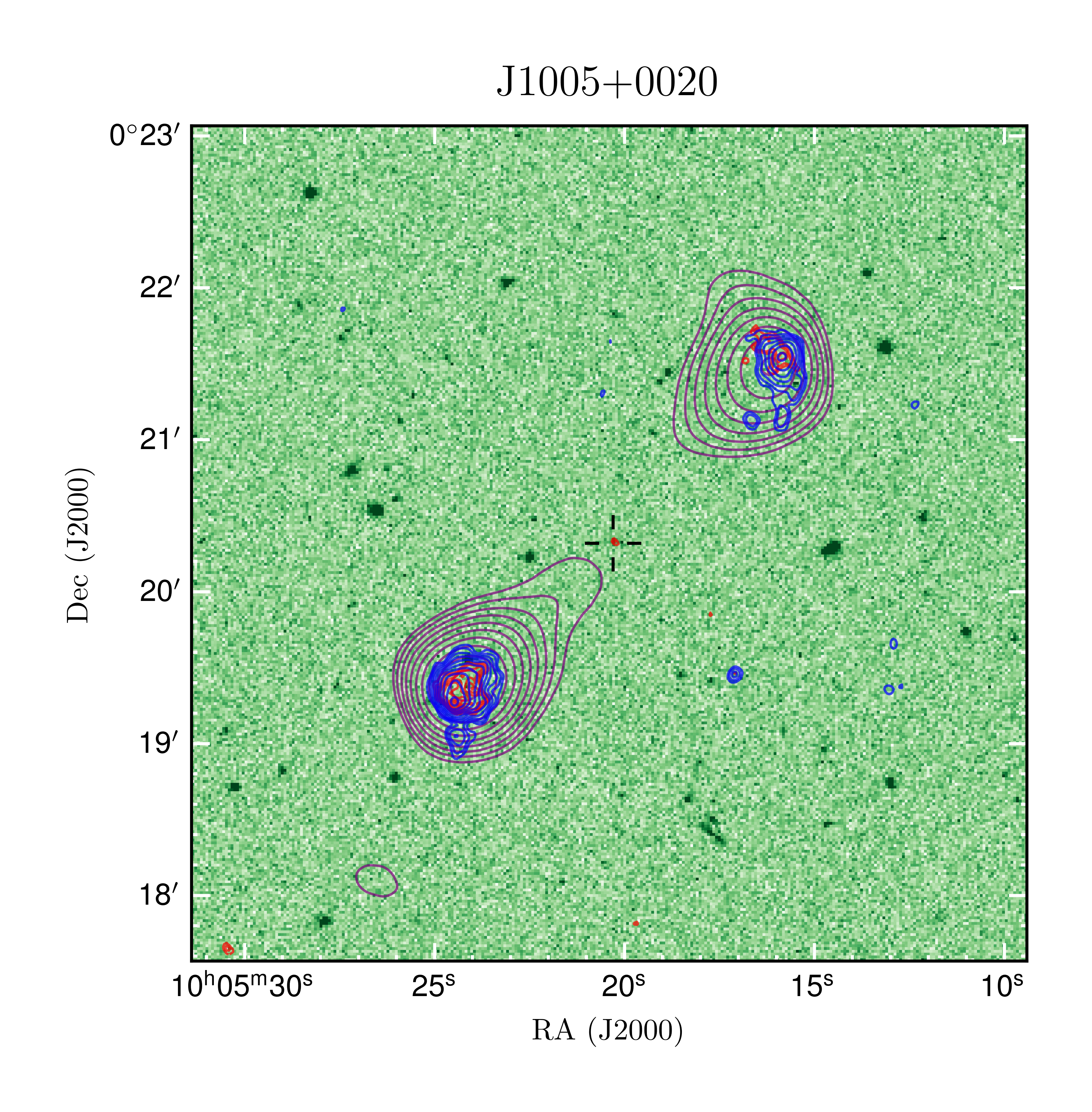}
\includegraphics[height=3.5cm,width=3.5cm,angle=0]{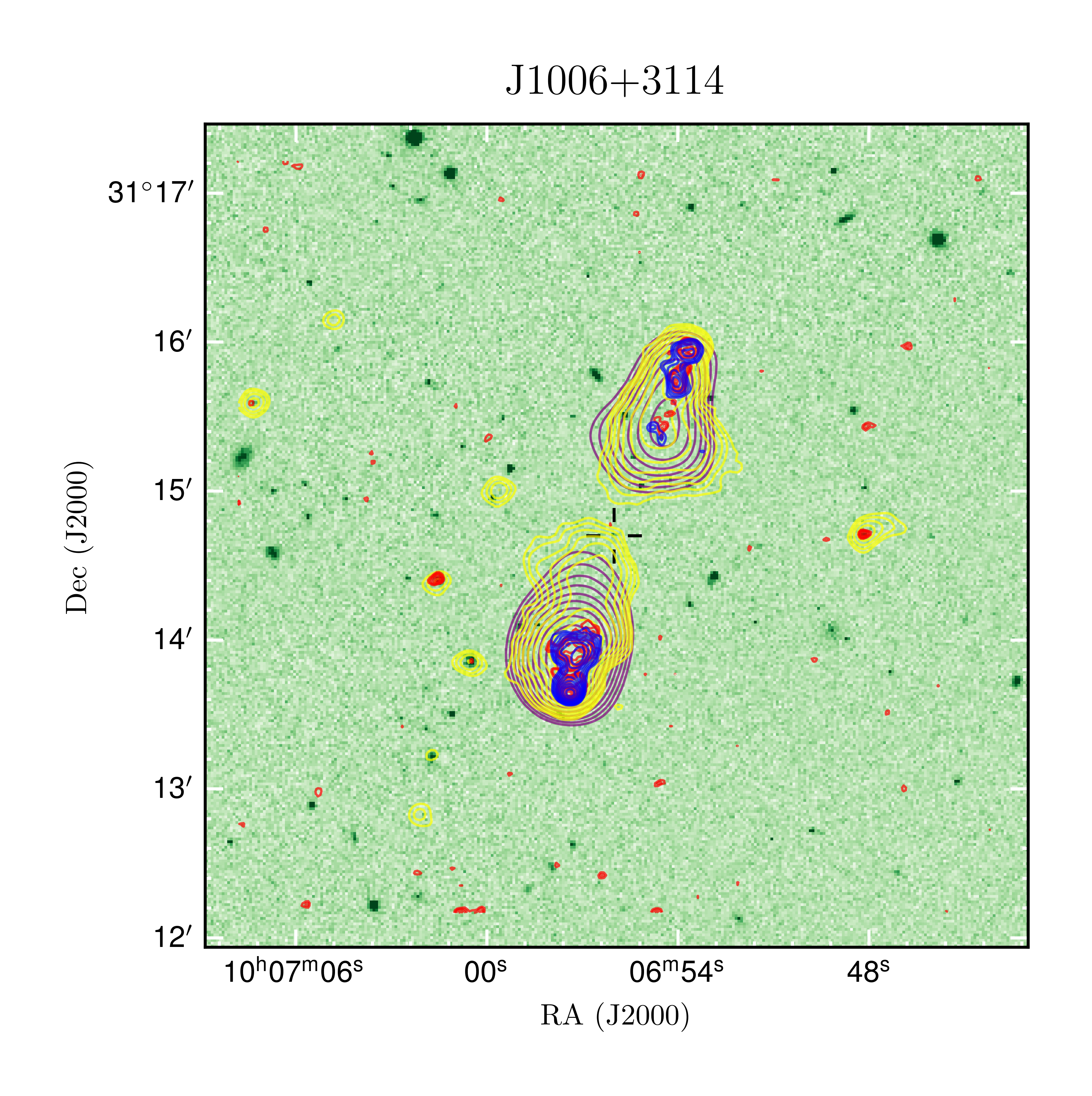}
\includegraphics[height=3.5cm,width=3.5cm,angle=0]{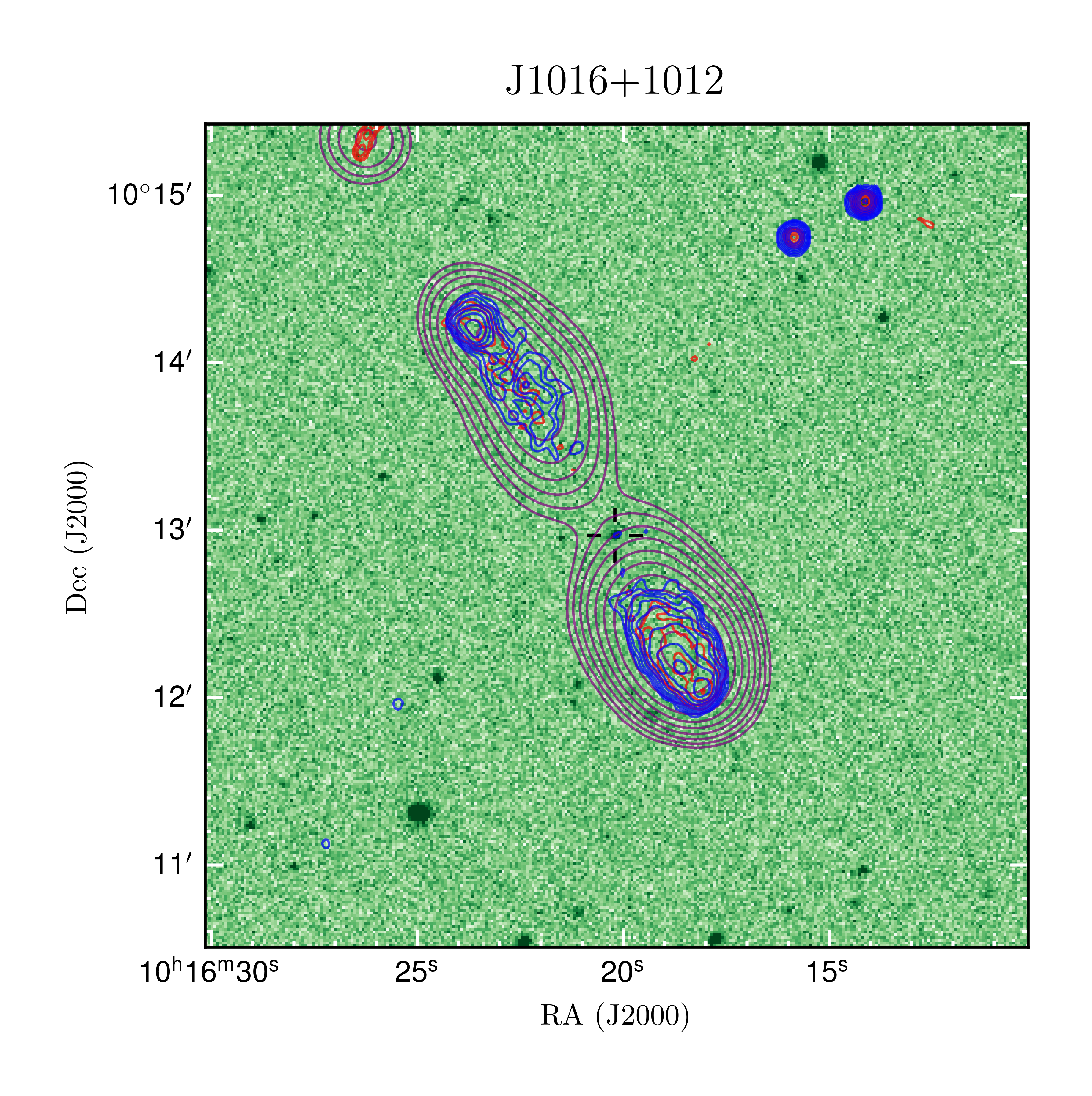}
\includegraphics[height=3.5cm,width=3.5cm,angle=0]{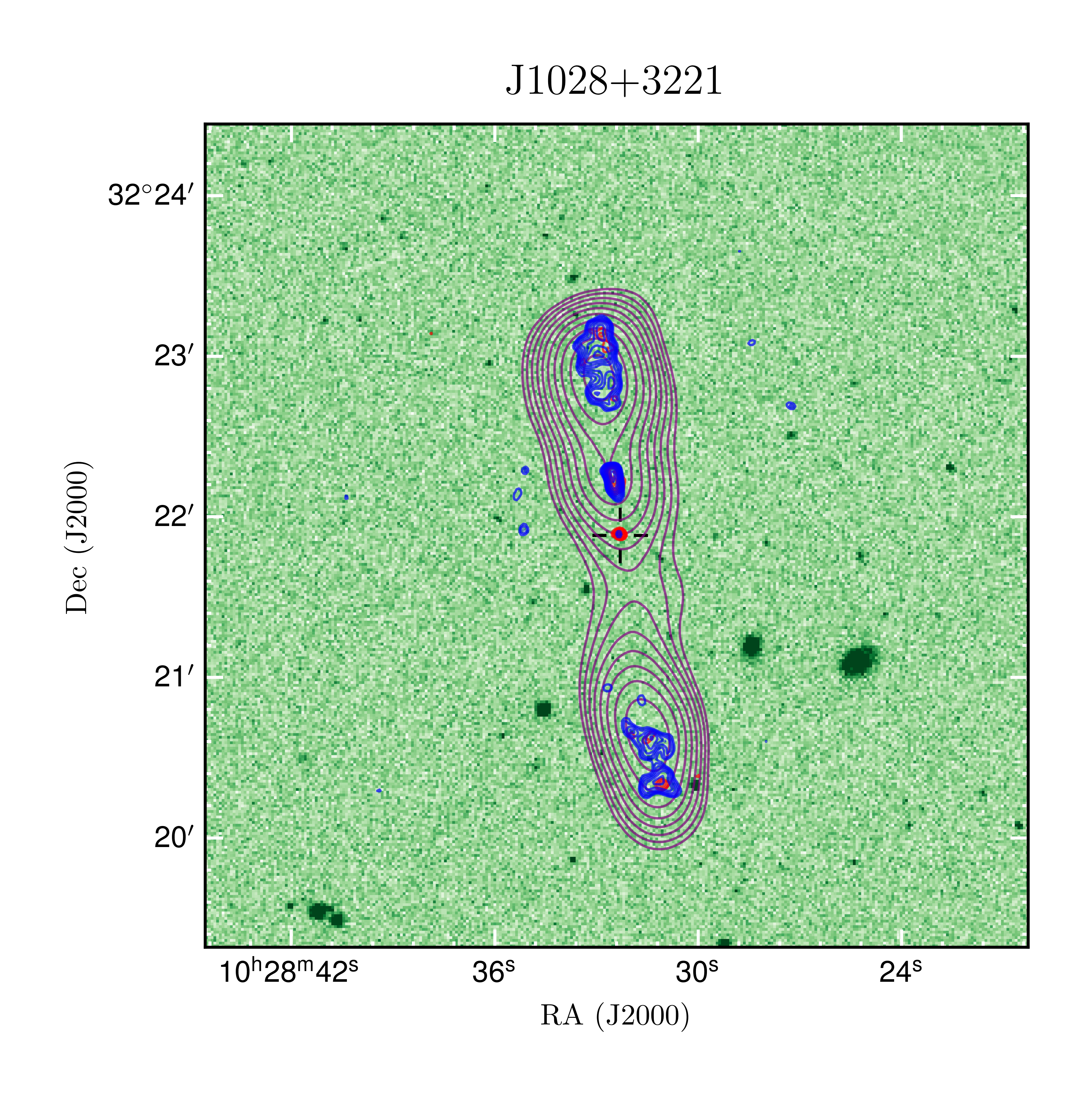}
\vskip 0.2cm
\includegraphics[height=3.5cm,width=3.5cm,angle=0]{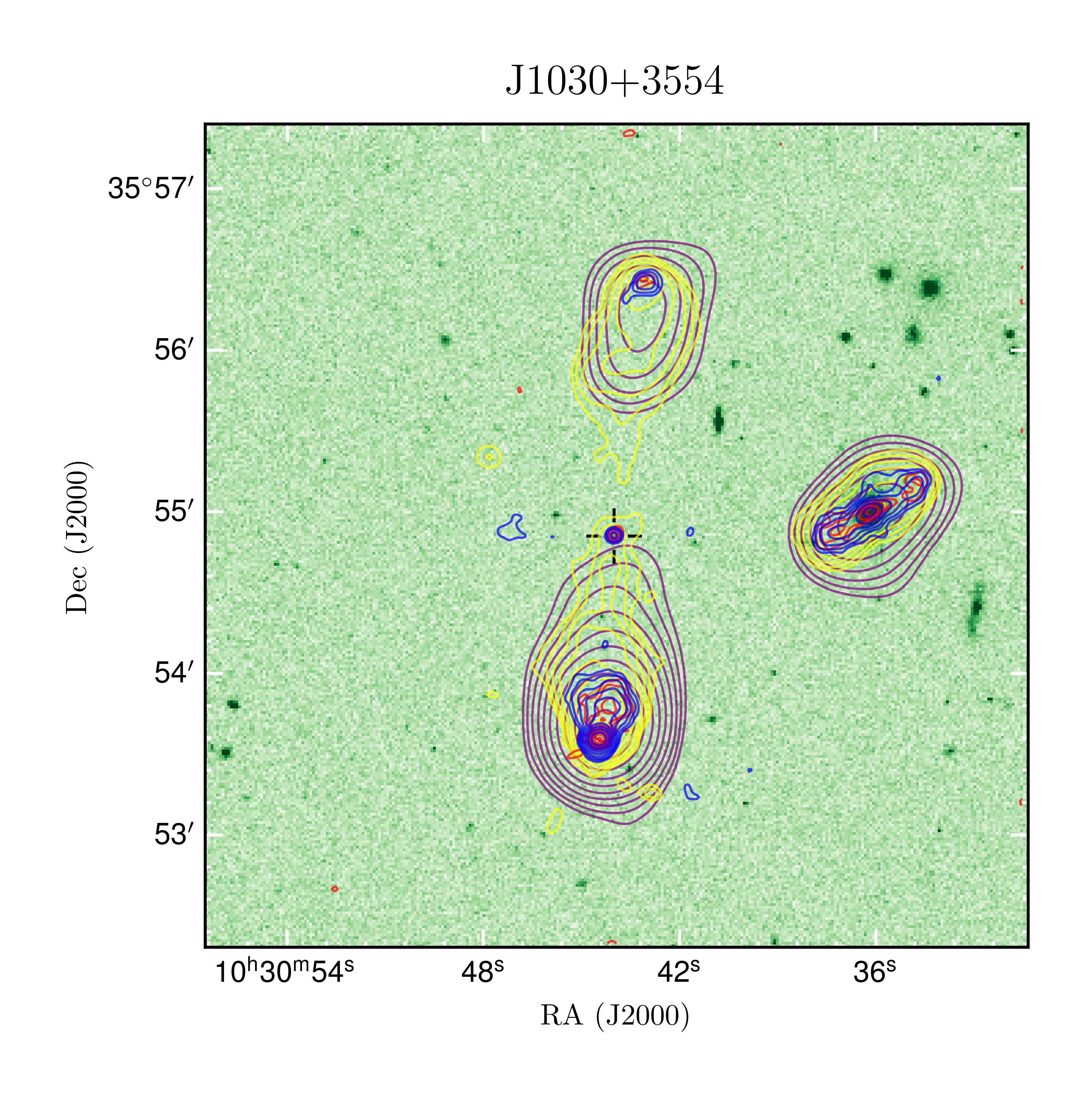}
\includegraphics[height=3.5cm,width=3.5cm,angle=0]{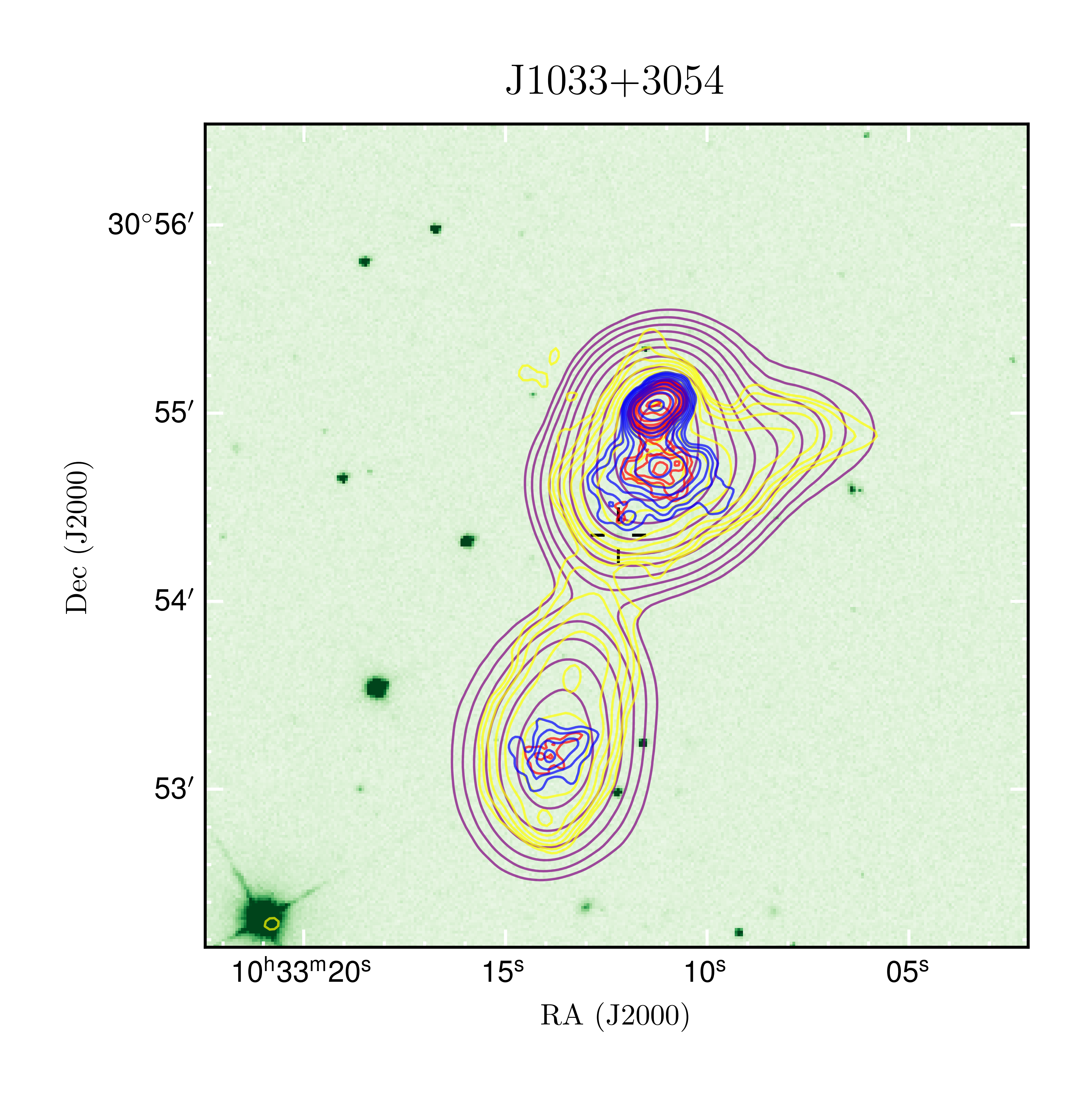}
\includegraphics[height=3.5cm,width=3.5cm,angle=0]{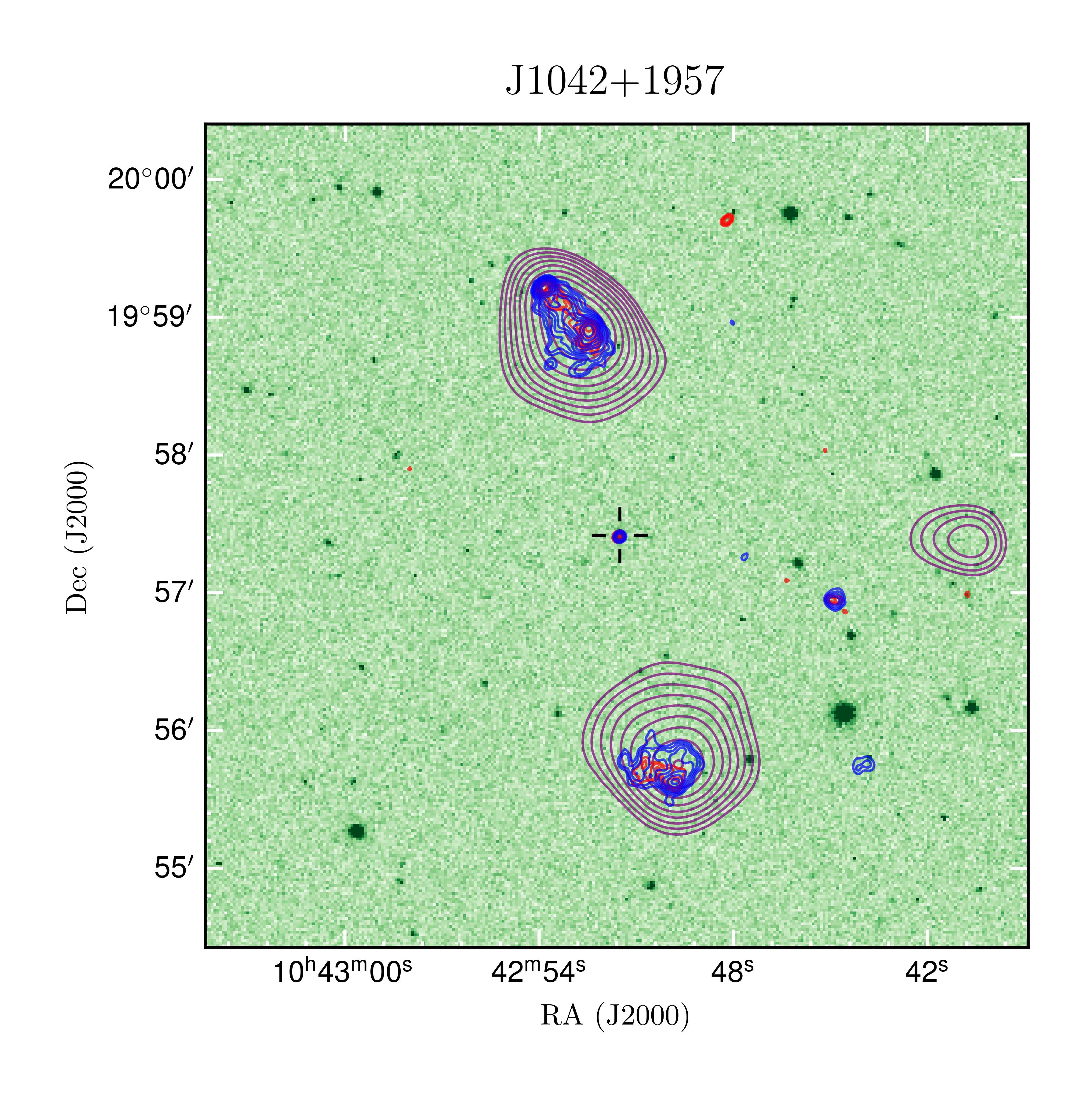}
\includegraphics[height=3.5cm,width=3.5cm,angle=0]{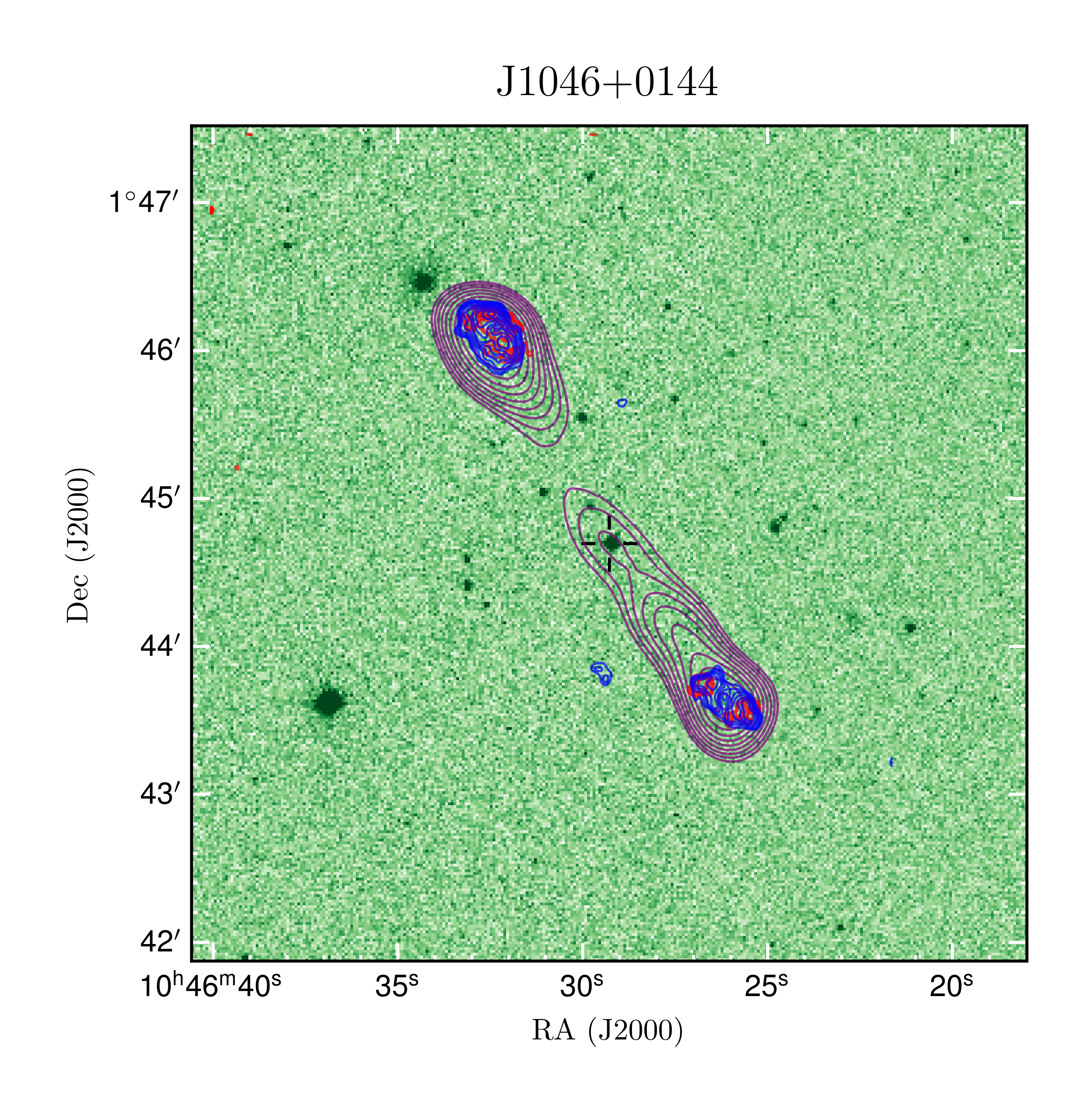}
\includegraphics[height=3.5cm,width=3.5cm,angle=0]{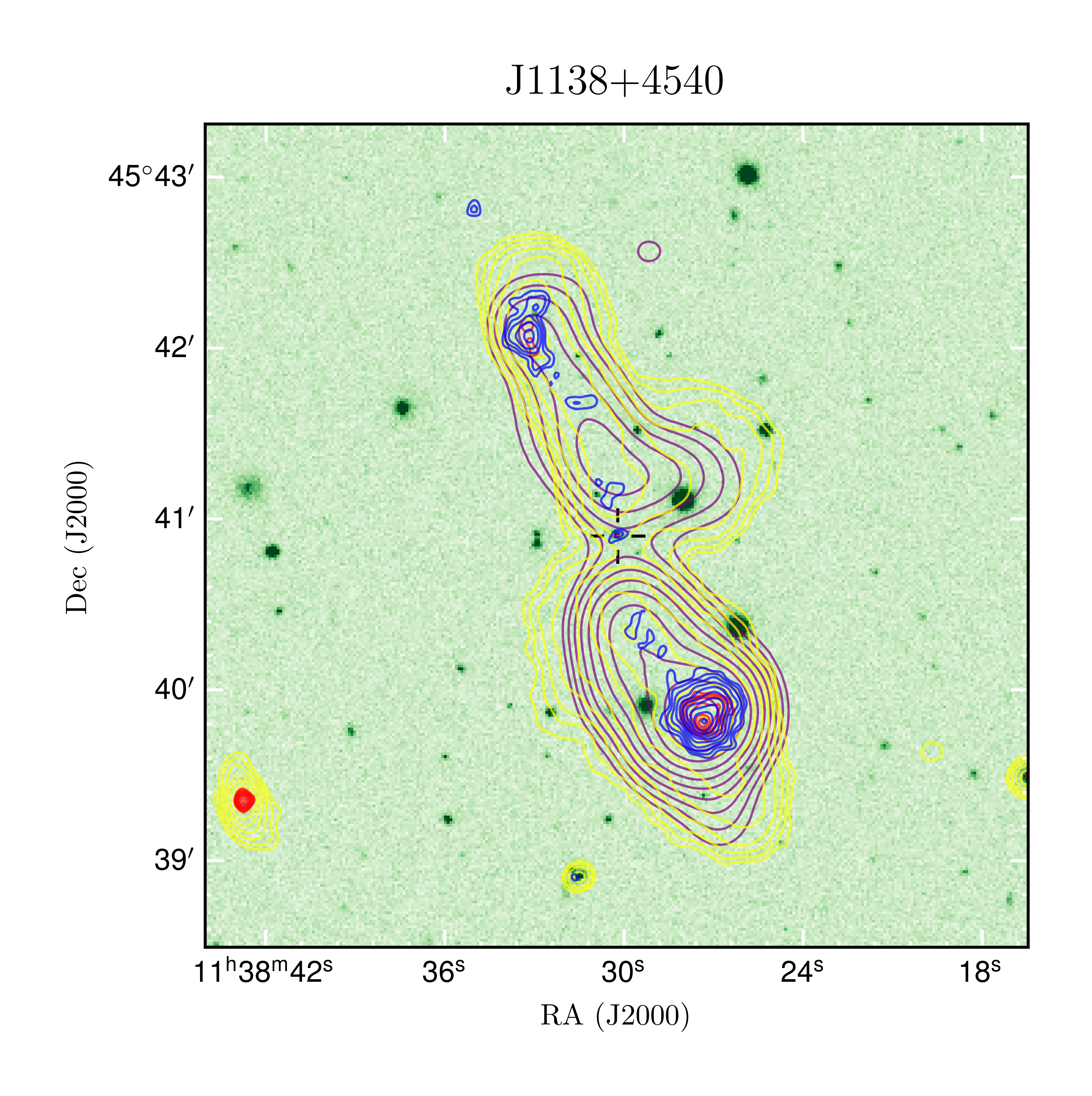}
\vskip 0.2cm
\includegraphics[height=3.5cm,width=3.5cm,angle=0]{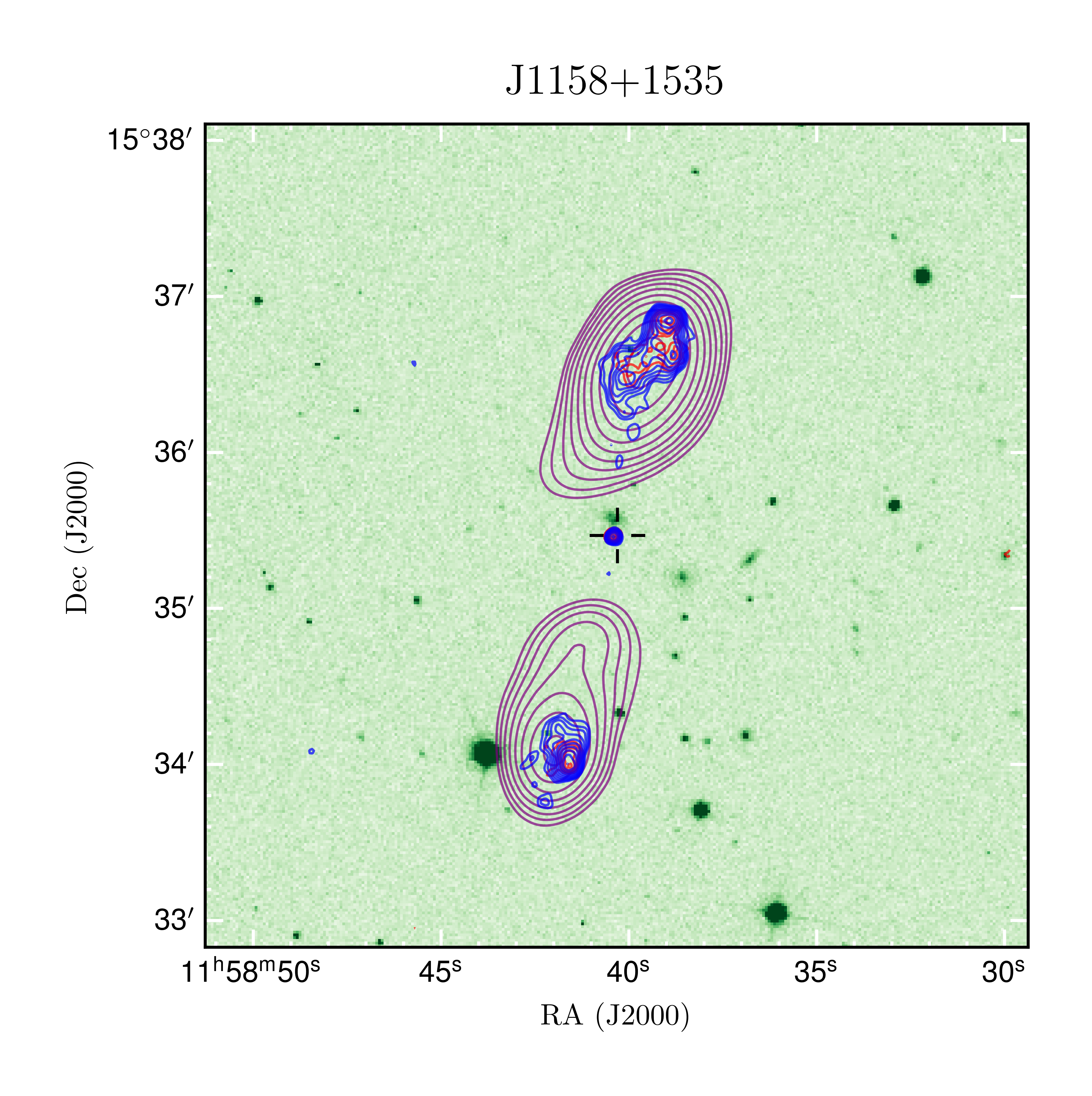}
\includegraphics[height=3.5cm,width=3.5cm,angle=0]{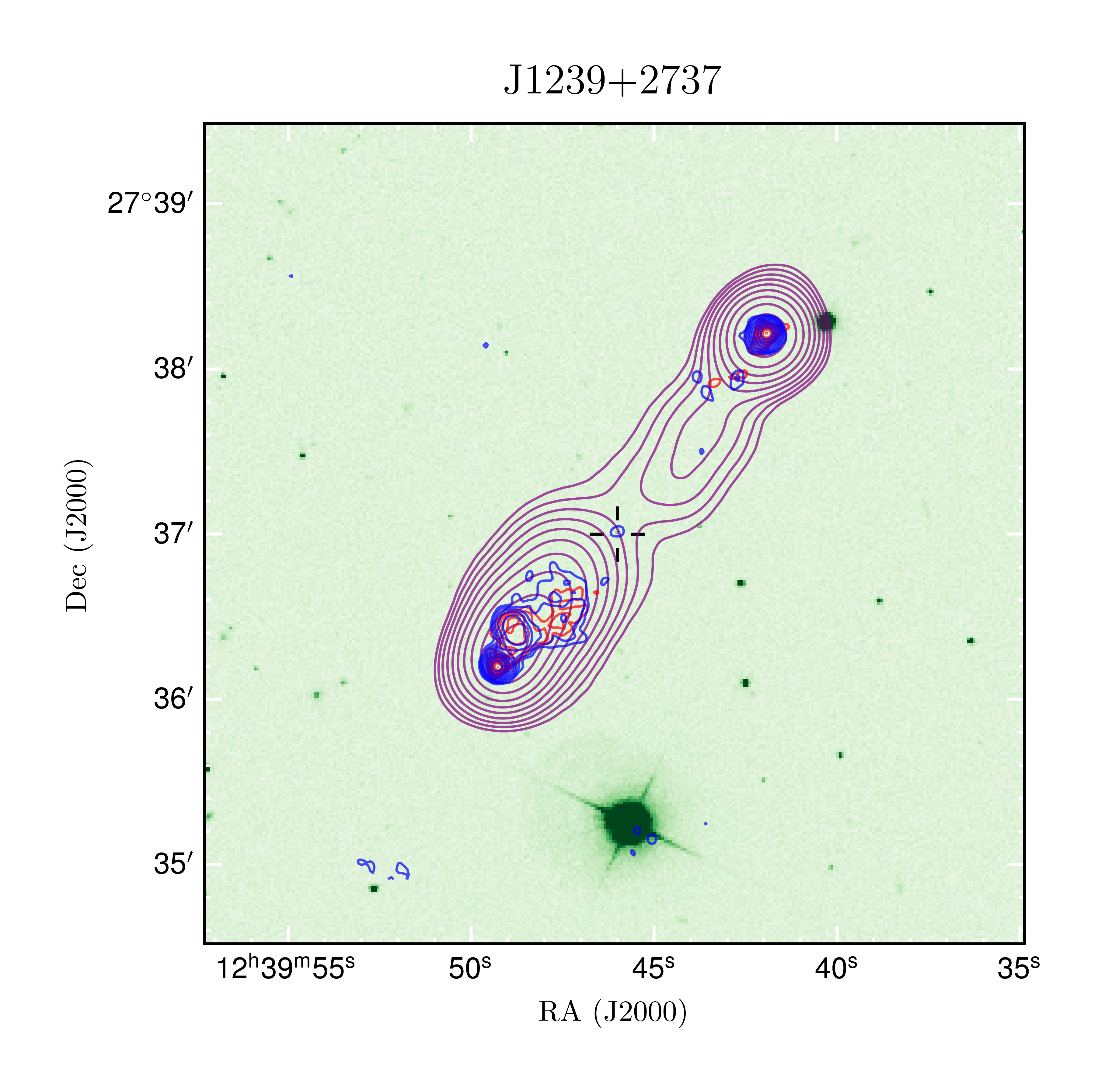}
\includegraphics[height=3.5cm,width=3.5cm,angle=0]{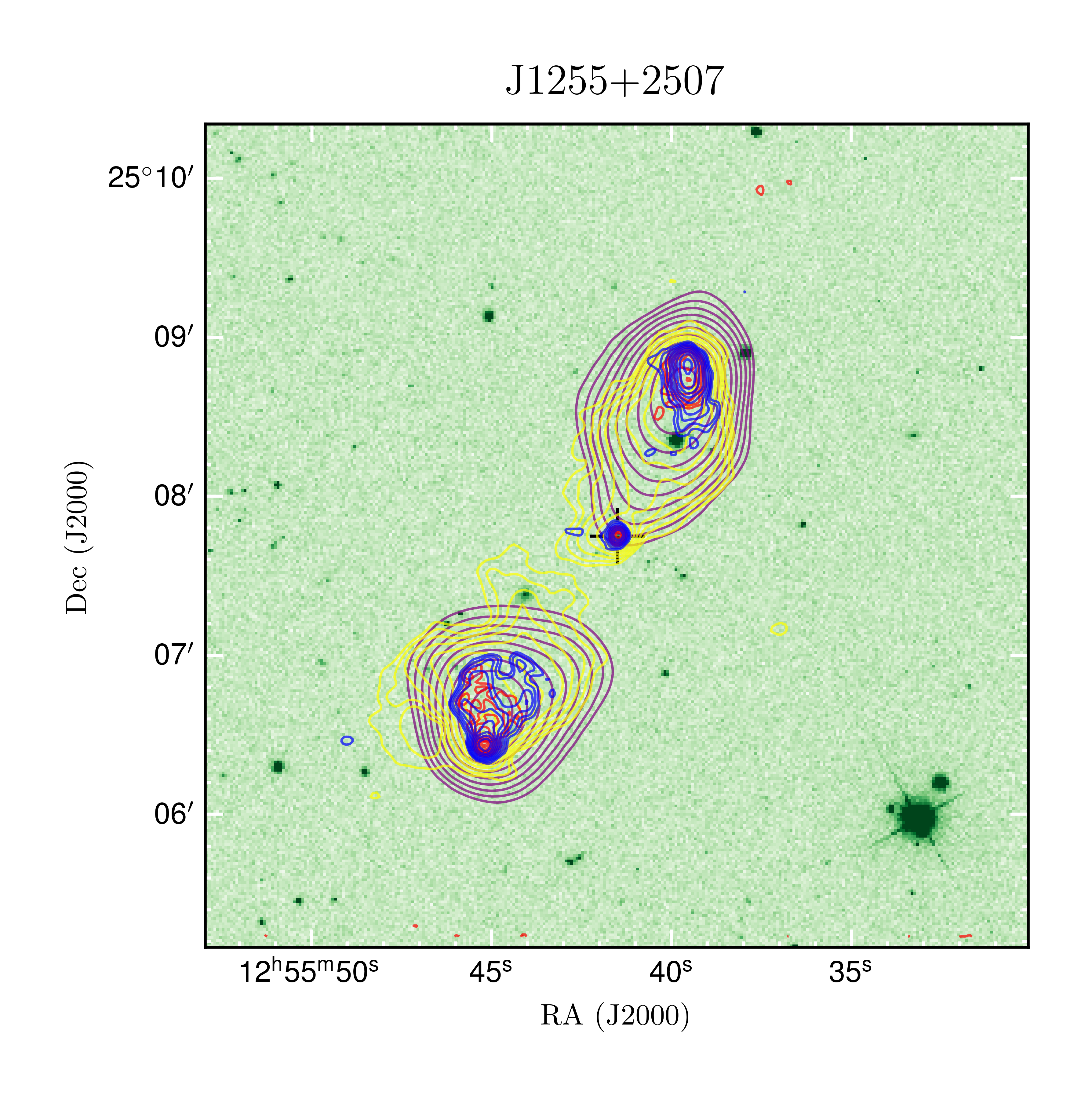}
\includegraphics[height=3.5cm,width=3.5cm,angle=0]{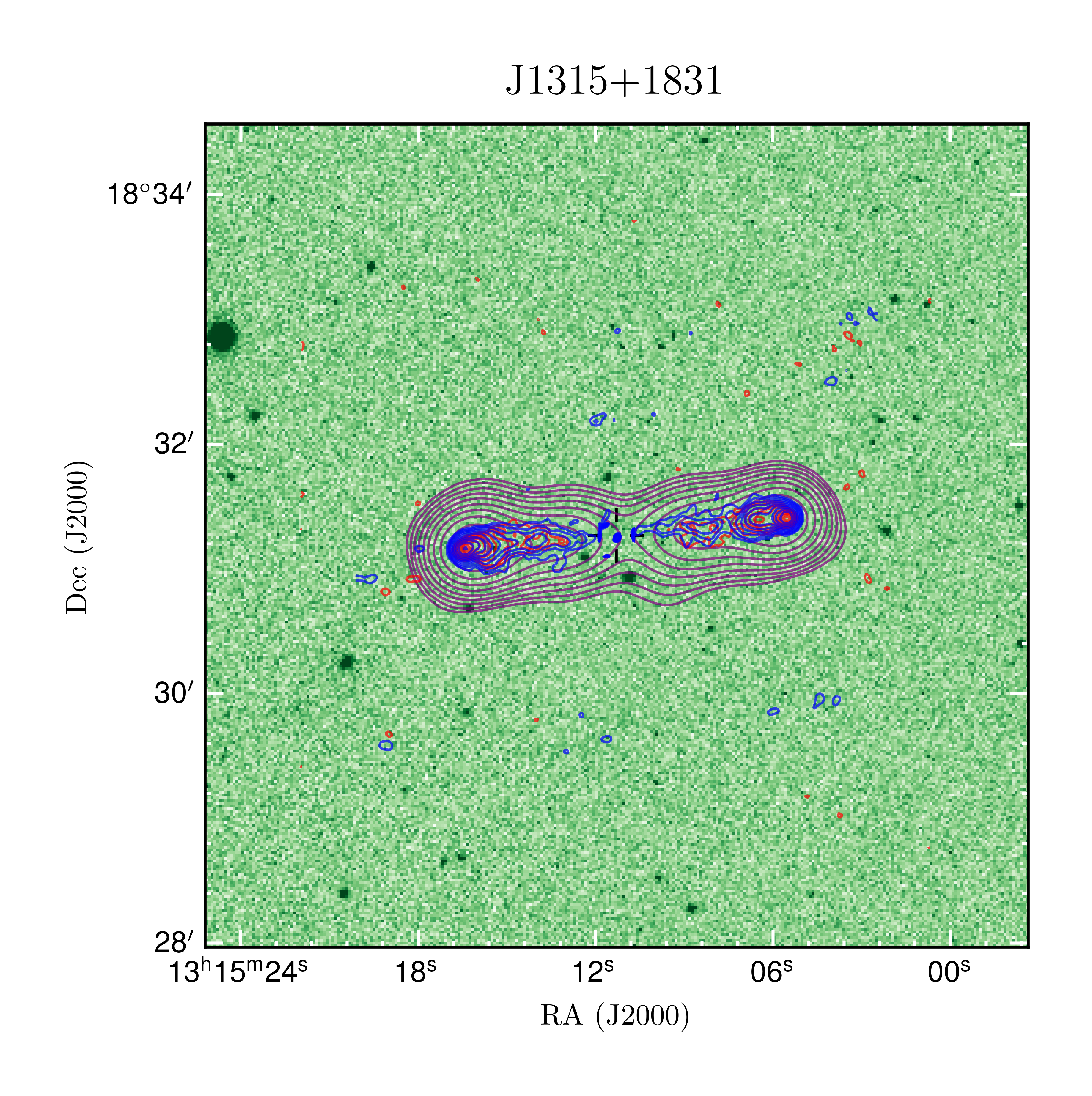}
\includegraphics[height=3.5cm,width=3.5cm,angle=0]{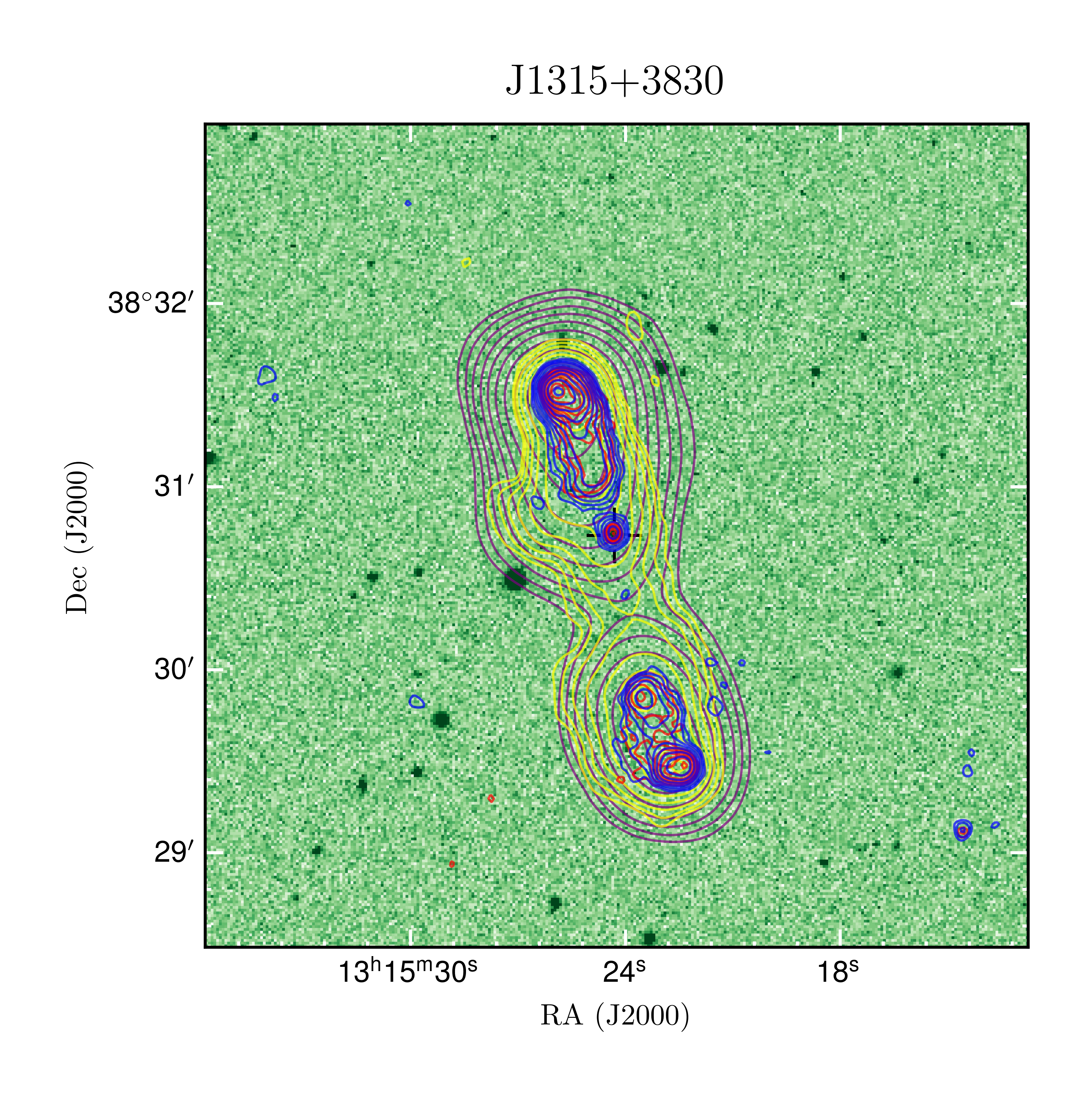}
\vskip 0.2cm
\caption {The Figure shows an multiwavelength overlay images of GRQs and GRGs. SDSS/Pan-STARRS r-band optical images are overlaid with TGSS (purple contours), LoTss-DR2 (yellow contours), FIRST (blue contours), and VLASS (where first coverage is not available) (red contours) images. For all images, contours are overplotted with ten levels spaced in log scale with the lowest level of 3$\sigma$. The black cursor mark represents the position of the host counterpart.}
\label{Fig:GRS}
\end{figure*}

\setcounter{figure}{12}
\begin{figure*}
\includegraphics[height=3.5cm,width=3.5cm,angle=0]{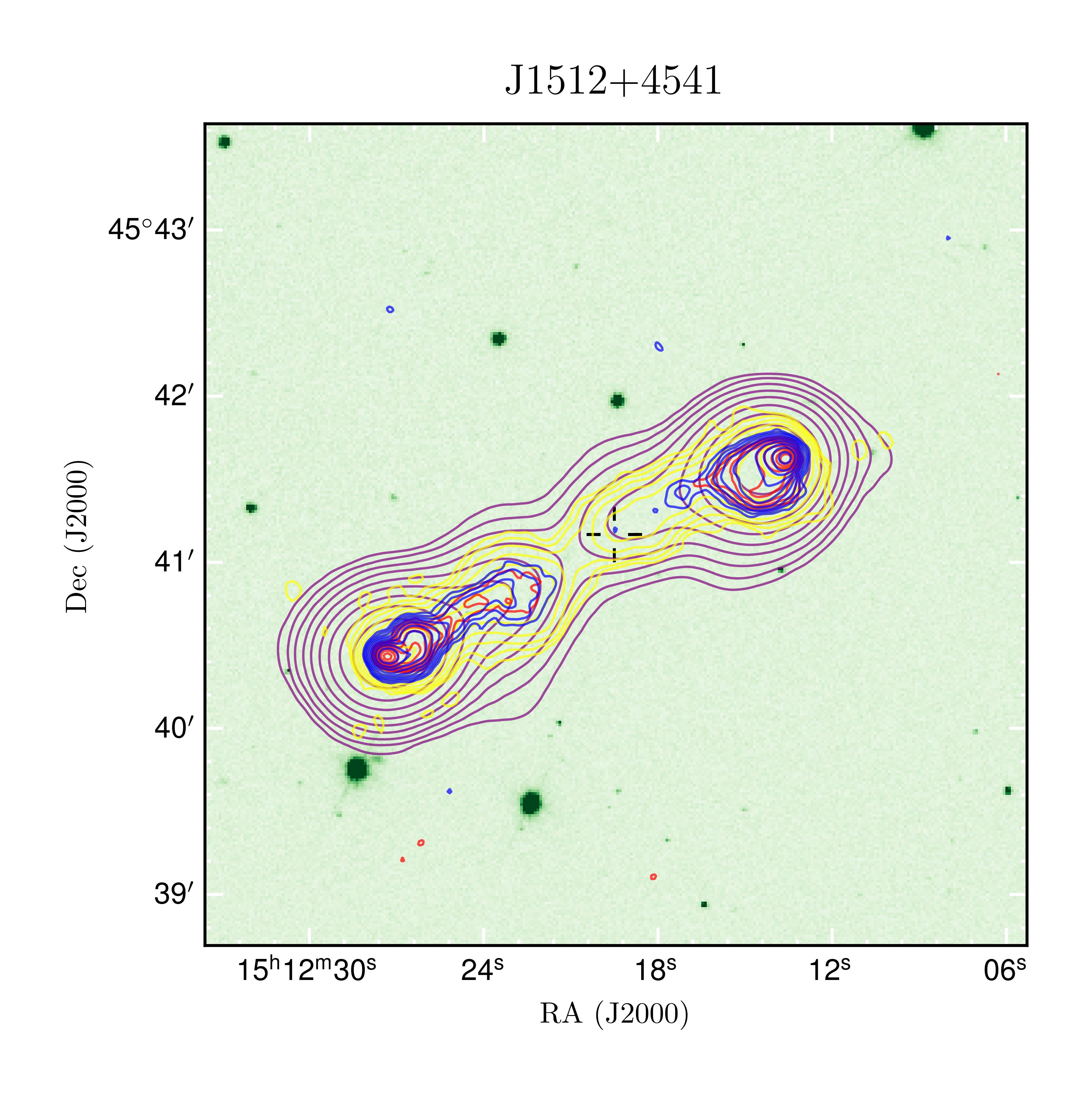}
\includegraphics[height=3.5cm,width=3.5cm,angle=0]{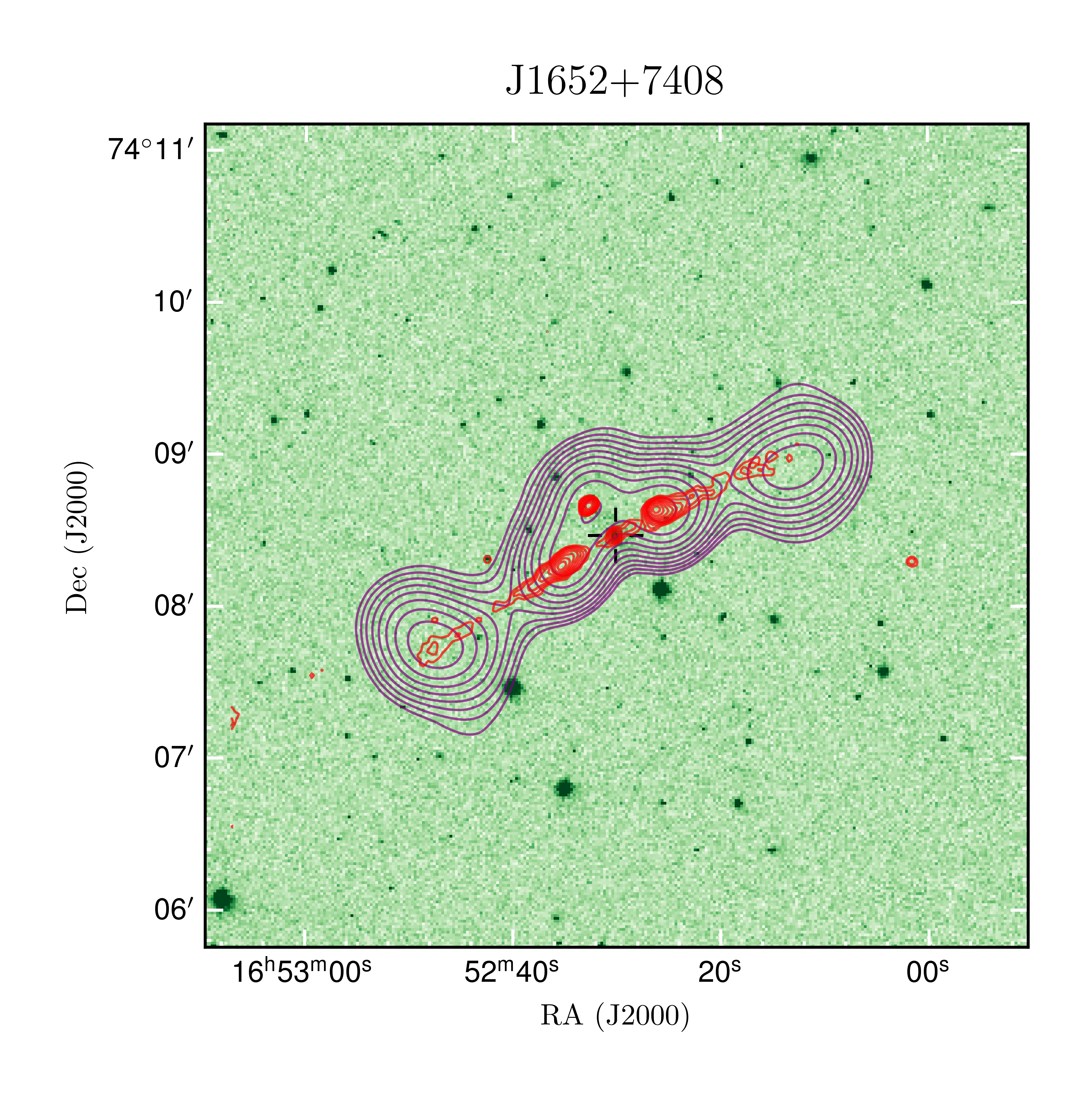}
\includegraphics[height=3.5cm,width=3.5cm,angle=0]{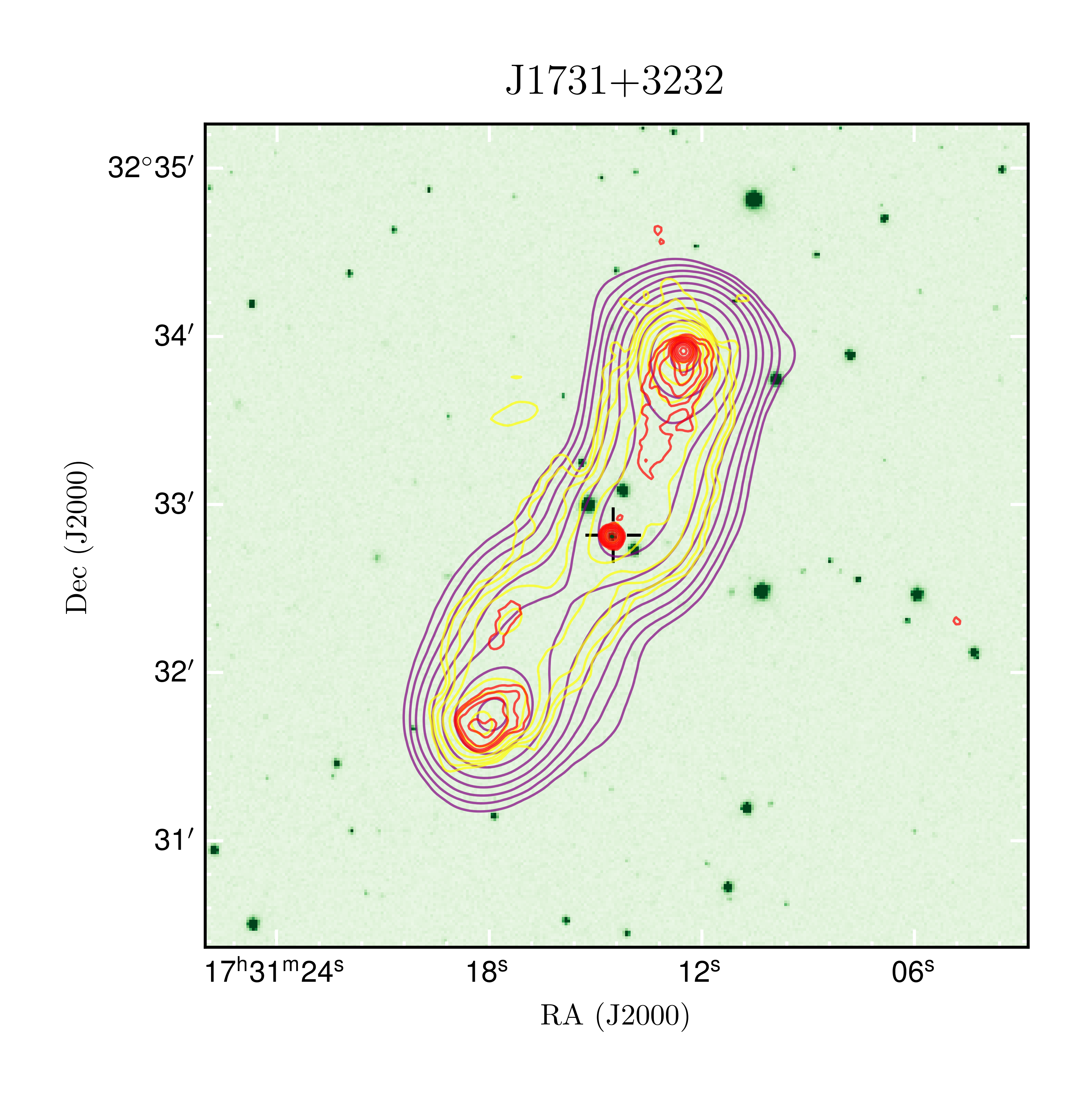}
\includegraphics[height=3.5cm,width=3.5cm,angle=0]{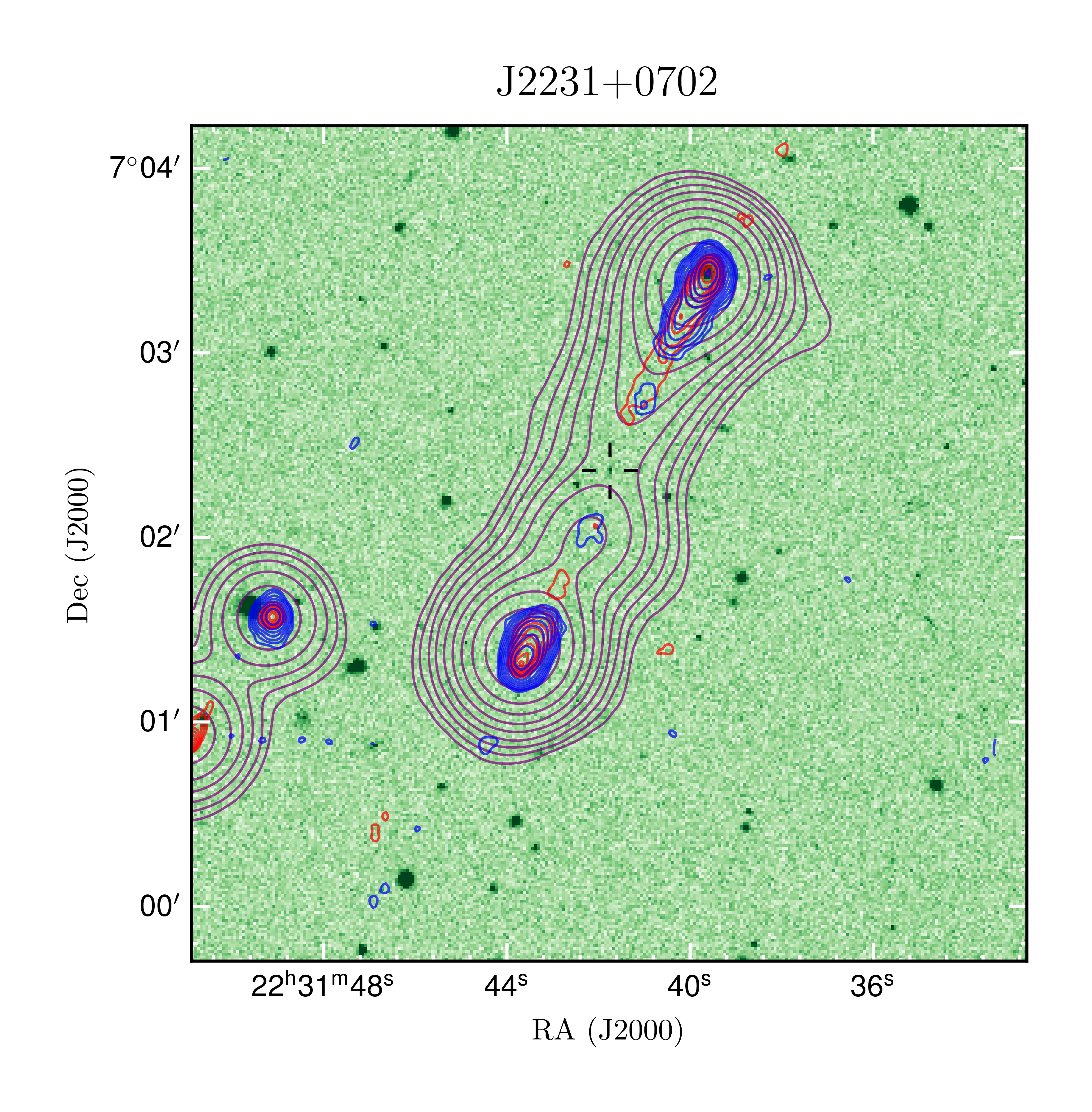}
\includegraphics[height=3.5cm,width=3.5cm,angle=0]{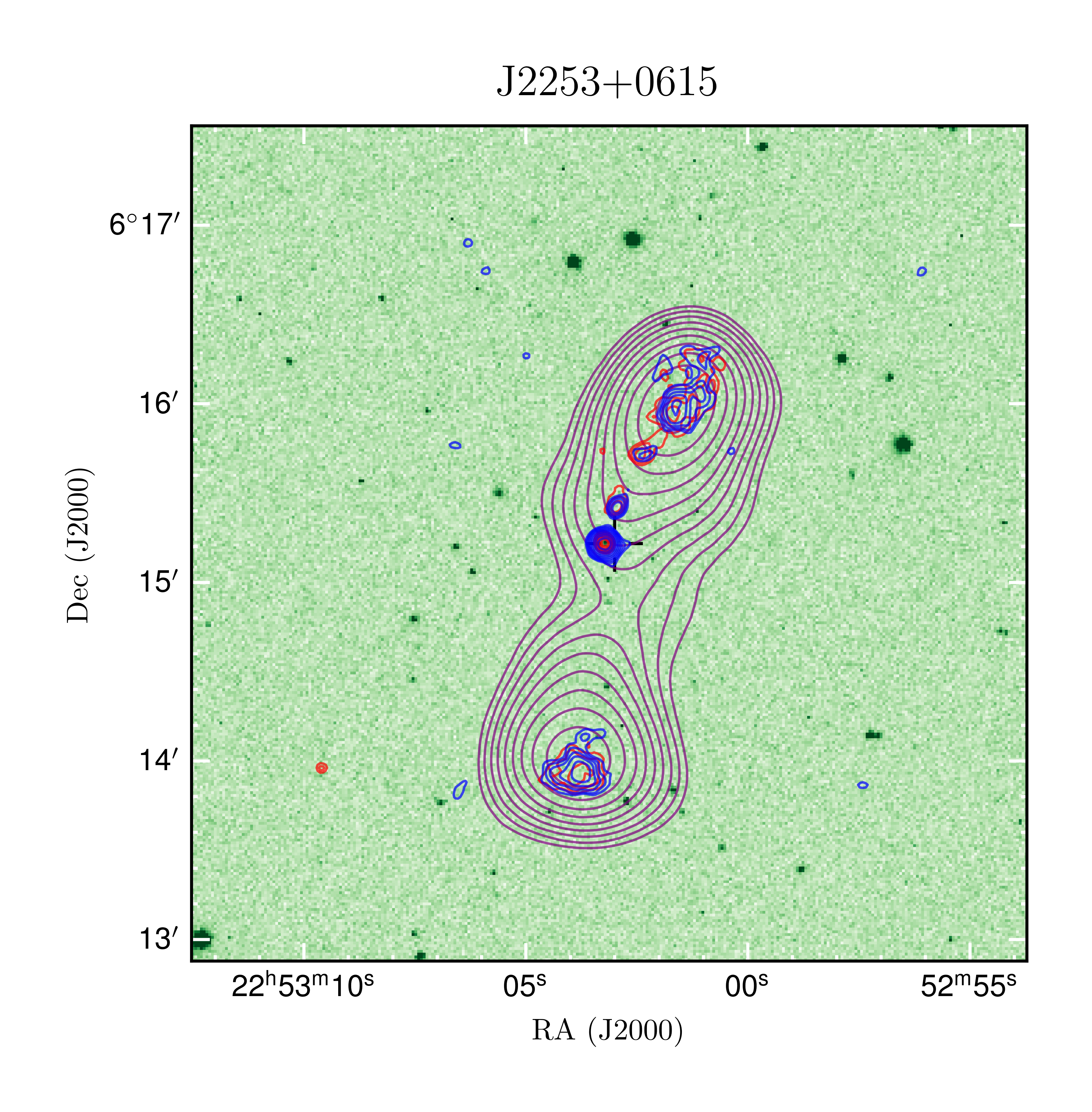}
\vskip 0.2cm
\includegraphics[height=3.5cm,width=3.5cm,angle=0]{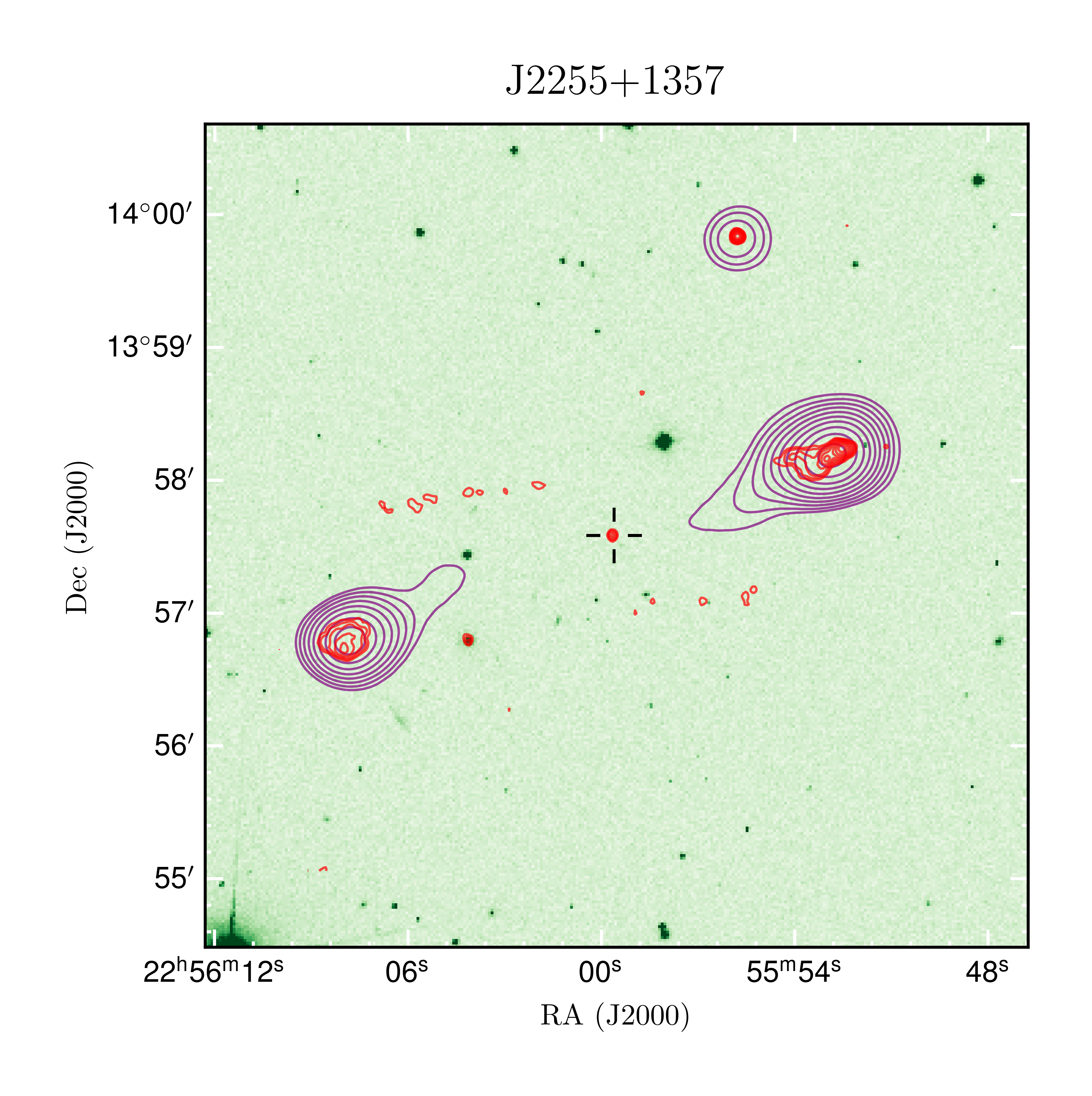}
\includegraphics[height=3.5cm,width=3.5cm,angle=0]{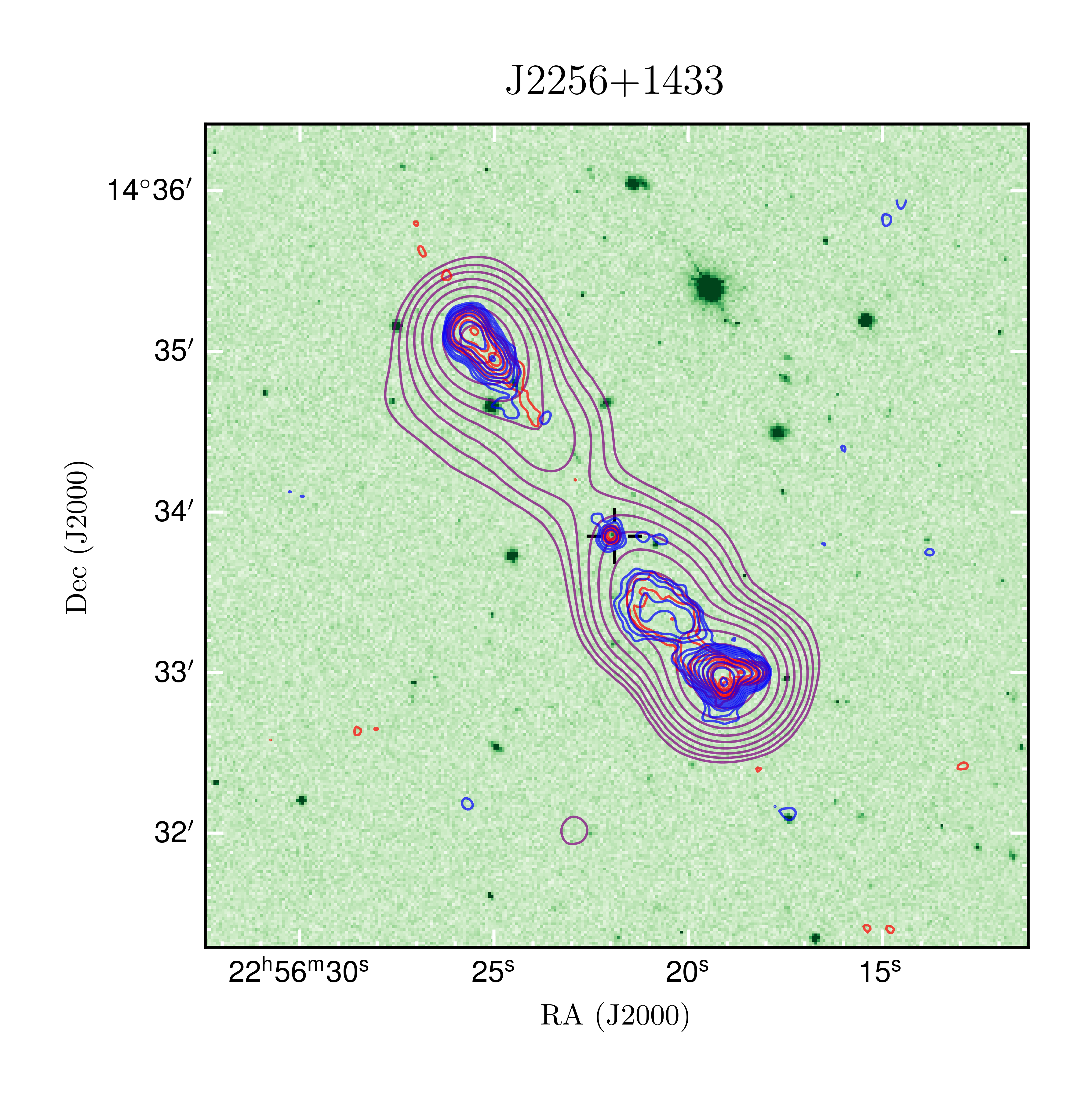}
\includegraphics[height=3.5cm,width=3.5cm,angle=0]{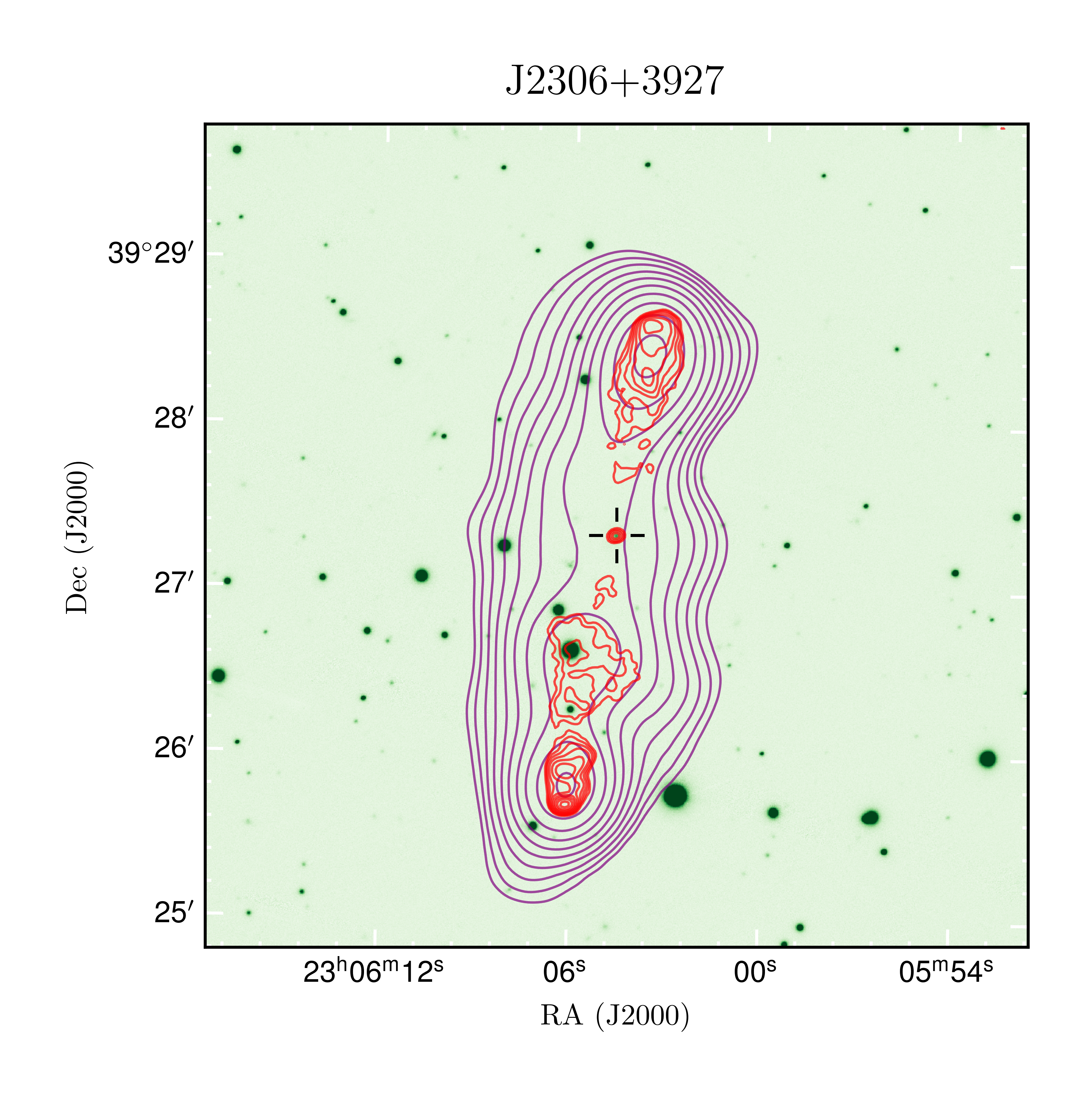}
\includegraphics[height=3.5cm,width=3.5cm,angle=0]{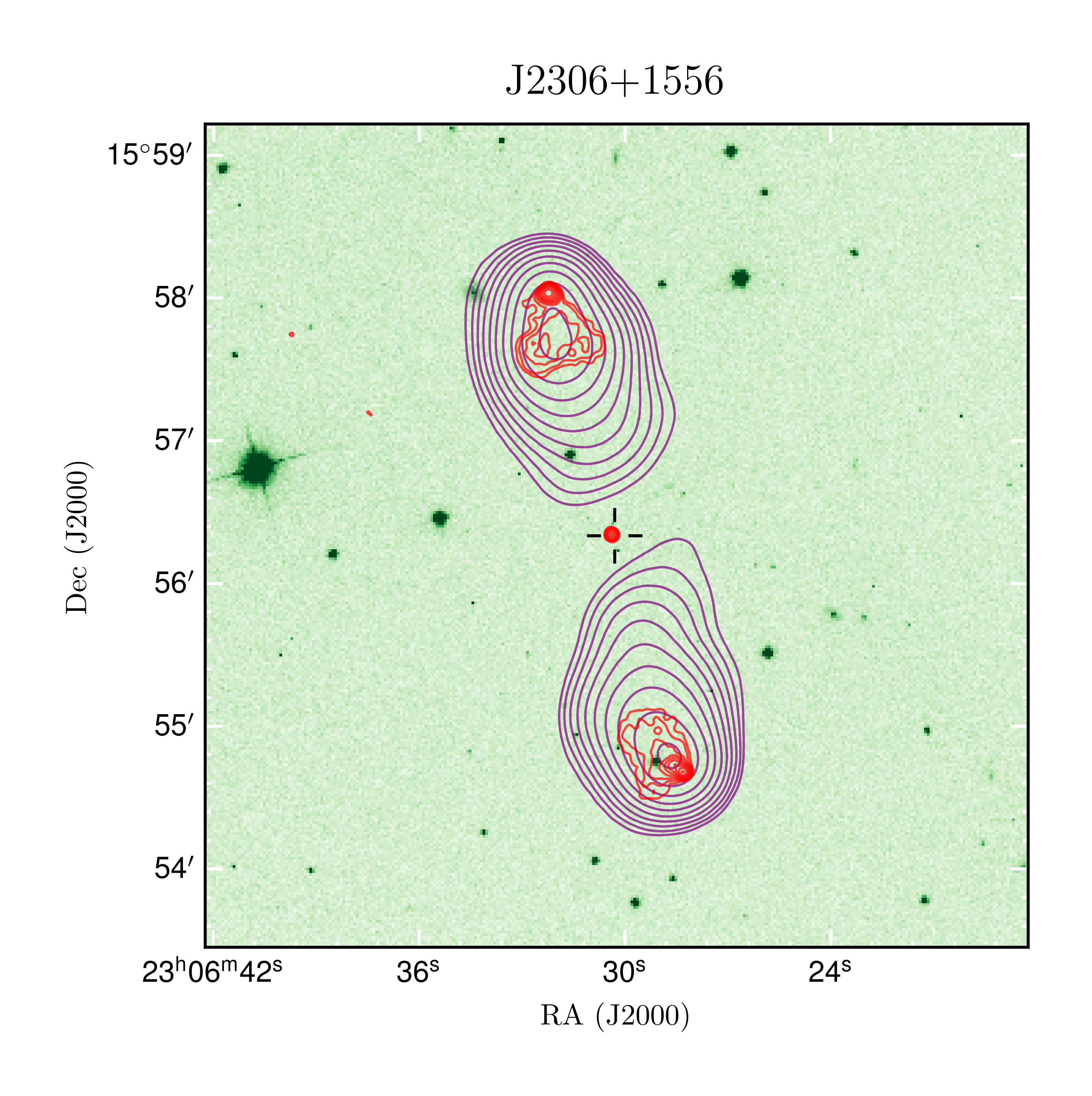}
\vskip 0.2cm
\caption{{\bf (Continued)}}
\label{Fig:GRS}
\end{figure*}

\end{document}